\documentclass[3p,10pt,unknownkeysallowed]{elsarticle}
\usepackage{ifpdf}
\usepackage{graphicx,amssymb,lineno,amsmath,epsfig}
\usepackage{natbib}

\begin{document}

\begin{frontmatter}



\title{Probing Magnetic Excitations and 
Correlations in Single and Coupled Spin Systems with Scanning Tunneling 
Spectroscopy}


\author[MP]{Markus Ternes\corref{cor1}}
\ead{m.ternes@fkf.mpg.de}

\cortext[cor1]{Corresponding author}

\address[MP]{Max-Planck Institute for Solid State Research, Heisenbergstr. 1, 
D-70569 Stuttgart, Germany}

\begin{abstract}

Spectroscopic measurements with low-temperature 
scanning tunneling microscopes have been used very successfully for studying 
not only individual atomic or molecular spins on surfaces but also complexly 
designed coupled systems. 
The symmetry breaking of 
the supporting surface induces 
magnetic anisotropy 
which 
lead to characteristic fingerprints in the spectrum of the 
differential conductance and can be well understood with simple model 
Hamiltonians. Furthermore, correlated many-particle states can 
emerge due to the interaction with itinerant electrons of the electrodes, making 
these systems ideal prototypical quantum systems. In this 
manuscript more complex bipartite and spin-chains will be discussed 
additionally. Their spectra enable to determine 
precisely the nature of the interactions between the spins 
which can lead to 
the formation of new quantum states which emerge by interatomic entanglement. 

\end{abstract}

\end{frontmatter}

\tableofcontents

\section{Introduction}

The transfer of electrons between metallic leads separated by vacuum or 
an insulating gap is classically forbidden. However, it becomes possible if the 
size of the gap is reduced to the scale of a few Angstroms due to the tunneling 
effect. This entirely quantum mechanical effect, which allows the electrons to 
cross the forbidden region, was already discussed in the early days of quantum 
mechanics \cite{Frenkel30} and first observed in the 1960s on planar 
superconducting -- oxide -- normal conducting junctions  \cite{Giaver60, 
Giaver60a}.

\begin{figure}[tbp]
\centering
\includegraphics[width=0.6\textwidth]{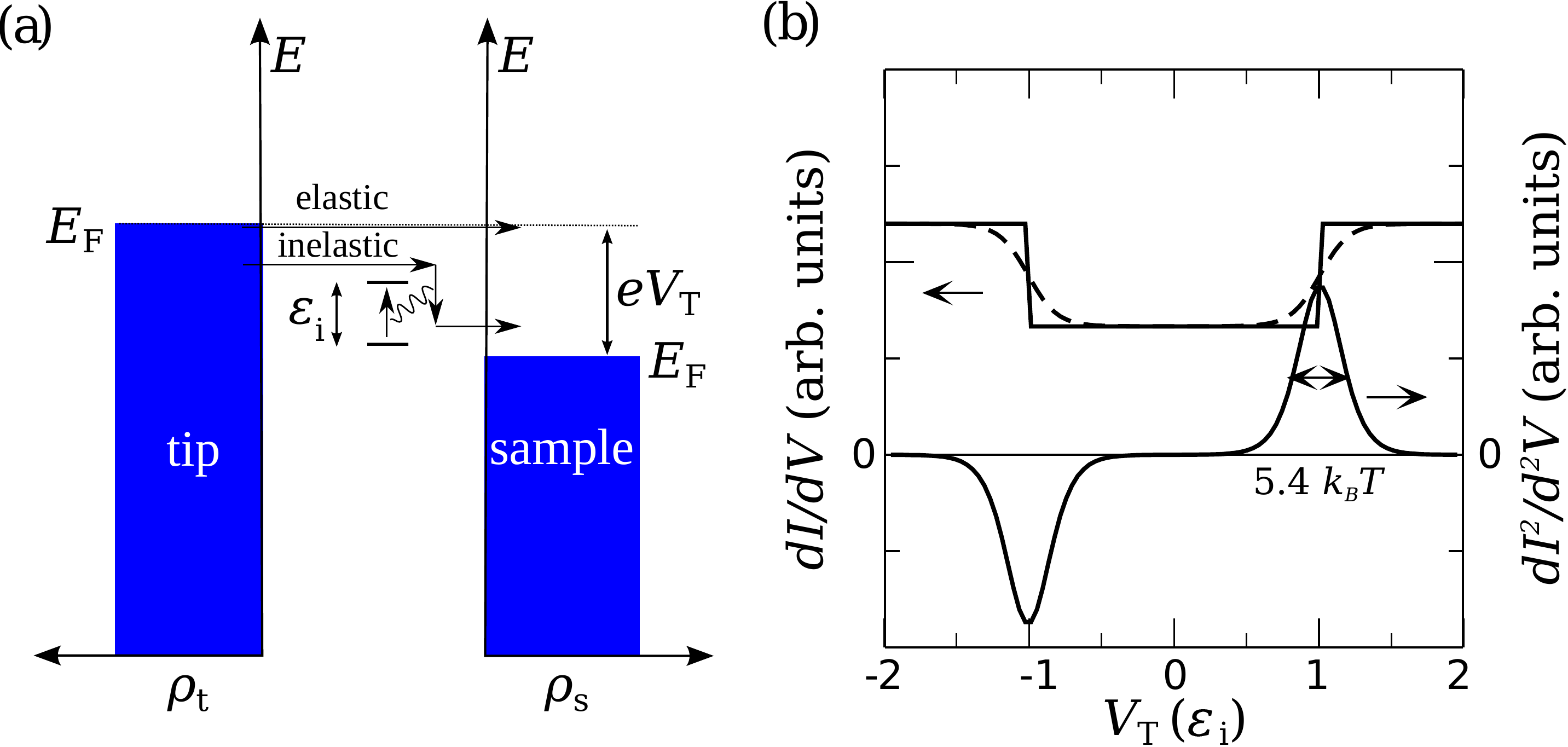}
	\caption{\textbf{(a)} Schematic view of the tunneling process between 
the electrodes of an idealized tip and sample with their constant densities of 
states $\rho_t$ and $\rho_s$, respectively: In 
addition to the elastic tunneling current, an inelastic channel may 
open at a bias $|eV_T|\geq \varepsilon_i$ with $\varepsilon_i$ as the energy 
difference between the ground and an excited internal state. Under this 
circumstances an electron crossing the barrier can change its 
quantum state by losing part of its energy and exciting an internal 
degree of freedom, for example, the spin orientation of an magnetic adsorbate on 
the surface. \textbf{(b)} Schematic differential 
conductance, $dI/dV$ (upper curve) and  $dI^2/d^2V$ (lower curve), spectra of 
an inelastic tunneling process. Symmetrically around $E_F$ a step like 
structure at a bias voltage $|eV_T|=\varepsilon_i$ 
is detected in the $dI/dV$ curve. This is smeared out at non-zero temperature 
(dashed line) leading to peaks with a width of 
$5.4 k_BT$ in the $dI^2/d^2V$ curve. Figure adapted from reference 
\cite{Ternes06}.}
	\label{fig:iets}
\end{figure}

In general, the electron transport in such tunnel junctions can be divided 
into two distinct classes: Elastic tunneling in which an applied 
bias drives electrons from the many states of one electrode to cross the 
junction without interaction with the local environment and inelastic 
tunneling, where the electrons interact with the junction 
environment and change their energy, phase, or angular momentum. 
These inelastic processes leave characteristic 
fingerprints in bias dependent conductance measurements. In particular, 
when discrete states are excited during tunneling and the tunneling 
electron looses partly its kinetic energy, a bias threshold voltage can be 
observed below which the inelastic process cannot occur (Figure~\ref{fig:iets}).
In this sense inelastic tunneling spectroscopy (IETS) is complementary 
to far-field methods such as high resolution electron energy-loss (HREELS), 
infrared reflection adsorption (IRRAS), Raman spectroscopy, or inelastic 
neutron spectroscopy \cite{Colthup75, Hippert06} 
and has been performed in planar tunnel junctions to detect vibrational 
excitation modes of molecules embedded in the junction for almost 50 years 
\cite{Jaklevic66, Klein68, Rowell68, Klein73, Hansma78, Weinberg78}. 
Remarkably, 
besides molecular vibrations, excitations due to the interaction of electrons 
with localized magnetic impurities have already been studied in these 
pioneering 
days leading to the discovery of anomalies in the density of states very close 
to the Fermi energy, which was shown to be due to the formation of a Kondo 
state and inelastic spin-flip excitations \cite{Wyatt64, Wolf70, Wyatt73, 
Wyatt73a, Bermon78}.  

While planar tunnel junctions have the disadvantage of averaging over an
ill-defined contact area, the development of the scanning tunneling microscope 
(STM) by Binning, Rohrer, Gerber, and Weibl in the early 1980s 
\cite{Binnig82, Binnig82a, Binnig83} opened a entirely new world for 
tunneling experiments. Very soon it became clear that the STM, with its 
capability to atomically resolve metallic and semiconducting surfaces, would
become a powerful tool for the analysis of surfaces and nanoscale structures 
down to the single molecule or atom level. Its discovery was 
awarded with the Nobel prize in a surprisingly short time of only 4 years 
after its first successful demonstration.

In the early days of STM, collective vibrational excitations at the 
surface of graphite were detected \cite{Smith86}, however, it was 
clear that the true capability of the STM would lie in combining its 
spectroscopic possibilities with its inherent atomic resolution. 
Nevertheless, it took about 15 years of technology development before
the mechanical stability and electronic sensibility of the STM at 
cryogenic temperatures and in ultrahigh vacuum was high enough to make this 
dream reality. 

The year 1998 brought two important experimental breakthroughs: The group 
around 
Wilson Ho at the Cornell University showed for the first time that IETS was 
possible on the single molecular level using the spatial resolution of the STM 
\cite{Stipe98}. In their experiment they detected the vibrational excitations 
of 
an isolated acetylene (C$_2$H$_2$) molecule adsorbed on a Cu(100) surface. The 
detection of mechanical excitation in molecular systems, has since been
applied to many quite different molecular systems ranging from diatomics like 
carbon monoxide \cite{Lauhon99, Heinrich02}, metal hydride molecules 
\cite{Pivetta07}, and molecular hydrogen \cite{Natterer13, Li13a, Natterer14} to 
complex molecules like porphyrins \cite{Wallis00} and C$_{60}$ bucky balls 
\cite{Franke12}. 
However, importantly for the work discussed in this manuscript, inelastic 
excitations can also be observed on individual spin systems as discovered by 
Andreas Heinrich and co-workers at IBM Almaden first on Mn atoms adsorbed on 
patches of 
Al$_2$O$_3$ on a NiAl surface \cite{Heinrich04}. As we will discuss
in the following, spin excitation spectroscopy gives unparalleled access to the 
quantum nature of individual and coupled spin systems enabling the 
determination and manipulation of their spin states, their magnetic anisotropy, 
and their coupling with the environment; properties which are actually 
mutually interdependent.  

The second breakthrough was the detection of the spectroscopic signature of the 
correlated many-particle Kondo state of individual magnetic atoms adsorbed on 
non-magnetic metal substrates. This discovery was made almost simultaneously by
the group around Wolf-Dieter Schneider at the University Lausanne 
\cite{Li98a}\footnote{Note, that due to 
newer measurements which revealed that single Ce adatoms on Ag(111) are even
at 5~K still very mobile \cite{Silly04, Silly04a, Ternes04, Ternes10}, the 
original publications was presumably measured on a small Ce cluster, which can 
indeed show a Kondo state \cite{Ternes09}.} and the group around 
Michael Crommie at the 
University of Boston \cite{Madhavan98}. 

\begin{figure}[tbp]
\centering
\includegraphics[width=0.7\textwidth]{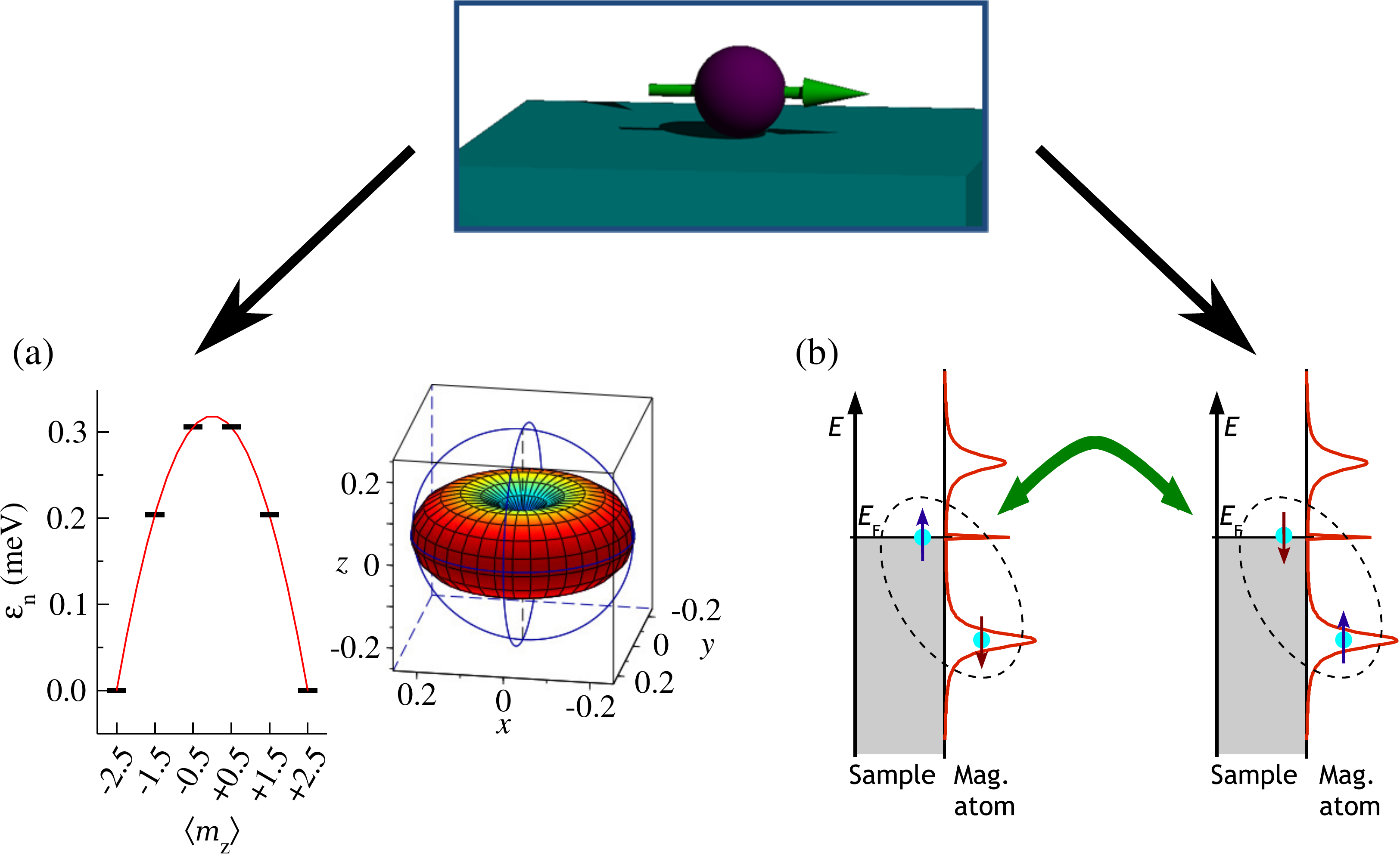}
	\caption{Schematic view of the emergent physical phenomena upon 
absorption of a paramagnetic spin onto a surface. \textbf{(a)} The surface 
breaks the symmetry and leads to magnetic anisotropy in which the spin prefers 
distinct alignments. As an example the energy level diagram for a Mn adatom on 
Cu$_2$N with axial anisotropy of $D=-40\mu$eV is shown (left). Here, the 
projected magnetic moment $m_z$ along the 
hard axis is a good quantum number allowing us to distinguish the 6 
eigenstates of the effective spin $S=5/2$ system. The visualization of the 
magnetic anisotropy 
shows the magnetic easy axis ($z$-axis) and the $x-y$ hard plane (right, scale 
in meV) \cite{Shick09, Ternes15}. \textbf{(b)} Degenerate ground states, 
as for example in spin $S=1/2$ systems (illustrated as singly occupied 
level below $E_F$), can interact with the electrons of the sample 
leading to the formation of a new singlet state in which the localized magnetic 
moment is screened by the many-electrons of the bath. This Kondo state leaves a 
characteristic signature in the quasiparticle excitation spectrum close to the 
Fermi energy $E_F$. Note that the processes (a) and (b) can compete against 
each other. Figure adapted from 
references \cite{Ternes09} and \cite{Ternes15}.}
	\label{fig:spin}
\end{figure}

These two hallmarking observations are the foundation on which 
the work presented here is based. Both rely on the interaction of individual 
spins, i.\,e.\ atoms or molecules which have a magnetic moment, with the local 
environment as is schematically illustrated in figure \ref{fig:spin}. 
A paramagnetic atom with a total spin $S>1/2$ has $2S+1$ eigenstates which are 
indistinguishable in the gas phase when no external magnetic field is 
applied. Upon absorption on the surface this situation changes. When the atom is
physisorbed, the out-of-plane direction forms a distinct axis, different 
from all other directions. This symmetry breaking is the origin of magnetic 
anisotropy which lifts the degeneracy and defines the stability of a spin in a 
preferred direction \cite{Gatteschi03} (figure \ref{fig:spin}a). In the 
case the atom is chemisorbed onto the surface, that means it forms covalent 
chemical bonds and is rather incorporated into the surface, complex molecular 
networks might form, further reducing the system's symmetry. Additionally, 
spin-orbit coupling, charge transfer, and delocalized spin polarization in 
the substrate influences the effective magnetism of the atom \cite{Gatteschi08}.

The influence of the magnetic anisotropy on the tunneling spectra 
will be briefly discussed in section \ref{sec:Anisotropy}. Here, we will 
additionally see how the strength of the direct exchange coupling between the 
localized  magnetic moment and the itinerant electron bath of the substrate 
modifies the magnetic anisotropy via virtual coherences between the 
eigenstates \cite{Oberg13, Jacobson15}.

While the magnetic anisotropy removes the degeneracies of high spin systems, 
for half-integer spins Kramers theorem prevails the full lifting of all 
degeneracies \cite{Kramers26}. In these systems every energy level is at least 
doubly degenerate at zero field. The ground state degeneracy has, in particular 
for $S=1/2$ systems, dramatic consequences which leads to an entirely new 
area of physics in which correlations between the localized magnetic moment and 
the many electrons of the substrate form, at low enough temperature, a new 
singlet ground state creating a prominent resonance at the Fermi energy (see 
figure \ref{fig:spin}b) \cite{Kondo64, Hewson97}.
Section \ref{sec:Kondo} will discuss this Kondo effect in detail. 
Starting with the temperature and magnetic field dependence in the weak 
coupling limit which was first measured on an organic radical \cite{Zhang13} 
we will elaborate a perturbative scattering model up to 3rd order in the 
exchange interaction and show under which circumstances the model breaks down 
and other, more sophisticated models have to be used \cite{Ternes15}. Here, 
individual Co atoms on different substrates can act as examples for this strong 
coupling regime \cite{Otte09}.  
Intriguingly, Co atoms on a Cu$_2$N substrate possesses a spin of $S=3/2$ and 
are thereby also influenced by the magnetic anisotropy leading to a 
directionally 
dependent magnetic field behavior \cite{Otte08a}. Additionally, on 
this system the spin polarization of the split 
Kondo state was determined \cite{Bergmann15}.

After the discussion of the emergence of anisotropy and correlations in single 
spin systems, section \ref{sec:Coupled_seystems} will discuss coupled 
systems containing more than one paramagnetic spin center. Starting with the 
prototypical molecular magnet Mn$_{12}$Acetate$_{16}$ which has a total 
ground state spin of $S_T=10$ we will observe that $S_T$ is not a conserved 
quantity and that spin excitations can change the total spin $S_T$ leaving 
characteristic fingerprints in the IETS spectrum \cite{Kahle12}. Afterward, 
spin dimer systems will be inspected with a particular focus on the 
description in the perturbative transport model \cite{Ternes15, Spinelli15} 
and the appearance of different quantum phases in the two-impurity Kondo 
system. Last, coupled spin chains are discussed which are of particular interest 
due to the emergence of entanglement that can be directly observed in the 
zero-energy peak \cite{Choi15}. Finally, section \ref{sec:summary} summarizes 
the manuscript and outlines possible routes for future research.





\section{The magnetic anisotropy in single spins}
\label{sec:Anisotropy}

Magnetic anisotropy defines the preferred directions in which the magnetic 
moment of a spin of strength $S$ likes to be aligned and is crucial 
for the lifetime in which the direction of magnetization is maintained 
\cite{Gatteschi03, Gatteschi08, Jacob16}. While a spin carrying atom in gas 
phase can not have any magnetic anisotropy due to its spherical symmetry, the 
situation changes when the spin is embedded into a crystal structure. The 
crystal field and the spin-orbit coupling lead to a lifting of the degeneracies. 
In general the crystal field spin Hamiltonian $\hat{H}_{\rm ani}$ can be 
perturbatively constructed from an 
infinite series of quadratic, cubic, etc.\ spin operators, which are usually 
expressed in Stevens operators $\hat{O}_N^k$ to easily connect to point group 
symmetries \cite{Rudowicz04}:
\begin{equation}
\hat{H}_{\rm ani}= \sum_{\substack{N=2,4,\ldots,2S\\ -N\leq k\leq N}} B_N^k 
\hat{O}_N^k,
\label{equ:Stevens}
\end{equation}
with $B_N^k$ as the parameters. As we will see, it is often a good 
approximation 
to assume the crystal 
field spin Hamiltonian to be only quadratic in the spin operators with a 
symmetric coupling matrix~{\bf D}:
\begin{equation}
\hat{H}_{\rm ani}=\hat{\bf S}\cdot{\bf D}\cdot\hat{\bf S}\equiv
D\hat{S}_z^2+E\left( \hat{S}_x^2-\hat{S}_y^2
\right),
\label{equ:SDS}
\end{equation}
where the scalar parameter $D$ determines the axial and $E$ the
transverse anisotropy, and $\hat{\bf S}=(\hat{S}_x, \hat{S}_y, \hat{S}_z)^T$ is 
the total spin operator.

For adatoms on surfaces, a low coordination number and 
changes in hybridization can lead to a dramatic enhancement of magnetic 
anisotropy \cite{Gambardella03, Miyamachi13, Heinrich13, Rau14} which 
have been shown to reach values of up to $\approx60$~meV for Co adatoms on MgO 
\cite{Rau14}. 
Additionally, different surface adsorption sites or the bonding to hydrogen 
alter the magnetic anisotropy or even the total spin 
of the adatoms 
\cite{Khajetoorians13a, Dubout15, Donati13, Jacobson15, 
Khajetoorians15, Jacobson16}. Furthermore, the exchange interaction with the 
substrate can affect the observed magnetic anisotropy as it has been found for 
3d metal adatoms on Cu$_2$N islands on Cu(100) \cite{Oberg13, Bryant13, 
Delgado14} or metal-hydrates on $h$-BN on Rh(111) \cite{Jacobson15}. For single 
molecule magnets containing $3d$ or $4f$ spin centers it is well known that 
chemical changes to the ligands surrounding the spin can affect the magnetic 
anisotropy \cite{Jurca11, Wegner09, Parks10}. However, the most important factor 
for creating and maintaining magnetic anisotropy in single molecule magnets 
remains a low coordination number and a high axial symmetry \cite{Rau14, 
Zadrozny13, Ungur14}. Under such conditions the symmetry-protected 
magnetic ground state of single Ho-atoms adsorbed on a double-layer of 
MgO on Ag(100) can lead to relaxation times of 1~h at a temperature of 2.5~K as 
recently observed in magnetic circular dichroism measurements \cite{Donati16}.



In this section we will discuss the use of scanning 
tunneling spectroscopy to measure IETS on individual magnetic spin systems. 
Such 
measurements enable the determination of the total spin as well as 
the orientation and strength of the magnetic anisotropies. 
Additionally, we will show that a perturbative scattering model \cite{Ternes15} 
can accurately reproduce the experimental observations enabling us to precisely 
measure the coupling to the underlying substrate. 
On hydrogenated metal complexes not only energetically low vibrational modes 
have been found \cite{Pivetta07, Hofer08} but also a wide range of magnetic 
excitations have been detected \cite{Dubout15, Donati13, Jacobson15, 
Khajetoorians15}. Therefore, we will focus on CoH$_x$ complexes coupled to the 
spatially varying template $h$-BN/Rh(111) moir\'{e} as an example of 
hydrogenated metal complexes. 

\subsection{Modifying the spin state and anisotropy in CoH$_x$ complexes}

\label{sec:CoH}

CoH$_x$ ($x=0-3$) complexes on the $h$-BN/Rh(111) moir\'{e} form when Co atoms 
from a metallic rod are deposited by an e-beam evaporator onto the cold 
($T\approx 30$~K)  surface together with residual hydrogen from the background 
vacuum \cite{Jacobson15}.
The $h$-BN monolayer, a two dimensional 
material with a wide band gap, decouples 
the CoH$_x$ from the underlying Rh(111) metal while the lattice mismatch 
leads to a spatial corrugation resulting in an enlarged supercell with 
$3.2$~nm periodicity corresponding to $13$ BN units on top of $12$ Rh atoms 
\cite{Herden14, Laskowski07}. 

\begin{figure}[tbp]
\centering
\includegraphics[width=0.7\textwidth]{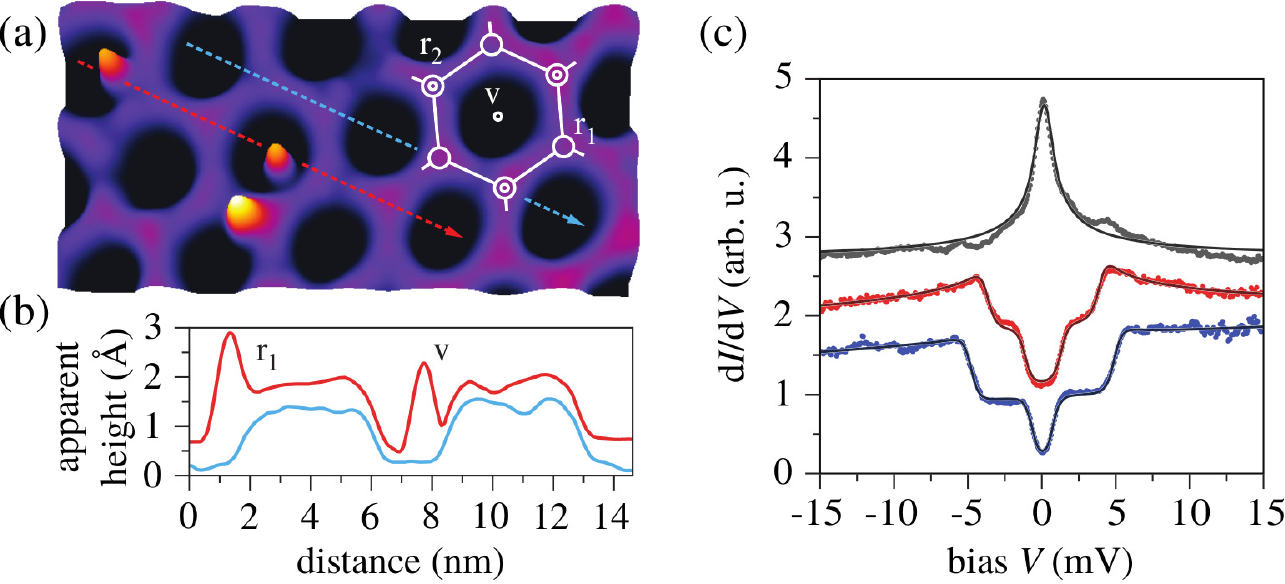}
	\caption{Cobalt hydrates adsorbed on a $h$-BN/Rh(111) surface.
\textbf{(a)} Constant current STM topography with three CoH$_x$ complexes 
(protrusions) adsorbed on different sites of the highly corrugated surface. 
(image size $15\times 4$~nm$^2$, $V=100$~mV, $I = 20$~pA, $T = 1.4$~ K). High 
symmetry points of the moir\'{e} are marked by the white overlay.
\textbf{(b)} Line profiles along the dashed lines indicated in (a) show two 
CoH$_x$ systems with adsorption sites r$_1$ and v (red line) and a $h$-BN 
reference cut (blue line), offset by 0.5 \AA. \textbf{(c)} Differential 
conductance $dI/dV$ curves measured on top of three different CoH$_x$ systems 
(stabilization setpoint: $I = 500$~pA, $V = -15$~mV, $T = 1.4$~K,
curves vertically offset for clarity). The upper curve (grey)
shows a spin $1/2$ Kondo resonance (see section \ref{sec:Kondo}) centered at 
zero bias. The two lower curves (red and blue) show step-like conductance
increases symmetric around zero bias indicating a $S=1$ system.
Solid black lines are least-square fits using the perturbative
transport model. Figure adapted from 
reference \cite{Jacobson15}.}
\label{fig:CoH-topo}
\end{figure}

Figure \ref{fig:CoH-topo}a shows a typical STM constant-current 
topography of the $h$-BN/Rh(111) moir\'{e} with isolated CoH$_x$ ($x=1,2$) 
complexes imaged as protrusions. Line profiles 
indicate that CoH$_x$ can adsorb at multiple positions within the moir\'{e} 
(Figure \ref{fig:CoH-topo}b) \cite{Natterer12, Jacobson15}. 
The differential conductance spectra, $dI/dV$, measured at low-temperature ($T 
= 1.4$~K) and zero magnetic field ($B = 0$~T) on these CoH$_x$ complexes can be 
divided into two broad classes: a sharp peak centered at zero bias or two 
symmetric steps of increasing conductance at well-defined threshold energies 
(Figure \ref{fig:CoH-topo}c). The peak at zero bias is consistent with a 
spin $S=1/2$ Kondo resonance which will be discussed in detail in section 
\ref{sec:Kondo} while the steps correspond to the onset of inelastic 
excitations from the 
magnetic ground state to excited states. The observation of two steps hints 
towards an effective $S=1$ system with zero-field splitting. The two lower 
spectra 
(Figure \ref{fig:CoH-topo}c, red and blue curves) are measured on CoH at 
different parts of the moir\'{e} and share the same overall characteristics but 
the step positions vary.

Employing density functional theory (DFT) performed by Oleg Brovko and 
Valerie Stepanyuk from the MPI Halle enables us to correlate the magnetic 
properties of the CoH$_x$ with 
the local adsorption configuration \cite{Jacobson15}. The calculations show 
that adsorption in 
the BN hexagon, i.\,e.\ hollow site, is preferable for bare Co leading to a 
magnetic moment of $2.2$ Bohr magnetons ($\mu_B$).
The addition of a hydrogen atom shifts the preferred adsorption site to N and 
concomitantly lowers the magnetic moment to $2.0 \mu_B$, equivalent to a 
$3d^8$ configuration. The second hydrogen changes the picture significantly, 
with the $sp-d$ hybridization sufficient to bring the 
Co $d$-levels closer together, reducing the magnetic moment to $1.2 \mu_B$ 
resulting in a $3d^9$ configuration \cite{Jacobson15}. 

\begin{figure}[tbp]
\centering
\includegraphics[width=0.7\textwidth]{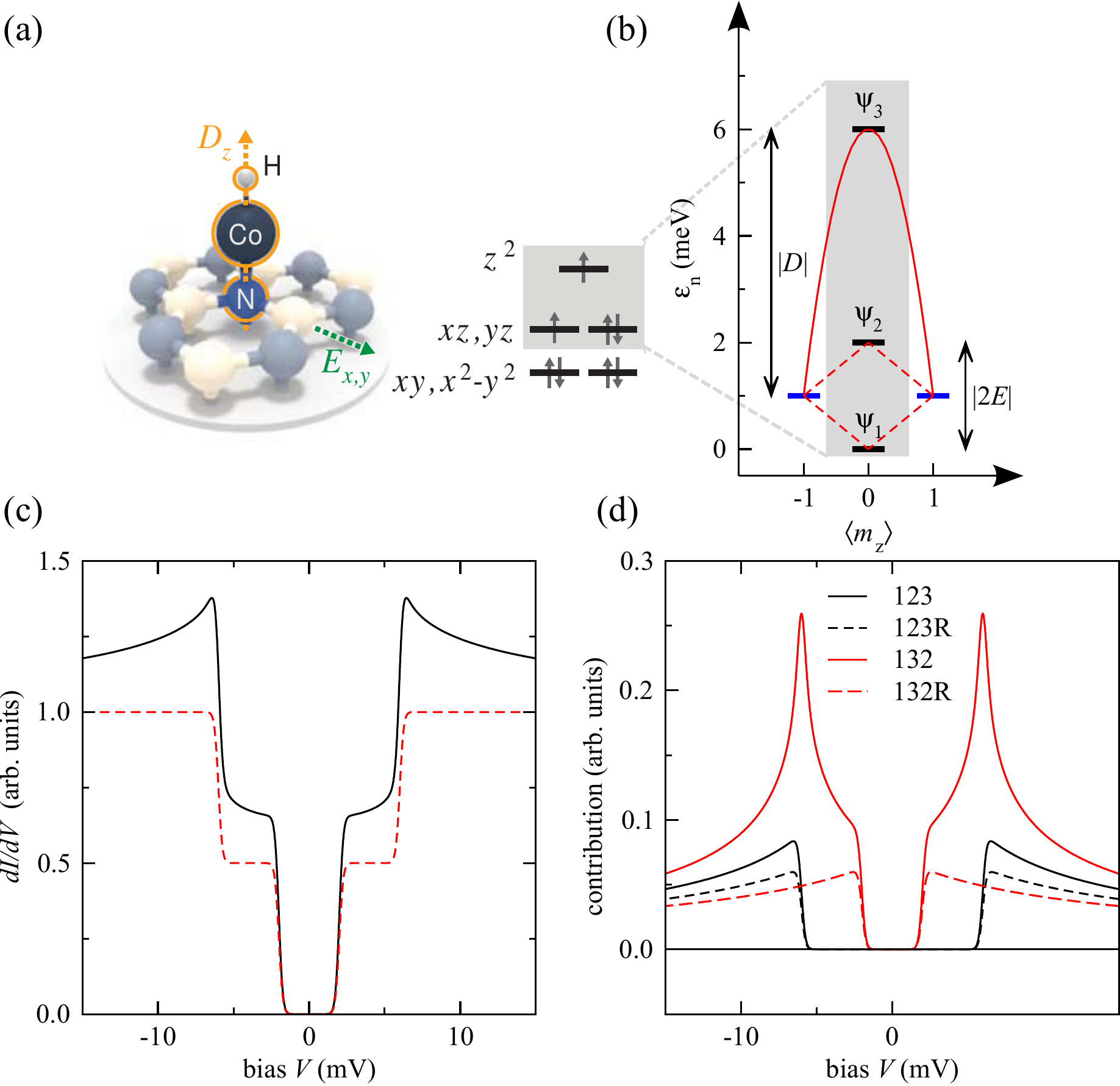}
\caption{The magnetic properties and the tunneling spectra of CoH. \textbf{(a)} 
Ball and stick model of the adsorption of CoH on $h$-BN. The linear 
adsorption geometry of CoH on the N atom is emphasized and marks the main 
(axial) magnetic anisotropy ($D$) along the $z$-axis. \textbf{(b)} Additional 
transverse anisotropy ($E$) in the $x-y$ plane lifts all degeneracies 
at zero field. The plot shows the state energy over the $m_z$ expectation value 
$\langle m_z \rangle = {\rm tr}\left( \hat{S}_z \left|\Psi_i 
\right\rangle\left\langle \Psi_i\right|\right)$ with $D=-5$~meV and
$E=1$~meV as an example. \textbf{(c)} The total 
spectrum with $J\rho_0 = 0$ (dashed red line) and  $-0.1$ (full black line) 
calculated for $T = 1$~K, $B = 0$~T.
\textbf{(d)} The significant higher order processes where the digits refer 
to the visited eigenstates during the scattering process. Figure adapted from 
reference \cite{Jacobson15}.}
\label{fig:Spin1}
\end{figure}

An important consequence of the N adsorption site is the linear crystal field 
acting on the cobalt (i.\,e.\ N--Co--H) removing the 5-fold degeneracy of the 
$d$-levels. 
From these results and the spectroscopic 
observations we can identify CoH as an effective $S=1$ and CoH$_2$ as an 
$S=1/2$ Kondo system.
The strong vertical bond between Co and N can be expected to provide the system 
with an out-of-plane magnetic anisotropy (Figure \ref{fig:Spin1}a). While the 
hydrogen is not rigidly pinned to the cobalt, tilting of the hydrogen 
combined with the underlying lattice mismatch reduces the $C_{3v}$ symmetry and 
introduces small shifts in the $d_{xz}$, $d_{yz}$ levels producing a 
non-negligible in-plane component of the anisotropy lifting all 
degeneracies of the spin system (Figure \ref{fig:Spin1}b).

To analyze the experimental data we use an impurity Hamiltonian which includes 
the Zeeman energy and the magnetic anisotropy (equation \ref{equ:SDS}), and 
which is sufficient to fully explain the spectroscopic features observed in our 
scanning tunneling spectroscopy measurements \cite{Hirjibehedin07, Otte08a,
Lorente09, Oberg13, Bryant13, Ternes15}:
\begin{equation}
\hat{H}_{\rm imp}=g\mu_B\vec{ B }\cdot\hat{\bf S}+
D\hat{S}_z^2+E\left( \hat{S}_x^2-\hat{S}_y^2
\right).
\label{eq:Atom-Hamilonian}
\end{equation}
In this equation is $g$ the gyromagnetic factor, $\mu_B$ the Bohr magneton,
$\vec{ B}$ is the external applied magnetic field and $D$, $E$, and $\hat{\bf 
S}$ the axial and transverse anisotropy, and the total spin operator for 
the $S=1$ spin with the components 
($\hbar=1$):
\begin{equation}
\hat{S}_x=\left(\begin{matrix}
  0 & \frac{1}{\sqrt{2}} & 0 \\
  \frac{1}{\sqrt{2}}  & 0 & \frac{1}{\sqrt{2}}  \\
  0 & \frac{1}{\sqrt{2}}  & 0
 \end{matrix}\right),\quad
\hat{S}_y=\left(\begin{matrix}
  0 & \frac{-i}{\sqrt{2}} & 0 \\
  \frac{i}{\sqrt{2}}  & 0 & \frac{-i}{\sqrt{2}}  \\
  0 & \frac{i}{\sqrt{2}}  & 0
 \end{matrix}\right),\quad
\hat{S}_z=\left(\begin{matrix}
  {~}1 & {~}0 & {~}0 \\
  {~}0 & {~}0 & {~}0 \\
  {~}0 & {~}0 & -1
 \end{matrix}\right).
\end{equation}

In the absence of a magnetic field the three eigenvectors $|\Psi\rangle_i$ 
and eigenenergies $\varepsilon_i$ of equation \ref{eq:Atom-Hamilonian} are 
calculated in the $m_z$ basis to
%
\begin{align*}
\varepsilon_1&=0, & |\Psi_1\rangle&=+\frac{1}{\sqrt{2}}|+1\rangle-\frac{1}{
\sqrt{2}}|-1\rangle,\\
\varepsilon_2&=2E, & |\Psi_2\rangle&=+\frac{1}{\sqrt{2}}|+1\rangle+\frac{1}{
\sqrt{2}}|-1\rangle,\\
\varepsilon_3&=E-D, & |\Psi_3\rangle&=|0\rangle,
\end{align*}
%
as shown in figure \ref{fig:Spin1}b for hard axis anisotropy ($D<0$) and non 
negligible transverse anisotropy ($E\neq 0$).

To calculate the tunneling spectrum we use a model based on the perturbative 
approach first established by Appelbaum, Anderson, and Kondo \cite{Kondo64, 
Appelbaum66, Anderson66, Appelbaum67} in which spin-flip scattering processes 
up 
to the 2nd order Born approximation are accounted for. While we 
will outline the main components of this model, a more detailed approach can 
be found in \cite{Ternes15}.

In this model the transition probability $W_{i\rightarrow f}$
for an electron to tunnel between tip and sample and concomitantly change the 
spin state of the CoH complex from its initial ($i$) to its final ($f$) state is
\begin{equation}
 W_{i\rightarrow f}\propto\left( 
|M_{i\rightarrow
f}|^2+\rho_0J\sum_{m} \left( \frac{M_{i\rightarrow
m}M_{m\rightarrow
f}M_{f\rightarrow i}}{\varepsilon_i-\varepsilon_m} +\mbox{c. c.}\right) \right)
\delta(\varepsilon_i-\varepsilon_f),
\label{equ:W}
\end{equation}
with $M_{i\rightarrow j}$ as the matrix elements given by the Kondo-like 
interaction of the scattering electron with the wavevector $|\varphi\rangle$ 
with the localized spin of the CoH complex
\begin{equation}
M_{i\rightarrow f}=\sum_{i',f'}\left\langle \varphi_{f'}, \Psi_f \right| 
\frac12\hat{\boldsymbol{\sigma}}\cdot\hat{\bf S} |\varphi_{i'}, \Psi_i \rangle.
\label{equ:M}
\end{equation}
In this equation $|\varphi_i, \Psi_i \rangle$ is the combined state vector of 
the localized $S=1$ spin and the interaction electron. 
$\hat{\boldsymbol{\sigma}}=(\hat{\sigma}_x, \hat{\sigma}_y, \hat{\sigma}_z)^T$ 
is the total spin operator for the spin-1/2 electrons, with 
$\hat{\sigma}_{x,y,z}$ as the standard Pauli matrices. 

The first term in equation \ref{equ:W} is responsible for the conductance steps 
observed in our spectra. When we assume zero magnetic field and no 
spin-polarization in the two electron reservoirs of tip and sample, the matrix 
elements are easily calculated to 
$ |M_{i\rightarrow j}|^2= 0.5$ for $i\neq j$ and $|M_{i\rightarrow i}|^2=0$ 
otherwise. This leads, at low temperature, i.\,e.\ $k_BT\ll\varepsilon_2$, when 
only the ground state $|\Psi_1\rangle$ is significantly occupied, to two,  
increasing steps in the differential conductance $dI/dV$ with identical 
amplitude at the energies $\pm\varepsilon_2$ and $\pm\varepsilon_3$ (red dashed 
line in Figure~\ref{fig:Spin1}c):
\begin{equation}
\sigma_1(eV)=\frac{1}{2}\sigma_0\left[
\Theta\left(\tfrac{\varepsilon_2+eV}{k_BT}\right)+\Theta\left(\tfrac{
\varepsilon_2-eV}
{k_BT } 
\right) +\Theta\left(\tfrac{\varepsilon_3+eV}{k_BT}\right)+\Theta\left(\tfrac{
\varepsilon_3-eV}
{k_BT } \right)\right],
\label{equ:sigma1}
\end{equation}
with $\Theta(\varepsilon)=\left[1+(\varepsilon-1)\exp(\varepsilon)\right]\left[
1-\exp(\varepsilon)\right]^{-2}$ \label{p:theta}
as the thermally broadened step function \cite{Lambe68}, and $\sigma_0$ as the 
total conductance in the limit of high bias.

The second term of equation \ref{equ:W} is due to the 2nd order 
Born approximation and accounts for scattering processes involving an
intermediate state $|\Psi_m\rangle$. At the bias voltage where 
this process changes from being virtual to real, the denominator approaches 
zero which leads to a temperature broadened logarithmic divergence in 
the spectrum:
\begin{equation}
g(\varepsilon)=-\int_{-\infty}^{+\infty}d\varepsilon'' 
\int_{-\omega_0}^{+\omega_0}d\varepsilon' 
\frac{1-f(\varepsilon',T)}{\varepsilon'-\varepsilon'' }  
f'(\varepsilon''-\varepsilon,T),
\label{equ:g}
\end{equation}
with $f(\varepsilon, T) = [1 + \exp(\varepsilon/(k_BT))]^{-1}$ as the 
Fermi-Dirac 
distribution and $f'(\varepsilon, T) = \partial f/\partial\varepsilon =
(k_BT)^{-1}\times {\rm sech}^2[\varepsilon/(2k_BT)]$ as its derivation 
\cite{Wyatt73, Ternes15}. For the tunneling spectra the correct value of 
the cut-off 
energy $\omega_0$ is uncritical, but is of crucial importance for the 
energy renormalization, as we will see in the section \ref{sec:Bloch-Redfield}. 
The 
dimensionless scaling factor $-J\rho_0$ accounts for the fact that either the 
scattering into the intermediate or the final state is performed with electrons 
originating and ending in the substrate. Here, $J$ is the coupling strength 
between substrate electrons and the localized spin and $\rho_0$ the substrate 
electron density at $E_F$.

In the case discussed here, with an effective spin $S=1$ and all state 
degeneracies broken, the 
real parts of the matrix elements at zero field are 
calculated to $\Re(M_{i\rightarrow m}M_{m\rightarrow f}M_{f\rightarrow 
i})=-1/4$ for the processes which go over all states and otherwise zero. 
Assuming again that solely the ground state is thermally populated, only the 
processes 
$1\rightarrow2\rightarrow3$ and $1\rightarrow3\rightarrow2$ can account 
to the tunneling transport leading to an additional conductance of:
\begin{eqnarray}
\sigma_2(eV)=-\frac{1}{4}\sigma_0 J\rho_0\!\!\!\!\!\!\!\!& \left\{ 
\left[g(\varepsilon_2+eV)+g(\varepsilon_2-eV)\right]\left[
\Theta\left(\tfrac{\varepsilon_3+eV}{k_BT}
\right)+\Theta\left(\tfrac{
\varepsilon_3-eV}{k_BT } \right)\right]\right.\nonumber \\
& \left.  
+\left[g(\varepsilon_3+eV)+g(\varepsilon_3-eV)\right]\left[
\Theta\left(\tfrac{\varepsilon_2+eV}{k_BT}
\right)+\Theta\left(\tfrac{
\varepsilon_2-eV}{k_BT } \right)\right]\right\}.
\label{equ:sigma2}
\end{eqnarray}

Interestingly, the conductance $\sigma_2$ changes in a very particular fashion 
the observed spectra which is the sum of $\sigma_1$ and $\sigma_2$: Additional 
peak-like structures arise at the energy $\varepsilon_3$ due to the scattering 
processes via the states $1\rightarrow3\rightarrow2$ which allow us to 
determine $J\rho_0$ very precisely from fits of equations \ref{equ:sigma1} and 
\ref{equ:sigma2} to the spectra measured at zero field. However, the 
scattering processes $1\rightarrow2\rightarrow3$, which has at $B=0$ the same 
weight to equation \ref{equ:sigma2}, does not lead to significant peaks at 
$\varepsilon_2$ due to the cut-off for electrons with a kinetic energy 
$|eV|<\varepsilon_3$ (Figure \ref{fig:Spin1}d).

\subsection{Renormalization of the eigenstate energies}
\label{sec:Bloch-Redfield}

\begin{figure}[tbp]
\centering
\includegraphics[width=0.85\textwidth]{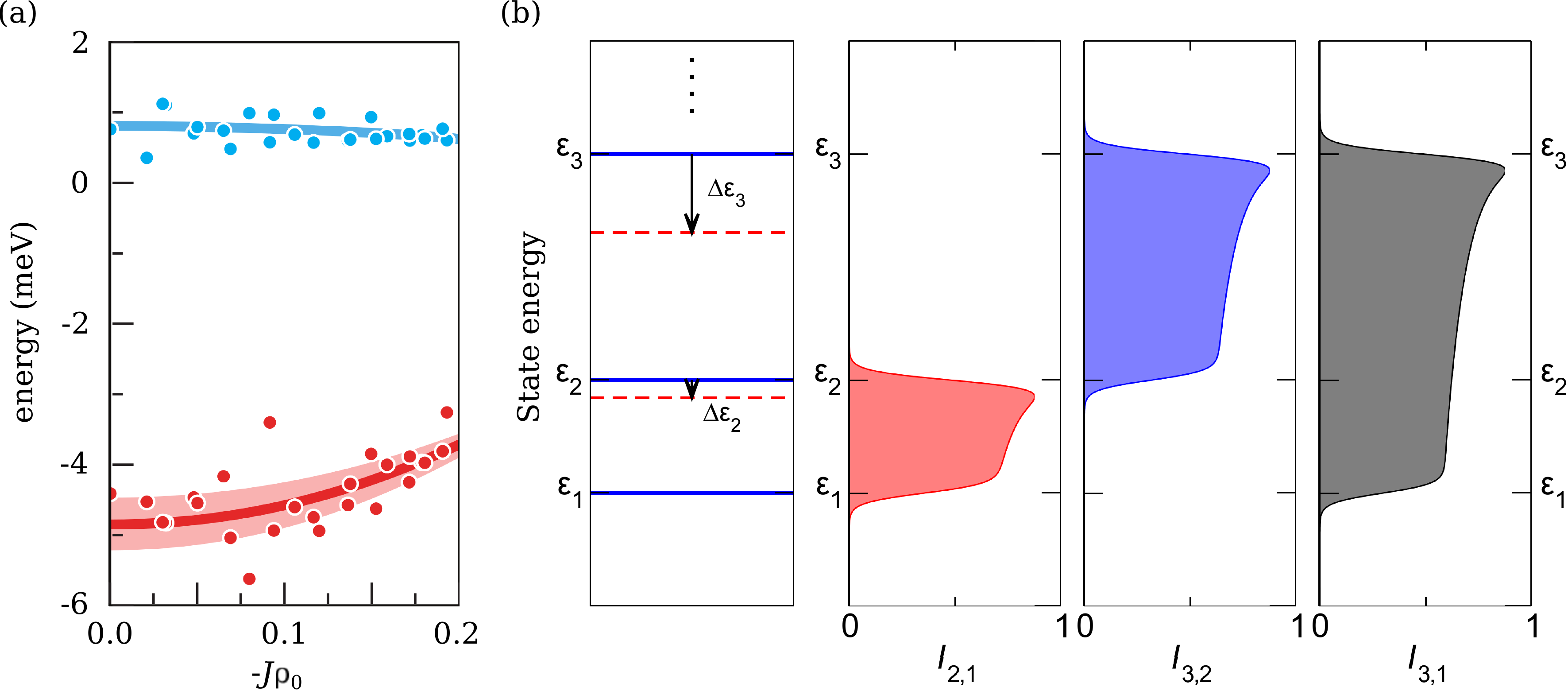}
	\caption{Magnetic anisotropy renormalization due to the 
coupling to the substrate. \textbf{(a)}~Experimentally determined $D$ and $E$ 
(red and blue dots) parameters plotted versus the coupling strengths 
$-J\rho_0$. 
Full lines show the expected renormalization of $D$ and $E$ due to virtual 
coherences calculated with a Bloch-Redfield approach taking exchange scattering 
with
the dissipative substrate electron bath into account. Shaded region shows the
experimental uncertainty. \textbf{(b)} Principle scheme of the shift of the 
state energies. The integral contributions $I_{i,j}$ to the energy shift 
$\Delta\varepsilon_i$ are displayed in the right hand side graphs revealing 
that for the low lying state at the energy $\varepsilon_2$ only $I_{2,1}$ has 
weight in the equation \ref{eq:deltae}, while for $\varepsilon_3$ the weights 
$I_{3,2}$ and 
$I_{3,1}$ have to be accounted for. Figure adapted from 
reference \cite{Jacobson15}.}
	\label{fig:Spin1-renorm}
\end{figure}

As shown in figure \ref{fig:CoH-topo}c different CoH 
complexes on $h$-BN have different step energy positions which 
correspond to different anisotropy parameters $D$ and $E$. Statistical analysis 
neither lead to a sharp distribution nor to a correlation with the adsorption 
site of the CoH on the corrugated $h$-BN substrate. However, by 
considering the values of $-J\rho_0$ from the fits to the IETS spectra, we 
observe a correlation between the magnetic anisotropy and the coupling with the 
substrate, $-J\rho_0$ \cite{Jacobson15}. The red branch in figure 
\ref{fig:Spin1-renorm}a shows 
that as the substrate coupling increases, the axial magnetic anisotropy 
decreases. For this analysis we restricted the evaluation to complexes with a 
clear out-of-plane anisotropy \cite{Gatteschi08} determined by the criterion 
$\frac{|D|}{3E}>1.5$. 

To rationalize this finding, we treat the quantum mechanical system of the 
impurity $\hat{H}_{\rm imp}$ (equation~\ref{eq:Atom-Hamilonian}) not as a 
separated system but as coupled to the dissipative bath of the substrate 
electrons. We then employ a 
Bloch-Redfield approach to account for the decay of excited states and 
coherences in the density matrix \cite{Cohen89}. 
Interestingly, this approach leads for the off-diagonal elements of the reduced 
density matrix $\chi=|\Psi\rangle\langle\Psi|$ of $\hat{H}_{\rm imp}$ 
not only to a fast decoherence but additionally to an energy shift of the 
eigenstates due to the interaction between $\hat{H}_{\rm imp}$ and the 
reservoir. We will 
restrict ourselves to the Kondo-like scattering between the substrate 
electrons and the localized spin, as described by equation~\ref{equ:M},  up 
to second order leading to a correction term of the 
form \cite{Cohen89, Oberg13}:
\begin{equation}
\Delta\varepsilon_{\alpha}=(J\rho_0)^2\sum_{n}\sum_{n',
\alpha'} \frac {\left|\left\langle \varphi_{n'}, \Psi_{n} \right| 
\frac12\hat{\boldsymbol{\sigma}}\cdot\hat{\bf S} |\varphi_{\alpha'}, 
\Psi_{\alpha} \rangle\right|^2}{ 
\varepsilon_{\alpha}-\varepsilon_n+\varepsilon_{\alpha'}-\varepsilon_{n'} }.
\end{equation}
Knowing the scattering matrix elements and making use of equation~\ref{equ:g} 
we can rewrite the energy shift as:

\begin{equation}
\Delta\varepsilon_{\alpha}=\frac{(J\rho_0)^2}{2}\sum_{n}\int_{-\infty}^{+\infty
} d\varepsilon\ g(\varepsilon_ { \alpha}-\varepsilon_n+\varepsilon) 
f(\varepsilon_n-\varepsilon)\left[ 1-f(\varepsilon_{\alpha}-\varepsilon) 
\right].
\label{eq:deltae}
\end{equation}
Figure \ref{fig:Spin1-renorm}b illustrates the effect of the energy 
renormalization. The energetically higher excited state at $\varepsilon_3$ is 
stronger affected than the low lying state at $\varepsilon_2$.
For the magnetic anisotropy parameters $D$ and $E$ of the CoH system the shift 
can be approximated as:
\begin{equation}
D(J\rho_s)\approx D_0\left(1-\alpha (J\rho_0)^2\right),\quad\mbox{ and }\quad 
E(J\rho_s)\approx E_0\left(1-\beta (J\rho_0)^2\right),
\label{equ:alpha}
\end{equation}
with the coefficients $\alpha$ and $\beta$ given by the integrals of equation 
\ref{eq:deltae}.

The solid red line in figure \ref{fig:Spin1-renorm}a shows the best fit when 
employing this model onto our data and follows the trend of equation 
\ref{equ:alpha}. The shaded regions accounts for the possible range of 
$\alpha$ and $\beta$ by considering an effective 
bandwidth of $\omega_0= 0.4 - 1.2$~eV \cite{Jacobson15}.

\subsection{The anisotropies of Fe and Mn embedded in CuN}
\label{sec:Ani-Fe-Mn}

After the $S=1$ CoH on $h$-BN system, 
we will now focus on the experimentally and theoretically intensively 
studied single $3d$ transition metal atoms Fe and Mn adsorbed  on a monolayer 
of Cu$_2$N on Cu(100). It was on these two systems that the magnetic anisotropy 
of individual, single atoms was measured by IETS for the first time 
\cite{Hirjibehedin07}.

\begin{figure}[tbp]
\centering
\includegraphics[width=0.7\textwidth]{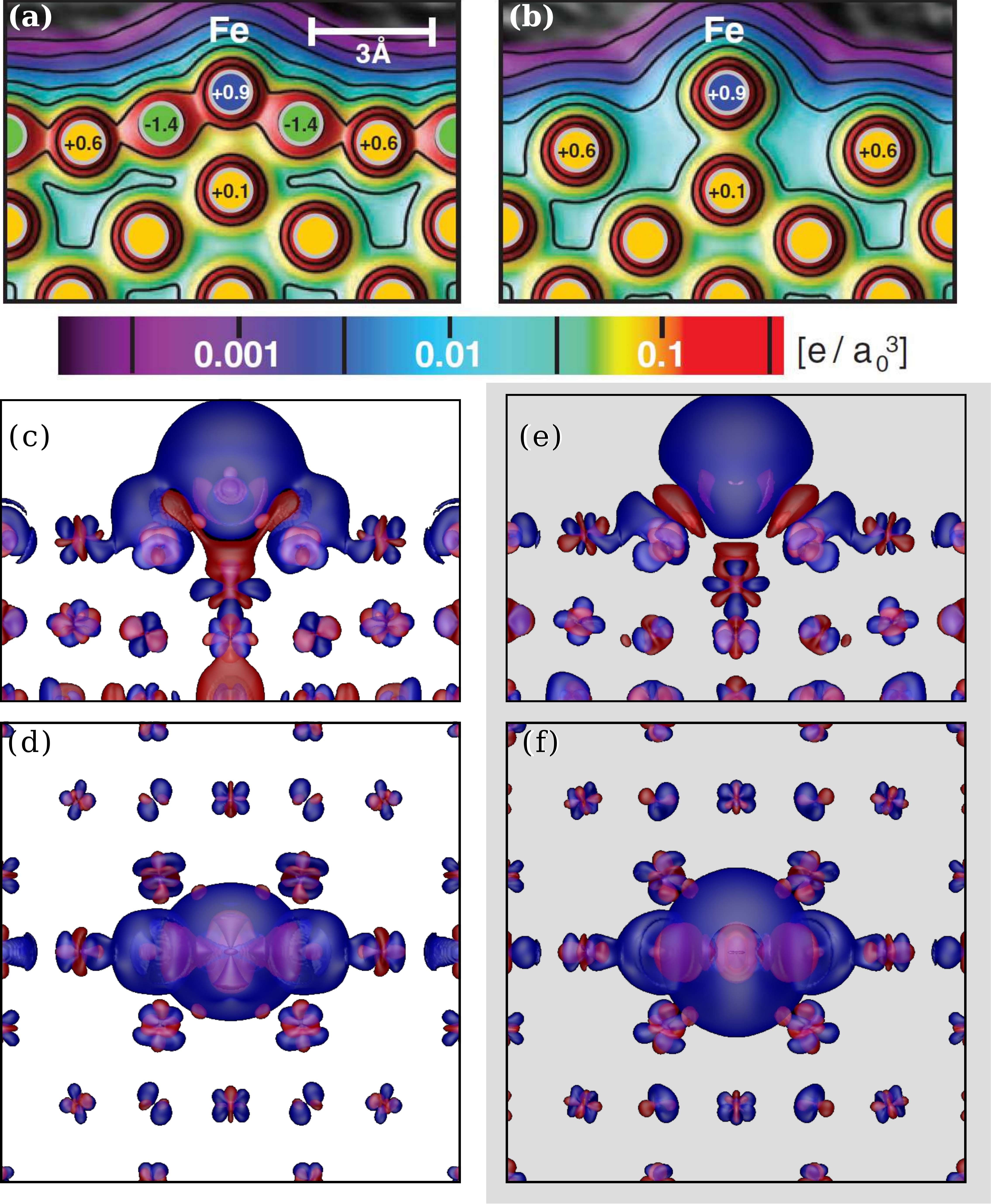}
	\caption{Adsorption of Fe and Mn on Cu$_2$N. \textbf{(a)} The charge 
density for a Cu$_2$N surface on Cu(100) with an Fe atom (blue) adsorbed on top 
of a surface Cu site along the N direction. Solid yellow and green circles 
label the centers of the Cu and N atoms, respectively. The numbers inside the 
circles indicate the net charge on selected atoms in units of $e$. \textbf{(b)} 
Same as (a) along the vacancy direction. \textbf{(c--f)} Calculated 
spin-density distribution for Fe (c, d) and Mn (e, f) on CuN. Contours in blue 
and red show the majority and minority constant spin density ($0.01 e/a_0^3$) as 
calculated by DFT. Abscissa is along the N direction. Ordinate 
in panel (c, e) [(d, f)] is the out of plane [vacancy] direction. 
$a_0=52.9$~pm is the Bohr radius. Figure adapted from reference 
\cite{Hirjibehedin07}.}
	\label{fig:Spin-dens-Fe-Mn}
\end{figure}

\begin{figure}[tbp]
\centering
\includegraphics[width=0.9\textwidth]{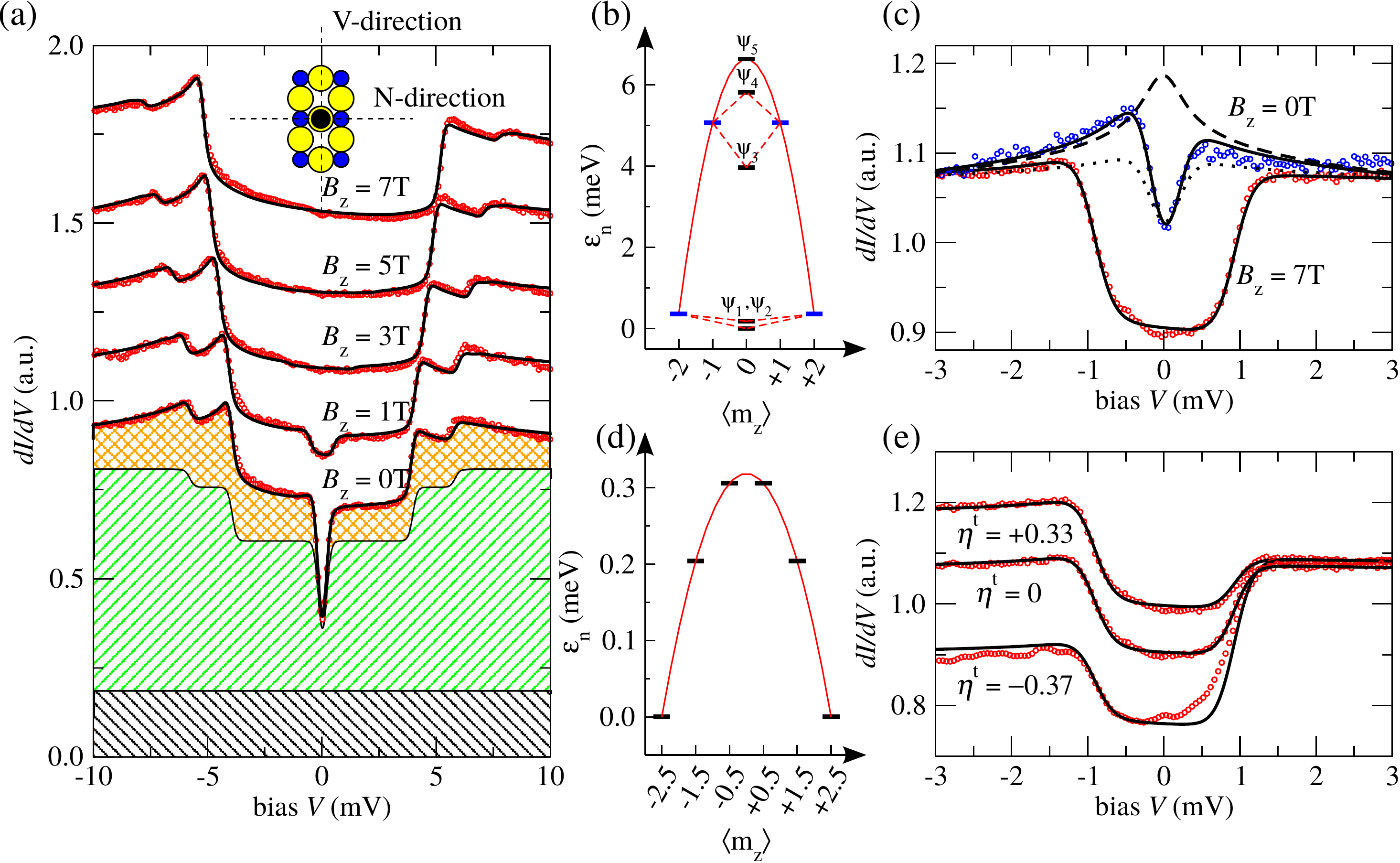}
	\caption{Comparison of experimental and calculated spectra on Fe 
and Mn atoms adsorbed on Cu$_2$N. \textbf{(a)} Experimental data measured on a 
single Fe atom at increasing field $B_z$ along the easy axis (N-direction) and 
at a temperature of $T=550$~mK (red circles). The simulations (black lines) for 
all plots are obtained with one set of parameters: $g=2.11$, $D=-1.57$~meV, 
$E=0.31$~meV, $J\rho_0=-0.087$, $U=0.35$, and $T_{\rm eff}=740$~mK. 
Additionally, a constant offset of $\approx20$\% of the total conductance has 
been added (black shaded area). For $B_z=0$ the second (green 
shaded area) and third order (orange hatched area) contributions to the 
conductance are indicated. The spectra at field are vertically shifted for 
better visibility. The inset shows the adsorption site of the $3d$ atoms (black 
circle) on the Cu$_2$N (Cu yellow, N blue circle) and the two distinct surface 
directions. \textbf{(b)} Schematic state diagram for Fe. 
\textbf{(c)} Experimental data of two different Mn atoms at $B_z=0$ and 
$B_z=7$~T (colored circles). The fits (full lines) for the $B_z=0(7)$T data 
results in $J\rho_0=-0.029 (-0.0091)$, $U=1.35 (1.28)$, $D=-51(-39)\ \mu$eV, 
$g=1.9$, and $T_{\rm eff}=790(930)$~mK. The dashed line shows 
simulated data for zero field and the absence of any anisotropy. The dotted 
line shows simulated data with the 7~T parameters in the absence of a magnetic 
field. \textbf{(d)} Schematic state 
diagram for Mn. \textbf{(e)} The 7~T atom as in (c) probed with tips of 
different spin polarizations $\eta^{\rm t}$ (Data from 
ref.~\cite{Loth10b}). Figure adapted from reference \cite{Ternes15}.}
\label{fig:spec-Fe-Mn}
\end{figure}

When Fe or Mn atoms are placed on top of a Cu site they form strong 
covalent bonds with the neighboring N atoms, as revealed by DFT calculations 
performed by Chiung-Yuan Lin and Barbara Jones and shown in 
figure~\ref{fig:Spin-dens-Fe-Mn}a, b \cite{Hirjibehedin07, Lin11}. This highly 
anisotropic adsorption geometry leads to three distinct symmetry axes that are 
perpendicular to each other: The direction out-of-plane and two 
in-plane directions along the Cu-N bonds and perpendicular to it, along the 
so called vacancy rows (Figure~\ref{fig:spec-Fe-Mn}a inset).

Single Fe atoms adsorbed on this surface have been found to be in the effective 
$S=2$ state with a magnetic easy-axis along the N rows ($z$-direction) and a 
magnetically hard-axis along the vacancy row ($x$-direction). These 
main anisotropy axes are directly visible in spin-resolved DFT calculations as 
shown in figure \ref{fig:Spin-dens-Fe-Mn}c and d. Using anisotropy values of 
$D=-1.55$~mV and $E=0.31$~mV, and a gyromagnetic factor of 
$g=2.11$  described the experimental data well using the spin Hamiltonian of 
equation~\ref{eq:Atom-Hamilonian} and a second order tunneling model 
\cite{Hirjibehedin07, Lorente09, Fransson09}. Note that possible 
forth order anisotropy parameters as $B_4^0$, $B_4^2$, $B_4^4$ (see equation 
\ref{equ:Stevens}) have been found to be $<10$~$\mu$eV \cite{Yan15a}. The 
Hamiltonian has as solution 
five non-degenerate eigenstates and, due to $D<0$, favors, at zero field, 
ground states with weights at high $m_z$ values. Similarly to the $S=1$ system 
the transverse anisotropy breaks the degeneracies leading to a symmetric and 
antisymmetric solution with the main weights at $\left|\pm2\right\rangle$ as 
ground and first excited state and weights in $\left|\pm1\right\rangle$ for the 
second and third excited state (Figure~\ref{fig:spec-Fe-Mn}b).

In second order, spin-flip scattering is allowed between the 
groundstate and the three lowest excited states but a transition to the highest 
state is forbidden because this would require an exchange of $\Delta m=\pm2$.
Experimental $dI/dV$ measurements on this system show, in addition to 
the conductance steps, peak-like structures at the second and third step but 
not at the lowest one (Figure~\ref{fig:spec-Fe-Mn}a). Additionally, they show 
an asymmetry between positive and negative bias. To 
rationalize these observations we can follow a similar argument as in the 
$S=1$ case (see section \ref{sec:CoH}): In third order, transitions like (121) 
are not possible and processes like (123) or (124) are 
strongly cut off due to the high energy difference between $\varepsilon_2$ and 
$\varepsilon_{3}$ or $\varepsilon_{4}$. In contrast, the processes (132) and 
(142) scale with $J\rho_s$ leading to the peak features in the differential 
conductance. The additional asymmetry hints at a non-negligible potential 
scattering with matrix elements of the form
\begin{equation}
M_{i\rightarrow f}=\sum_{i',f'}\left\langle \varphi_{f'}, \Psi_f \right| 
U|\varphi_{i'}, \Psi_i \rangle= U\delta_{if}.
\label{equ:MU}
\end{equation}
This \emph{elastic} scattering term can interfere with the exchange 
scattering matrix elements (equation \ref{equ:M}) leading to bias asymmetries 
in the spectrum (see also section \ref{sec:Kondo-fano}) \cite{Ternes15}. 
As the computed curves in figure \ref{fig:spec-Fe-Mn}a reveal, this model 
almost perfectly fits the magnetic field data without any adaption of the 
parameters. The coupling strength in these simulations is $J\rho_s=-0.087$, 
close to the 
$-0.1$ found in a similar perturbative approach \cite{Hurley11a}.
A potential scattering term of $U=0.35$ is necessary to reproduce the 
asymmetry. This value is significantly smaller than the $U\approx 
0.75$ found in experiments where the magneto-resistive elastic 
tunneling was probed \cite{Loth10b}. Part of this discrepancy can be 
understood by an additional conductance term that does not coherently interact 
with the spin-system and which would lead to an overestimation of $U$ in 
magneto-resistive measurements. Indeed we need a constant conductance 
offset of about $20$\%, which is added to the calculated conductance to 
reproduce the spectra. 

Switching from an integer to a half-integer spin system we now discuss 
individual Mn atoms on Cu$_2$N, which have a spin of $S=5/2$ and only a small 
easy-axis anisotropy of $D\approx -40\ \mu$eV along the out-of-plane 
direction and a negligible transverse anisotropy \cite{Hirjibehedin06, 
Hirjibehedin07}. 
Also here spin-resolved DFT can visualize the main anisotropy 
axis (Figure~\ref{fig:Spin-dens-Fe-Mn}e,~f). 
The easy-axis anisotropy prohibits the immediate formation of 
a Kondo state due to a Kramer's degenerate ground state doublet with 
$m_z=\pm5/2$, which would require a $\Delta m=5$ to flip
(Figure~\ref{fig:spec-Fe-Mn}c). At zero field a typical spectrum 
shows only one step, which belongs to the transition between the $\pm5/2$ and 
the $\pm3/2$ states that have superimposed asymmetric peak structures 
(Figure~\ref{fig:spec-Fe-Mn}d). 
The fit to the model yields $J\rho_s=-0.029$ and resembles a $S=1/2$ 
split-Kondo peak at small magnetic fields (see next 
section, figure~\ref{fig:M449}e). A 
different Mn atom investigated at $B_z=7$~T shows a significantly reduced 
$J\rho_s=-0.0091$. Interestingly, we find for both atoms a potential scattering 
value of $U\approx\frac{1}{2} S$, which allows one to describe the spectra 
without the need of any additional conductance offset. This high $U$ value that 
is the origin of the bias asymmetry has been independently found in 
spin-pumping experiments \cite{Loth10} and by measuring the 
magneto-resistive elastic tunneling contribution \cite{Loth10b}. The 
extraordinary agreement 
between model and experiment can be seen in measurements using different 
spin-polarized tips on the same atom (Figure~\ref{fig:spec-Fe-Mn}e). Here, the 
strong influence of the tip-polarization on 
the inelastic conductance  at bias voltages $|V|<1$~mV is evident while the 
differential conductance at $V>1.5$~mV stays constant for all tips.

\section{The Kondo effect }
\label{sec:Kondo}

The electrical resistance of pure metals usually deceases when they are cooled 
down because one of the main origin of the resistivity, the scattering of 
electrons on lattice vibrations, is frozen out at reduced temperature. 
However, already in the 1930s it was discovered that in 
some metals containing diluted magnetic impurities the electrical resistance 
increases again below a certain temperature \cite{Meissner30, deHaas34}. 

\begin{figure}[tbp]
\centering
\includegraphics[width=0.5\textwidth]{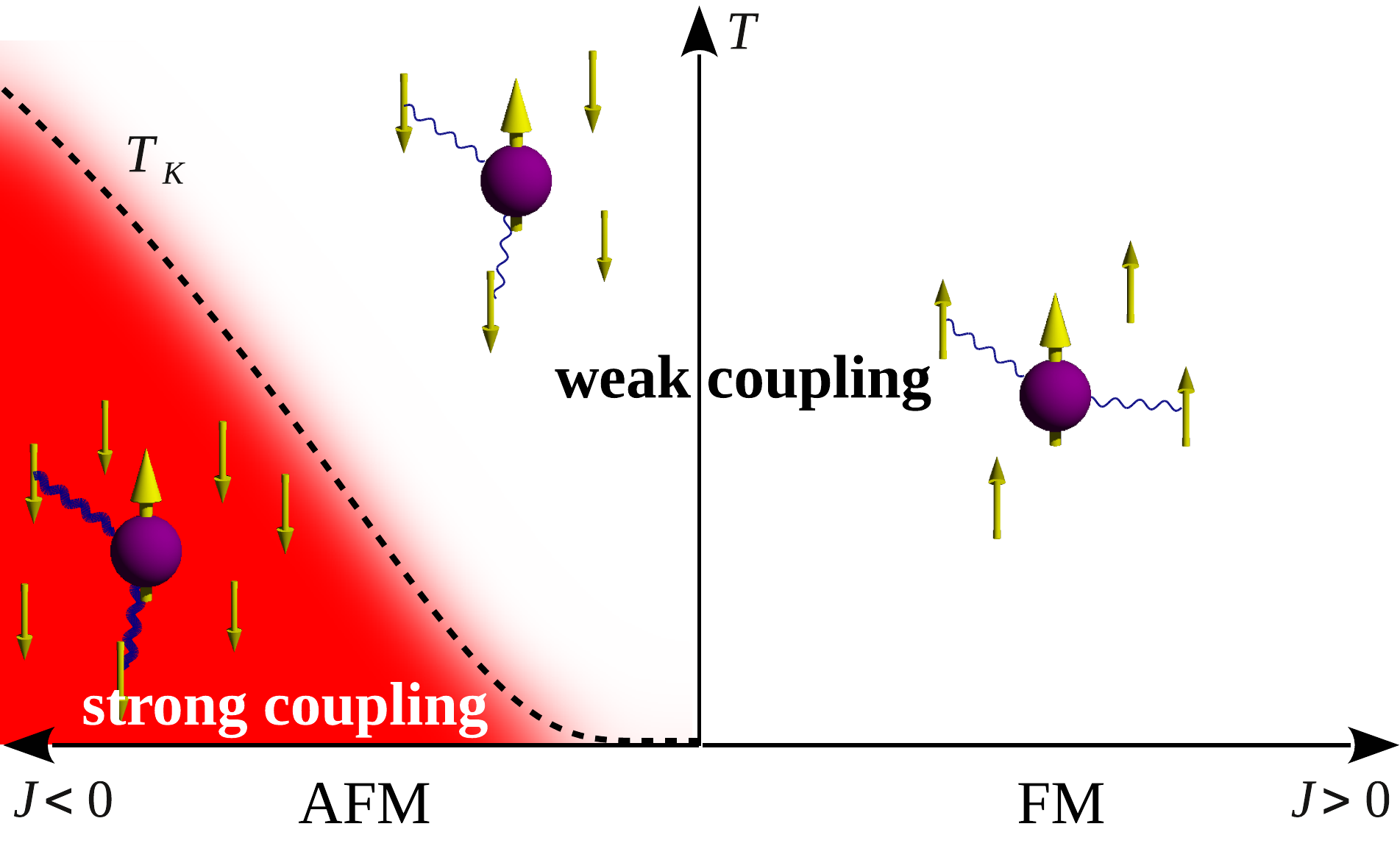}
	\caption{The different regimes of the Kondo effect. The Kondo
interaction $J\hat{\bf S}\cdot\hat{\boldsymbol{\sigma}}$ couples itinerant 
electrons of the host 
to a magnetic impurity. For exchange interaction $J<0$, the antiferromagnetic 
(AFM) coupling leads to an entangled many-body state, where the antiparallel 
alignment of the spins of the conduction electrons effectively screens the 
impurity spin. The ground state at temperatures $T$
below the characteristic Kondo temperature $T_K$ is a singlet with total spin
$S=0$ (red area), well protected from higher energy states. In contrast, for
$J>0$, the ferromagnetic (FM) coupling tends to create a screening cloud of
spins aligned parallel to the impurity spin, which becomes asymptotically
free at low temperatures \cite{Anderson70, Gentile09a}. For FM coupling or at 
temperatures $T\gg T_K$, the
system is in the weak coupling regime, which can be treated perturbatively. 
Figure adapted from reference \cite{Zhang13}.}
	\label{fig:Kondo}
\end{figure}

The origin of this effect remained obscure for a long time but was 
explained in 1964 by Jun Kondo \cite{Kondo64, Kondo68}. He showed that 
these experimental observations can be understood within a scattering model, 
which explicitly takes into account the interaction of the spins of the 
conduction electrons of the host metal with the localized spin of the
magnetic impurities. This interaction is usually considered to be 
antiferromagnetic (AFM), i.\,e.~the spin-spin exchange coupling $J$ is negative, 
and creates correlations between the localized magnetic moment and the 
surrounding host electrons. This leads to a screening of the impurity magnetic 
moment and to the formation of a new, non-magnetic, many-electron singlet 
ground-state below a critical temperature $T_K$ \cite{Hewson97} (Figure 
\ref{fig:Kondo}).

Interestingly, this so-called 'Kondo effect' emerges 
in quite a broad range of different physical contexts, such as the zero-bias 
anomalies observed in quantum dots and nanowires \cite{Goldhaber98, Nygard00, 
Kogan04, Paaske06, Potok07, Kretinin11, Keller14}, or the 
dynamical behavior close to a Mott transition \cite{Georges96, Nozieres05}. 
The simplicity of the underlying model Hamiltonian (equation 
\ref{equ:M}) contrasts the complex physics and the non-trivial solutions 
that occurs in the strong coupling regime. The origin of this Kondo 
problem lies in the appearance of logarithmic divergences which make 
perturbative models fail for $T\rightarrow 0$. Only the development of a 
completely new theoretical understanding clarified the nature of the 
many-electron ground state which manifest itself in a strong 
resonance close to the Fermi-energy \cite{Anderson70, Wilson75, Andrei80}.

The existence of the Kondo
resonance in dense Kondo systems like solid surfaces or thin
films of $\gamma$- and $\alpha$-cerium as well as in Ce-heavy-fermion
compounds has been experimentally confirmed by high-resolution photoemission 
electron spectroscopy (PES) \cite{Patthey85, 
Patthey87,Patthey90,Laubschat90,Weschke91,Ehm07}
and by inverse photoemission \cite{Wuilloud83,Gschneidner87}.
While these measurements always probe an ensemble of impurities due
to the limited spatial resolution in PES, STM opened the
unique opportunity to detect the Kondo effect in the smallest
conceivable Kondo system: a single magnetic impurity supported
on top of a nonmagnetic metal \cite{Li98a, Madhavan98, Nagaoka02, Knorr02, 
Schneider02, Wahl04, Limot04, Wahl05, Schneider05}. 

While most of these experiments where performed at $T\ll T_K$ much less 
attention has been paid to the weak coupling regime,
which is relevant either at elevated temperatures ($T\gg T_K$) or for 
ferromagnetic (FM) interactions ($J>0$). In the case of FM interaction, 
the impurity spin is always weakly coupled and becomes asymptotically
free in the limit of low temperature \cite{Anderson70, Gentile09}. A possible 
path for creation of such a state will be discussed in section 
\ref{sec:Coupled_seystems}. For AFM
interactions at high temperatures, thermal fluctuations destroy
the singlet state enabling the physics to be described by perturbation theory 
\cite{Anderson70}.

In this section we will start our discussion with the results obtained on a 
fully organic radical molecule which was the first detailed study of the Kondo 
effect in the weak coupling limit \cite{Zhang13}. 
From thereon, we will briefly summarize the externally controllable 
conditions, like temperature and magnetic field, that causes the perturbative 
approach to breakdown \cite{Ternes15}. Therefore, individual Co atoms on 
Cu$_2$N, with their high effective spin $S=3/2$ will link the Kondo physics with 
the magnetic anisotropy discussed in section \ref{sec:Anisotropy} 
\cite{Otte08a, Oberg13}. Additionally, this system is particularly interesting 
because it was the one used to determine the spin polarization of the field 
split Kondo state \cite{Bergmann15}.

\subsection{The weak coupling limit in an organic radical}

\begin{figure}[tbp]
\centering
\includegraphics[width=0.65\textwidth]{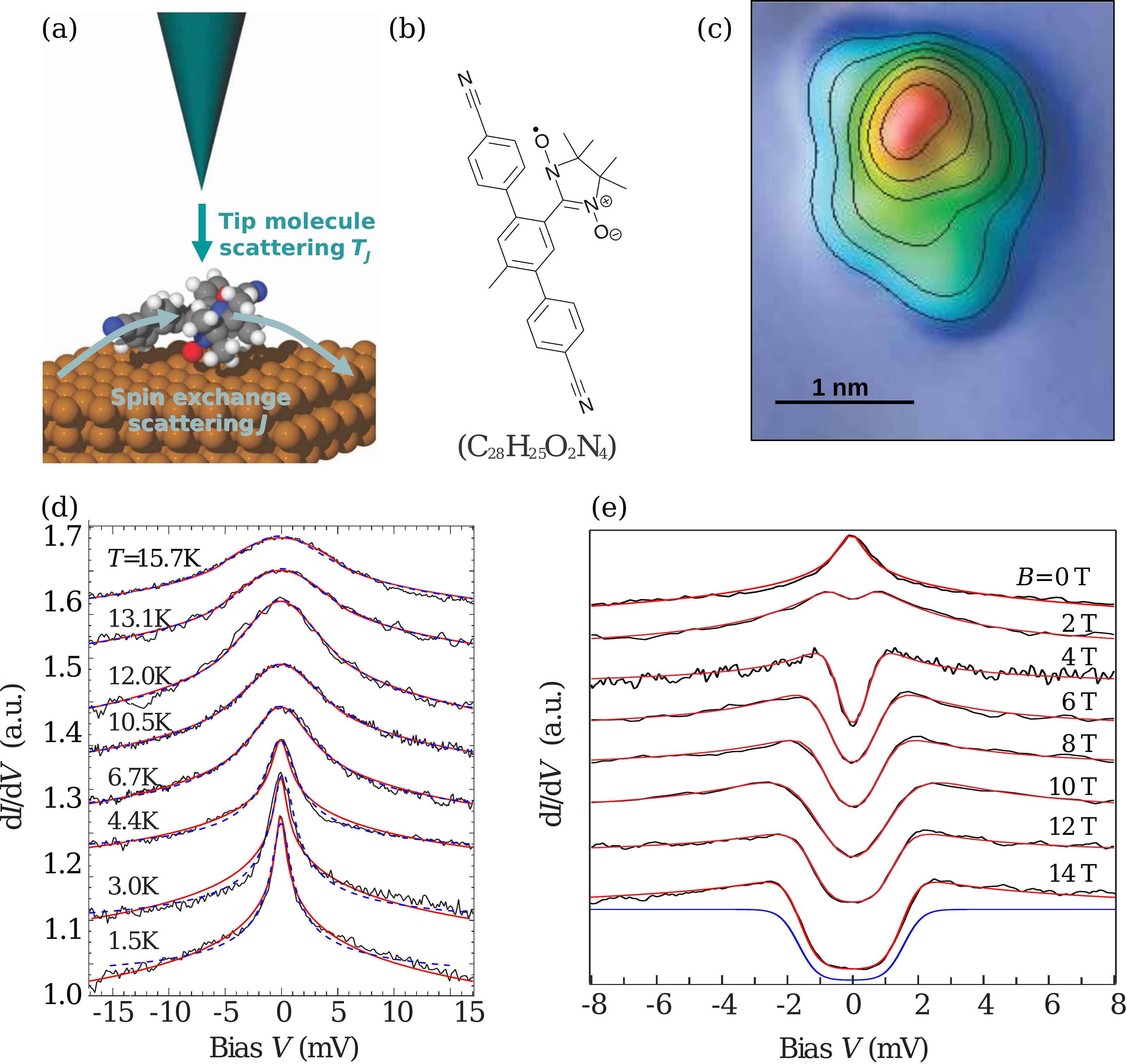}
	\caption{The weak coupling Kondo effect in an organic radical molecule. 
\textbf{(a)} Principle scheme of the experiment. \textbf{(b)} Chemical 
structure of the studied organic radical molecule (C$_{28}$H$_{25}$O$_2$N$_4$) 
with a nitronyl-nitroxide side group that contains a delocalized singly 
occupied molecular orbital. The molecule is drawn with a similar orientation as 
in the topography in (c). \textbf{(c)} Topography of one organic radical 
molecule ($V=100$~mV, $I=33$~pA). Contour lines are at height intervals of 
$50$~pm. \textbf{(d)} Typical $dI/dV$ spectra taken on the radical side group 
of 
the molecule (black) and simulated spectra using the perturbation model (red). 
\textbf{(e)} Spectra taken at successively increased magnetic fields at 
$T=1.8$~K. Blue line at 14~T shows simulation in 2nd order only. All 
spectra in (d) and (e) are normalized and offset for visual clarity. Figure 
adapted from reference \cite{Zhang13}.}
	\label{fig:M449}
\end{figure}

In this study we used a purely organic molecule, which has a radical
nitronyl-nitroxide side group \cite{Osiecki68}, adsorbed on a clean Au(111) 
surface as sketched in figure \ref{fig:M449}a.  Molecular crystals containing 
the same radical side group showed ferromagnetic coupling below a Curie 
temperature of $0.6$~K \cite{Tamura91}. In this molecule the unpaired electron 
is spatially delocalized over the O--N--C--N--O part of the side group instead 
of being localized on a specific atom (Figure~\ref{fig:M449}b). 
This delocalization stabilizes the unpaired electron against 
chemical reaction and charge transfer, which would lead to a spin zero
system. Constant-current STM images show the elongated 
molecular backbone and the radical side group, which is imaged as a 
$\approx0.3$~nm high and $1.0$~nm wide protrusion (Figure~\ref{fig:M449}c).

Figure \ref{fig:M449}d shows the differential conductance measured at 
zero-field on the side-group of the molecule revealing a strongly temperature 
dependent peak a zero bias. Field dependent measurements at low temperature 
($T=1.8$~K) show that this peak is split into two peaks as soon as the 
Zeeman-energy $g\mu_BB$, with $g\approx2$ for a free $S=1/2$ spin, is 
comparable to $k_BT$ (see figure \ref{fig:M449}e). At higher fields ($B \gtrsim 
10$~T) the spectra show a clear steplike structure symmetrically around zero 
bias and additional peak-like conductance increases at the step-positions.

To describe the excitation processes during tunneling we use a model based
on the perturbative approach established by Appelbaum, Anderson, and Kondo
\cite{Kondo64, Appelbaum66, Anderson66, Appelbaum67} in which spin-flip
scattering processes up to 3rd order in the spin-spin exchange coupling $J$ are
accounted for (Figure \ref{fig:M449}a). This model is equivalent to the one 
discussed in section \ref{sec:CoH} and leads to a tunneling conductance 
$\sigma(eV)=dI/dV(V)$ due to the electron-spin interaction with 
the following contributions:
\begin{align}
\sigma_1(eV)=&\frac{e^2}{h}T_J^2\left( 1+2\sum_{\substack{i,f\\ i\neq
f}}\varrho_i(T)\left[
\Theta\left(\frac{\Delta_{if}+eV}{k_{\rm
B}T}\right)+\Theta\left(\frac{\Delta_{if}-eV}
{k_{\rm B}T } \right)\right]\right)
\label{equ:sigma11}
\end{align}
%
\begin{align}
\sigma_2(eV)=\frac{e^2}{h}T_J^2J\rho_0\sum_{\substack{i,f\\ i\neq
f}} \varrho_i(T)& \Biggl\{
\left[\Theta\left(\frac{\Delta_{if}+eV}{k_{\rm B}T}\right)
+
\Theta\left(\frac{\Delta_{if}-eV}{k_{\rm
B}T}\right)\right] \times \nonumber \\
& \left[\underbrace{2\times g\left(eV\right)
}_{112 {\rm\ and\ } 112R}
+\underbrace{g\left(\Delta_{if}+eV\right)
+g\left(\Delta_{if}-eV\right)}_{122 {\rm\ and\ } 122R}
\right]\nonumber \\
&+ \underbrace{ g\left(\Delta_{if}+eV\right)
+g\left(\Delta_{if}-eV\right)}_{121 {\rm\ and\ } 121R} 
\Biggr\}
\label{equ:sigma21}
\end{align}
%
Here, $T_J$ is the coupling constant determining the overall 
conductance of the tunnel junction,
%
$\varrho_i(T)=Z^{-1}\times\exp[-\varepsilon_i/(k_{\rm B}T)]$, with 
$Z=\sum_i\exp[-\varepsilon_i/(k_{\rm
B}T)]$,
is the thermal occupation of the localized spin in the 
state~$\left|\Psi_i\right\rangle$,
$\Theta(\varepsilon)$
is the
thermally broadened step function (see page~\pageref{p:theta}) \cite{Lambe68},
$\Delta_{if}=\varepsilon_f-\varepsilon_i=\pm g\mu_BB$ is the energy difference 
(Zeeman energy) between the two spin states of the
molecule, and $g(\varepsilon)$ is the function originating from the divergence 
of 
the second term in equation~\ref{equ:W} as defined in equation \ref{equ:g}.

\begin{figure}[tbp]
\centering
\includegraphics[width=0.7\textwidth]{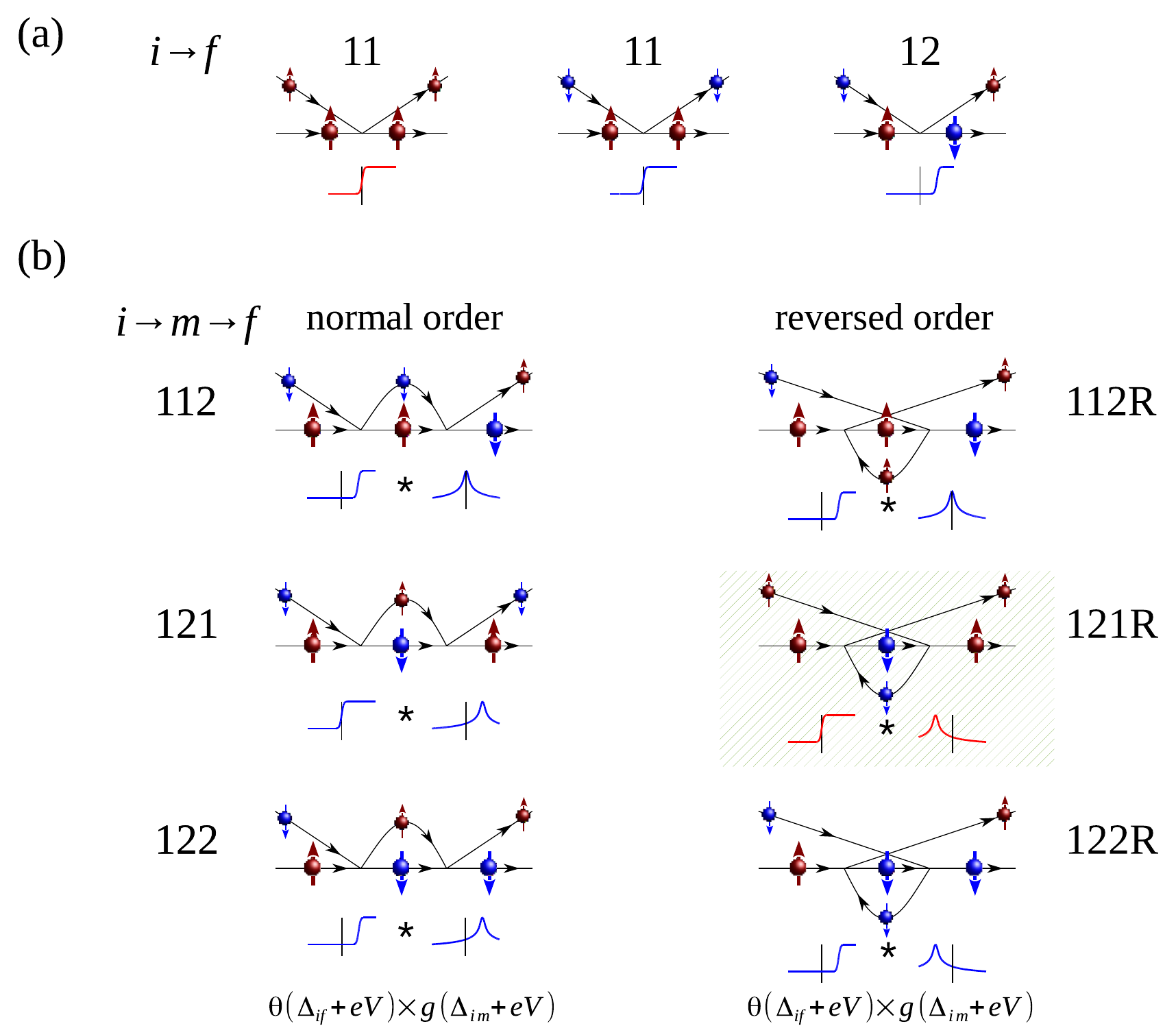}
	\caption{Interaction diagrams of order two \textbf{(a)} and three 
\textbf{(b)} for an electron tunneling from tip to 
sample into a two level $S=1/2$ spin system. The large spheres 
depict the state of the localized spin and the small spheres the state of 
the interaction electron. Schematic spectra show their contributions to the 
conductance 
at positive bias. The numbers label the processes with the state order of the 
localized spin-system. An appended `R' label processes in which the 
scattering 
into the intermediate state is performed \textit{before} the tunneling electron 
interacts (exchange diagrams). Note that the time order of the processes 
strongly influences the conductance spectra as schematically displayed in the 
small graphs (vertical line is $E_F$; the $\ast$ means multiplication). Figure 
adapted from reference \cite{Ternes15}.
}
	\label{fig:Feyman}
\end{figure}

At zero field the conductance simplifies to
$\sigma(eV)=\sigma_0-\alpha\times g(eV/(k_{\rm B}T))$, a temperature broadened 
logarithmic function, with the temperature
$T$ as the only relevant fit parameter. 
%
At $B\neq0$ the step-like structure is governed by ordinary inelastic spin-flip
scattering of $\sigma_1$ in equation \ref{equ:sigma11} as 
illustrated in figure \ref{fig:Feyman}a \cite{Otte08a, Loth10b, Ternes15}.
The additional logarithmic peaks in the conductance result from the different
possible higher order scattering paths described by $\sigma_{2}$ and labeled 
as Feynman diagrams illustrated in figure \ref{fig:Feyman}b. We see that there 
are in total 6 processes of order three per initial state which have to be 
accounted for and which are reflected in the therms of $\sigma_2$ in equation 
\ref{equ:sigma21}.

The model fits exceptionally well with our data (see red lines in figure 
\ref{fig:M449}d, e) using a coupling to the substrate electrons of  
$J\rho_0=-0.04\pm 0.01$. Different as for the case of the spin $S=1$ system 
discussed in section \ref{sec:Bloch-Redfield}, in this half-integer Kondo system 
energy renormalization occurs already in first order of the coupling strength 
$J\rho_0$ leading to an effective gyromagnetic factor of $g_{\rm 
eff}=g_0(1+J\rho_0)$ \cite{Wolf69}\label{p:renorm}. For the highest fields 
$B\geq 10$~T this is in good agreement with an experimentally observed $g_{\rm 
eff}=1.93\pm 0.02$ \cite{Zhang13}.
Note, however that this perturbative approach only 
holds as long as higher order contributions can be neglected, that is, 
as outlined in the next section, as long 
as the temperature is high compared to the Kondo temperature $T_{K}$ of 
the system.

\subsection{The limit of the perturbative approach	}
\label{sec:Kondo-limit}

\begin{figure}[tbp]
\centering
\includegraphics[width=1\textwidth]{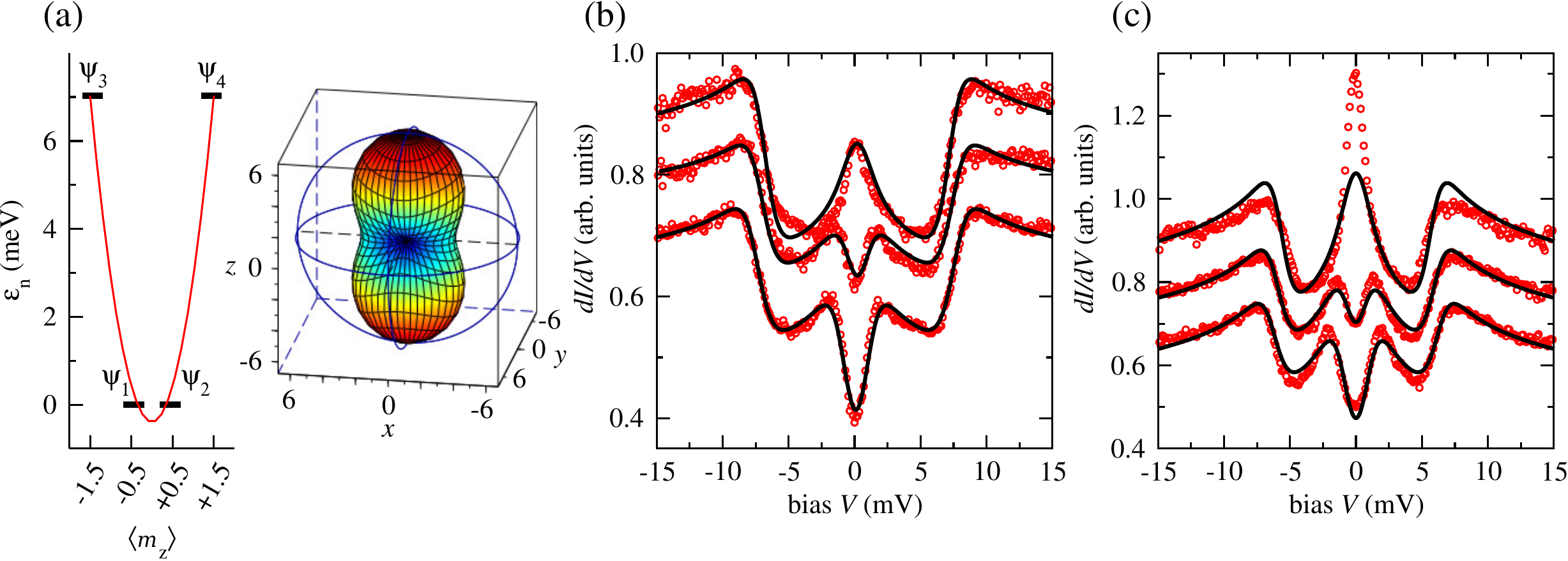}
	\caption{The tunneling spectra of Co atoms ($S=3/2$) on Cu$_2$N. 
\textbf{(a)} Schematic state diagram and visualizations of the magnetic 
anisotropy (in meV). At $B=0$ the states $|\psi\rangle_1$ and $|\psi\rangle_2$ 
are degenerated 
and differ by $\Delta m_z=\pm1$. \textbf{(b)} and \textbf{(c)} Experimental 
data from reference \cite{Oberg13} of two different Co atoms at $B_x=0,4,6$~T 
(colored circles, top to bottom, shifted for clarity). The best fits (full 
lines) 
results in $T_{\rm eff}=4$~K, $D=3.3$~meV, $E=0.7$~meV, $g=2$,
$J\rho_0=-0.11$ for (b) and $T_{\rm eff}=3.8$~K, $D=2.7$~meV, $E=0.5$~meV, 
$g=2$, $J\rho_0=-0.25$ for (c). Figure adapted from 
reference \cite{Ternes15}.}
\label{fig:Co-CuN}
\end{figure}

When a half-integer spin with $S>1/2$ has easy-plane anisotropy $D>0$,
its ground state at zero field is a doublet with its main weights in 
$m_z=\pm1/2$. This enables an effective scattering with the substrate electrons 
and leads, at low enough temperatures, to the formation of a Kondo state. 
Experimentally, this has been observed for bare Co on 
$h$-BN/Rh(111) \cite{Jacobson15} and for Co atoms on Cu$_2$N 
\cite{Bergmann15, Otte08a, Choi09, Oberg13, Choi14, Bryant15, Toskovic16}. Both 
systems can be described as effective $S=3/2$ whereby the 
latter enters the correlated Kondo state with a characteristic 
Kondo temperature of $T_K=2.6$~K in 
experiments performed on small patches of Cu$_2$N at temperatures down to 
$T=550$~mK \cite{Otte08a}.
Apart from $D>0$, Co on Cu$_2$N also has a small in-plane anisotropy 
($E\neq0$) 
which creates an easy axis ($x$) along the nitrogen row and a hard axis ($z$) 
along the vacancy rows (Figure~\ref{fig:Co-CuN}a).

Interestingly, in this system the coupling to the substrate $J\rho_0$ changes 
with the position of the Co atom on larger Cu$_2$N patches, concomitant with 
a change in the anisotropy energies which separates the 
$\left|\pm1/2\right\rangle$ states from the $\left|\pm3/2\right\rangle$ 
states \cite{Oberg13, Delgado14}. This effect can be rationalized by virtual 
coherences as discussed in section \ref{sec:Bloch-Redfield}. 
For us this allows the study of the transition from the weak coupling to 
the strong coupling regime. In the case where the Co atom is relatively weakly 
coupled to the substrate ($J\rho_0\approx-0.1$) the experimental data can 
be consistently fitted to the 3rd order perturbation model even at different 
field strengths along $B_x$ (Figure~\ref{fig:Co-CuN}b) \cite{Ternes15}. 
We observe a zero-energy peak that splits at 
applied 
fields similar to the $S=1/2$ radical molecule shown in figure \ref{fig:M449}e. 
But while for $S=1/2$ the field direction does not play a role, here the peak 
splitting depends strongly on the direction due to the magnetic anisotropy 
\cite{Otte08a}. 
For this high-spin system we furthermore observe inelastic steps 
due to the transition to energetically higher excited states which are located 
at $|eV|=2|D|$ for $B=0$ and whose additional peak structure is well described 
within the scattering model \cite{Ternes15}.

For Co atoms adsorbed closer to the edges of the Cu$_2$N patches, the 
coupling to the substrate increases and the fit to the model worsens 
significantly (Figure~\ref{fig:Co-CuN}c). While the experimental data measured 
for non-zero fields are reasonably well described with $J\rho_0\approx-0.25$, 
at $B=0$ the experimentally detected peak at $E_F$ is stronger than the peak 
created by the model. 
Additionally, the experimental peak-width appears to be smaller than the 
temperature broadened logarithmic function. Furthermore, we observe that the 
calculated spectrum no longer accurately describes the steps at $2|D|$. At this 
energy the third order contributions are less pronounced in the experimental 
data, 
indicating that we reach the limit of the perturbative approach.

\begin{figure}[tbp]
\centering
\includegraphics[width=0.9\textwidth]{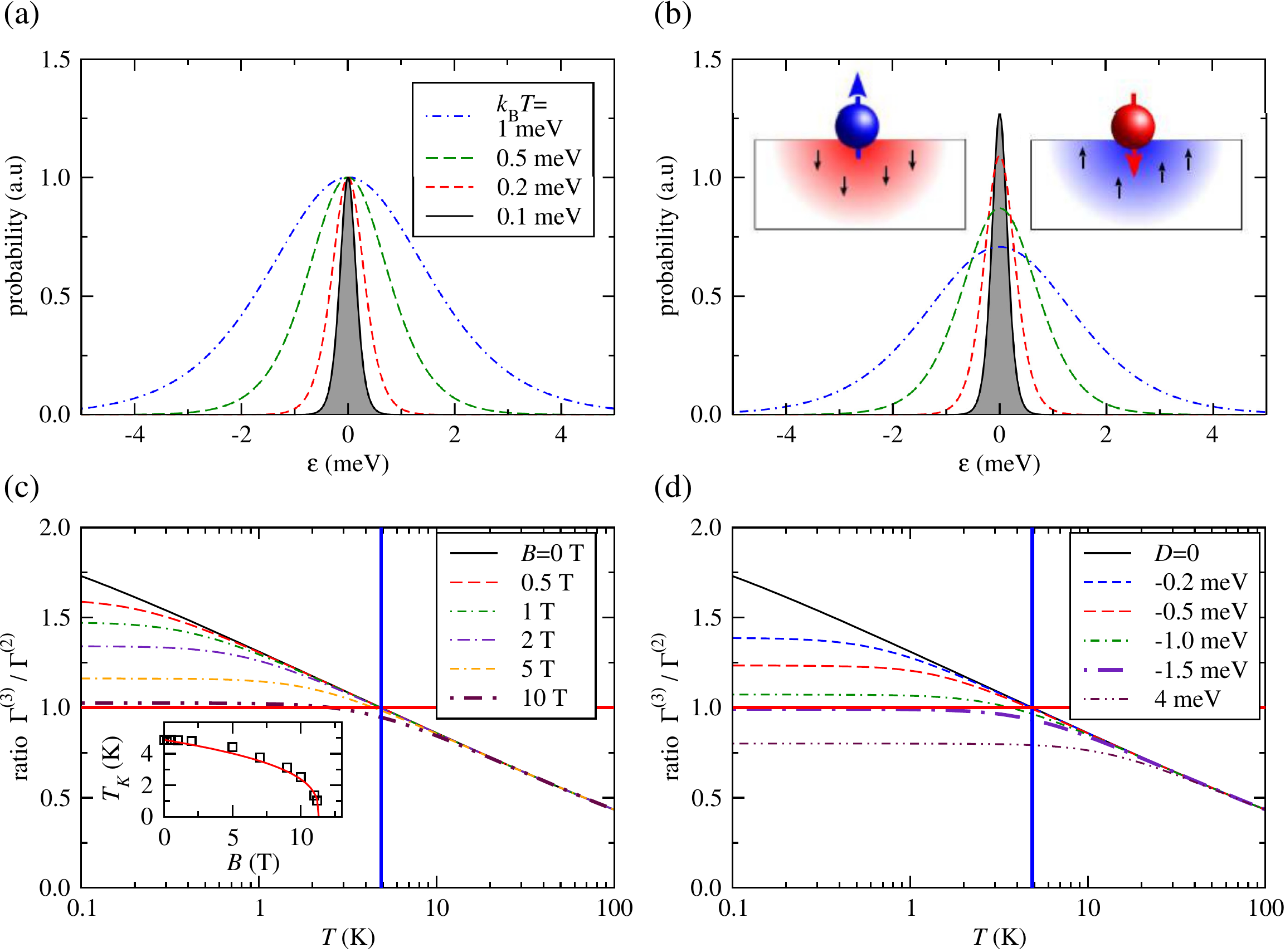}
	\caption{Correlations induced by substrate electrons.
\textbf{(a)} In second order the scattering probability of a substrate 
electron with energy $\varepsilon$ is given by the overlap of the electron 
and hole-like Fermi-Dirac distributions (area underneath the curve) and is for 
degenerate ground states directly proportional to the temperature. Third order 
scattering \textbf{(b)} gets logarithmically stronger than 
the second order processes with decreasing 
temperature. The graphical inset illustrate the entangled state at 
temperatures below the characteristic Kondo 
temperature. \textbf{(c)} Ratio of the 2nd and 3rd order contributions when 
an additional external field is applied ($S=1/2$, $g=2$, $J\rho_0=-0.1$,
$\omega_0=100$~meV). The inset show $T_K$ as defined by 
$\Gamma^{(2)}/\Gamma^{(3)}=1$ for different fields $B$. \textbf{(d)} Same as 
(c) but for a $S=1$ system with varying magnetic anisotropies $D$ and $E=0.2D$. 
At $D\approx-1.5$~meV, the strong correlation regime cannot be reached even at 
$T\rightarrow 0$. Figure adapted from 
reference \cite{Ternes15}.}
\label{fig:Sample_entanglement}
\end{figure}

The full description of a spin system in the strong coupling regime 
requires complex theoretical methods like numerical renormalization group 
theory \cite{Wilson75, Hewson97, Bulla08} which are beyond the scope 
addressed here. 
Nevertheless, we can discuss some of the physical consequences within the 
perturbative model. In contrast to the examples discussed in 
section \ref{sec:Anisotropy}, the two lowest degenerate ground states of the 
Co/Cu$_2$N system, as well as any $S=1/2$ system, have weights in states that 
are separated by $\Delta m=\pm1$. Thus, at 
zero field, electrons from the substrate can efficiently flip between these two 
states. The computation of the transition rate between 
the two degenerate states $|\Psi_i\rangle$ and $|\Psi_f\rangle$ of the spin 
system up to second order only is directly proportional to the temperature 
(Figure~\ref{fig:Sample_entanglement}a) \cite{Ternes15}: 
\begin{equation}
\Gamma_{12}^{(2)\rm s\rightarrow 
s}=\frac{2\pi}{\hbar}(J\rho_s)^2|M_{12}|^2k_BT.
\label{equ:Gamma_SS(2)},
\end{equation}
with the scattering matrix $M_{12}$ as defined in equation \ref{equ:M}.

These scattering processes tend to change the spin 
polarization of the electronic states in the sample \emph{near the adsorbate} 
to be correlated with the localized spin. Nevertheless, this local correlation 
will be quickly destroyed by decoherent scattering processes with the remaining 
electron bath, which we can assume to be large and dissipative. This 
decoherence rate is also proportional to the temperature, 
$\Gamma_{\rm decoh}\propto 
k_BT$ \cite{Delgado12}, but usually stronger so that no highly correlated state 
can form. This picture changes when we additionally
consider third order scattering processes which yield the probability 
\cite{Ternes15}:
%
\begin{eqnarray}
\Gamma_{if}^{(3)\rm s\rightarrow 
s}=\frac{4\pi}{\hbar}(J\rho_s)^3\int_{-\infty }^{ \infty}d\varepsilon\,\sum_m 
\Re(M_{fi}M_{mf}M_{im})\nonumber\\
\times\left[g(\varepsilon_{mi}-\varepsilon,T)+g(\varepsilon_{im}-\varepsilon,
T)\right ]
f(\varepsilon, T) [
1-f(\varepsilon-\varepsilon_ { if }, T)].
\label{equ:Gamma_SS(3)}
\end{eqnarray}

Due to the growing intensity of $g(\varepsilon= 0)$ at reduced temperatures, 
for 
temperatures $T\rightarrow 0$, the scattering 
$\Gamma^{(3)}$ decreases significantly more slowly than $\Gamma^{(2)}$ (see 
figure~\ref{fig:Sample_entanglement}b) so that their ratio steadily increases 
(see figure~\ref{fig:Sample_entanglement}c):
\begin{equation}
\frac{\Gamma^{(3)}}{\Gamma^{(2)}}\approx 
J\rho_s\ln\left(\frac{k_BT}{\omega_0}\right).
\label{equ:Gamma3_SS}
\end{equation}
In contrast, the decoherent scattering rate  
with the bath $\Gamma_{\rm decoh}$ lacks localized scattering centers and 
therefore has no significant third order 
contributions. Equation \ref{equ:Gamma3_SS} leads to a characteristic 
temperature, the Kondo temperature $(T_K)$, where $\Gamma^{(3)}$ and 
higher order scattering terms become the dominant processes and 
perturbation theory breaks down \cite{Kondo64, Suhl65, Nagaoka65, Hewson97}:
\begin{equation}
T_K\approx \frac{\omega_0}{k_B}\exp\left(\frac{1}{J\rho_s}\right).
\label{equ:TK}
\end{equation}
Below this temperature the assumption used up to now, i.\,e.\ that the 
electronic states in the sample are not influenced by the presence of the spin 
system, no longer applies. Using exact methods like the modified Bethe ansatz 
\cite{Andrei80, Wiegmann80} or numerical renormalization group theory 
\cite{Bulla08} reveals that the sample electrons in the vicinity rather form an 
entangled state with the impurity, i.\,e.\ 
\begin{equation}
\Psi^{\rm total}=\frac{1}{\sqrt{2}} 
\left(\left|\downarrow^{\rm s}\right\rangle\left|+\frac12\right\rangle-
\left|\uparrow^{\rm s}\right\rangle\left|-\frac12
\right\rangle\right),
\label{equ:Kondo-state}
\end{equation}
as illustrated in the inset of
figure~\ref{fig:Sample_entanglement}b. This combined state is quite 
complicated because the electronic states in the sample are continuous in 
energy 
and extend spatially, and therefore strongly alter the excitation spectrum of 
the adsorbate \cite{Costi00, Zitko09}. 

An externally applied magnetic field will counteract the formation of the 
singlet Kondo state. However, while the precise calculation at finite 
temperature and field is rather difficult, we can use our definition of $T_K$, 
which is the break-down of the perturbative approach, 
to estimate the behavior at applied field. Figure 
\ref{fig:Sample_entanglement}c shows the result for a prototypical $S=1/2$ 
system. We see that the increasing field reduces the Kondo temperature $T_K$ 
and 
that at $B\geq B_C$ the 3rd order scattering rate $\Gamma^{(3)}$ will not 
exceed 
the 2nd order scattering rate $\Gamma^{(2)}$ even at $T\rightarrow0$. In this 
example the ratio $g\mu_BB_C/k_BT_K\approx3$ differs from the usually cited 
ratio of $\approx0.5$ at which the splitting of the Kondo state can be observed 
\cite{Costi00}. Note also, that this approach has neglected the 
renormalization of the gyromagnetic factor due to the coupling 
with the substrate (see page~\pageref{p:renorm}) \cite{Wolf69}. The 
relationship between 
the Kondo temperature and the critical field can be described with a 
simple equation,
\begin{equation}
\frac{B_C(T)}{B_C(T=0)}=1-\left(\frac{T}{T_K(B=0)}\right)^{\alpha}
\label{equ:BC},
\end{equation}
with the exponent $\alpha\approx3$ (see inset figure 
\ref{fig:Sample_entanglement}c). Note, that equation \ref{equ:BC} is formally 
identical to the relation between the critical magnetic field and the 
transition 
temperature in classical, BCS-like, superconductors \cite{Tinkham75}.

We can use the same approach to determine under which conditions a 
non-degenerated high-spin system might enter the strongly correlated Kondo 
state. As an example we will use a spin $S=1$ system, similar to the one 
discussed 
in section \ref{sec:CoH}. While we have already seen that \emph{virtual} 
coherences lead to an energy 
renormalization (see section \ref{sec:Bloch-Redfield}), we now ask about the 
interplay between magnetic anisotropy and the high-spin Kondo screening phase. 
Figure \ref{fig:Sample_entanglement}d shows the result, which behaves similarly 
to the $S=1/2$ system in an external magnetic field. The introduction of an 
easy axis 
anisotropy reduces $T_K$, eventually prohibiting the formation of the highly 
correlated state Kondo state even at zero temperature. In the  example here, 
that uses parameters close to the ones discussed for CoH on $h$-BN, the 
critical 
anisotropy is $D\approx-1.5$~meV. Thus, the evaluation of the $S=1$ system with 
$D\leq-3$~meV (see figure \ref{fig:Spin1-renorm}a) in terms of a purely 
perturbative model is appropriate.

\subsection{The strong coupling regime}
\label{sec:Kondo-fano}

At temperatures $T\ll T_K$ the zero bias anomaly peak at the Fermi energy can 
no longer be well  reproduced by a temperature broadened logarithmic function 
which in any case must diverge for $T\rightarrow 0$. While the quantum system 
enters the Fermi liquid regime \cite{Nozieres74}, the zero-bias peak remains 
at finite width and amplitude. Such a bias dependent peak is much better 
described by either a Lorentzian function or the so-called Frota function 
\cite{Frota92, Prueser11, Zitko11a} which has additional weight at elevated 
biases to account for the logarithmic tails:
\begin{equation}
g_{\rm 
Frota}(\varepsilon)=\Re\left(\sqrt{\frac{i\Gamma_F}{i\Gamma_F+\varepsilon}}
\right).
\label{equ:Frota}
\end{equation}

This peak reveals itself as an apparent increase of 
the local density of states or, more precisely, as a weight increase of the 
spectral function of the many particle system in tunneling experiments. If an 
additional coherent tunneling channel exists, then interferences can change the 
differential conductance significantly, leading to strongly asymmetric peaks or 
even dips in the spectrum \cite{Ternes09}. This behavior is quite easy to 
understand; let us assume that the bare Kondo peak can be well described by a 
Lorentzian 
which has the transfer 
function $t_1=(1+i\varepsilon)^{-1}$, with $\varepsilon=(E-E_0)/\Gamma$ as the 
normalized energy. Additionally, we will assume an energy independent constant 
direct tunneling channel which we describe without any restrictions as $t_0 
=-(1-iq)^{-1}$.

Tunneling experiments detect the absolute square of the sum of all possible 
transfer channels. Using the above defined $t_0$ and $t_1$ results in the 
well known Fano equation \cite{Fano61}
\begin{equation}
\label{equ:Fano}
g_{\rm 
Fano}(\varepsilon)=\left|t_0+t_1\right|^2=\left|-\frac{1}{1-iq}+\frac{1}{
1+i\varepsilon } \right|^2=
\frac{1}{1+q^2}\frac{(q+\varepsilon)^2}{1+\varepsilon^2},
\end{equation}
in which $q$ determines the asymmetry. Equivalently, equation 
\ref{equ:Frota} can be generalized \cite{Ujsaghy00} leading to an asymmetric 
Frota function as:
\begin{equation}
g_{\rm 
Frota}(\varepsilon, 
q)=\frac{q^2-1}{q^2+1}\Re\left(\sqrt{\frac{i\Gamma_F}{i\Gamma_F+\varepsilon}}
\right)+\frac{2q}{q^2+1}\Im\left(\sqrt{\frac{i\Gamma_F}{i\Gamma_F+\varepsilon}}
\right).
\label{equ:Frota2}
\end{equation}
For bare metal adatoms on metal surfaces, typically $q\approx 0$ has been found 
\cite{Schneider02, Wahl04}, in which case $g(\varepsilon)$ becomes a dip. For 
$q \approx 1$, the line shape becomes strongly asymmetric, and for 
$q\rightarrow\infty$, the peak is recovered.

At $T\ll T_K$ the half-width $\Gamma$ of the Lorentzian (or the effective 
half-width of the Frota function $\Gamma \approx 2.54\Gamma_F$ \cite{Frota92}) 
is directly 
related to the Kondo temperature and the correlation energy of the 
Kondo state: $k_BT_K=\Gamma$. This zero-temperature result can be expanded in 
the Fermi-liquid framework to elevated temperatures using corrections in 
first leading order \cite{Nagaoka02}: 

\begin{equation}
\label{equ:Kondo-FL}
\Gamma(T)=\sqrt{(k_B T_K)^2+\alpha (k_B T)^2}.
\end{equation}
Here, one expects theoretically $\alpha=2\pi$, close to the 
experimentally observed $\alpha=4.5$ in quantum dots \cite{Cronenwett98}, 
$\alpha\approx5$ for individual Ti atoms on Cu(100) \cite{Nagaoka02}, and 
$\alpha=5.4\pm0.1$ in low temperature measurements on (relatively) strongly 
coupled Co atoms on small patches of Cu$_2$N \cite{Otte08a}.
\begin{figure}[tbp]
\includegraphics[width=\columnwidth]{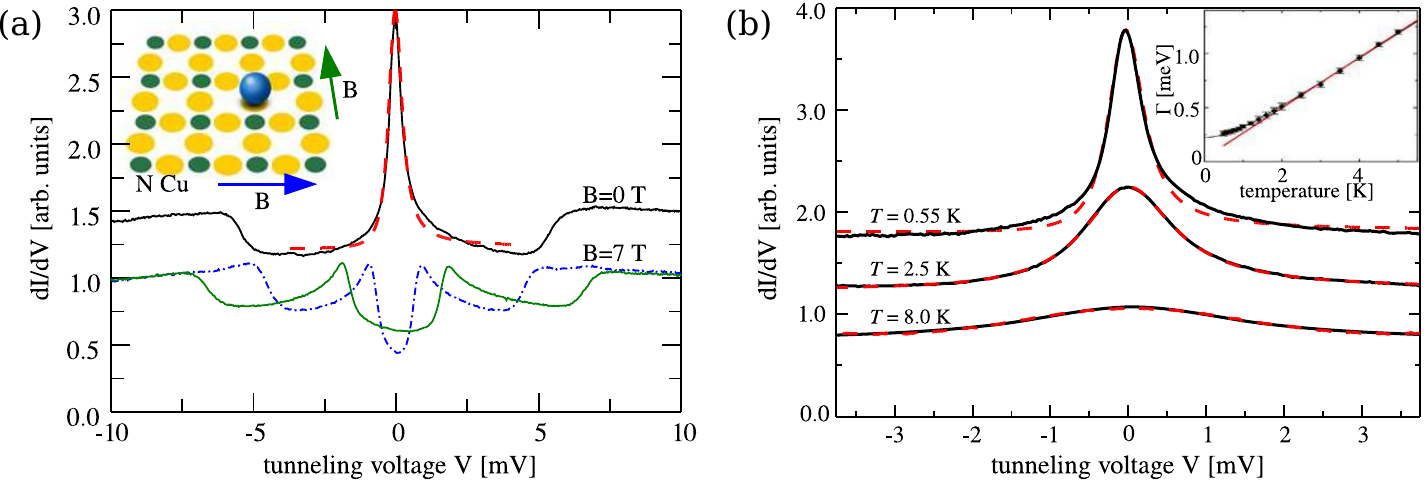}
\caption{\textbf{(a)} Spectra of a Co atom on Cu$_2$N/Cu(100) measured at a 
temperature of $T = 0.5$~K and an external magnetic field of $B = 0$ and
$7$~T. The zero-field spectrum reveals a Kondo peak which is fitted to a 
temperature-broadened Fano function (red dashed line) and IETS steps
at $V\approx\pm6$~mV due to spin-flip transitions \cite{Otte08a}. At 
$B = 7$~T, the spin-flip step positions have moved and the Kondo
peak has split. This splitting depends strongly on the direction of the 
magnetic 
field. \textbf{(b)} The central Kondo peak measured
at different sample temperatures (black lines) and fitted with a
temperature-broadened Fano function (red dashed lines) Inset:
intrinsic half-width $\Gamma$ extracted from the fits. The red line shows the
linear behavior of $\Gamma$ at high temperature and has a slope of
($5.4\pm0.1)k_B$. The black line is a fit to equation \ref{equ:Kondo-FL}. 
Figure 
adapted from reference \cite{Ternes09}.}
\label{fig:Co-CuN-Strong}
\end{figure}

The latter system is of particular interest because here the many-body Kondo 
effect and the magnetic anisotropy, usually described within a single-particle
approximation, are of similar strength. Figure \ref{fig:Co-CuN-Strong}a shows 
such spectra measured at $T=0.5$~K at zero magnetic field $B$ and at $B = 
7$~T \cite{Otte08a}. The spectra can be well described with an asymmetric 
Lorentzian and additional conductance steps at voltages that enable scattering 
of the spin system. Surprisingly, these steps are well described using only a 
second order perturbation spin-flip model and do not show any third order 
logarithmic contributions. This means that the probability 
of the third order scattering channels must be closed due to the ground-state 
correlation between the localized spin and the substrate electrons.

Under an applied magnetic field, the peak splits and the  
spin-flip excitation step positions shift in energy. 
Interestingly, the single-particle magnetic anisotropy 
Hamiltonian of equation \ref{eq:Atom-Hamilonian} not only describes the 
energy shift of the inelastic conductance steps accurately, but also the 
positions of the split Kondo peak. The peak width at zero-field corresponds to 
a Kondo temperature of $T_K=2.3\pm0.3$~K and the broadening of the peak at 
elevated temperatures follows equation \ref{equ:Kondo-FL} as expected from 
Fermi 
liquid theory \cite{Nozieres74, Nagaoka02} and as shown in the inset of 
figure \ref{fig:Co-CuN-Strong}b. Note, that at even higher temperatures, where 
$T\gg T_K$, we reach again the weak-coupling limit and the spectrum is better 
described with the perturbative model and temperature broadened logarithmic 
divergences as shown in figure \ref{fig:Co-CuN}. Due to the limitations 
of equation \ref{equ:Kondo-FL}, which is only valid at $T\lesssim T_K$, fitting 
temperature dependent experimental data in the weak coupling limit with 
Lorenzian or Frota functions can lead to unphysically high $\alpha$ values 
\cite{Zhang13}. In such cases, $T_K$ is presumably much smaller than the 
experimentally 
accessed temperatures.

\subsection{Spin polarization of the split Kondo peak}
\label{sec:Kondo-split}

As we have seen in the last section, when external magnetic fields which exceed 
the Kondo correlation energy, i.\,e. $g_{\rm eff}\mu_B B > k_BT_K$, are applied 
to a strongly coupled Kondo system the zero-bias resonance splits into two 
distinct parts (Figure \ref{fig:Co-CuN-Strong}a) \cite{Otte08a}. However, 
while this has been observed also for other systems \cite{Chen08, Dubout15}, the 
spin-resolved properties of such a split Kondo state and, in 
particular, the amount of spin polarization of the two resulting peaks remains 
elusive \cite{Patton07, Seridonio09}. While there is one early spin-resolved 
measurement of a split Kondo state \cite{Fu12}, the asymmetry of the peaks was 
not studied systematically and a comprehensive picture was only found recently 
using individual Co adatoms on Cu$_2$N as a test system \cite{Bergmann15}.

In this experiment a spin-polarized tip was prepared by picking up Mn
atoms from the Cu$_2$N surface with a conventional STM tip \cite{Loth10}. 
Since the
measured asymmetry in the spectrum is the product of sample and tip spin 
polarization, $\eta^{\rm eff}=\eta_{\rm sample}\times\eta_{\rm tip}$, it is
crucial to characterize the degree of spin polarization of the
tip. This was done by spin-resolved measurements of individual Mn
atoms on the same surface. 
As discussed in section \ref{sec:Ani-Fe-Mn} and displayed in figure 
\ref{fig:spec-Fe-Mn}c and \ref{fig:spec-Fe-Mn}e, Mn atoms on Cu$_2$N show one
spin-flip excitation at about 1 mV. When measured with a spin-polarized tip, 
the heights of the inelastic steps at
positive and negative voltage ($h^+$ and $h^-$) differ \cite{Loth10, Loth10a}. 
The asymmetry of the step heights $\eta^{\rm eff}=(h^--h^+)/(h^-+h^+)$ can now 
be measured as a function of the external magnetic field. Because of the small 
magnetocrystalline anisotropy of Mn the nominal spin polarization of the step
is $\eta_{\rm Mn}=1$ and therefore the experimental $\eta^{\rm 
eff}$ is a quantitative measure of the tip spin polarization $\eta_{\rm tip}$.
The magnetic-field dependence of the tip's spin
polarization was found to be consistent with a paramagnetic behavior of 
the Mn atoms on the sample and the metallic tip. Hence the
magnetic field-dependent spin polarization of the tip can be well described 
by the Brillouin function:
\begin{equation}
\eta_{\rm tip}=\eta_{\rm max}\left(\frac{2S+1}{2S}\coth\left(\frac{2S+1}{2S}
x\right)-\frac { 1 } { 2S } \coth\left(\frac{1}{2S}x\right)\right),
\label{equ:brillouin}
\end{equation} 
with $x=\frac{g\mu_BS|\vec{B}|}{2k_BT}$, $g=2$, $S=5/2$, and a maximal 
polarization $\eta_{\rm max}=0.5\pm0.05$. 

\begin{figure}[tbp]
\begin{center}
\includegraphics[width=0.6\columnwidth]{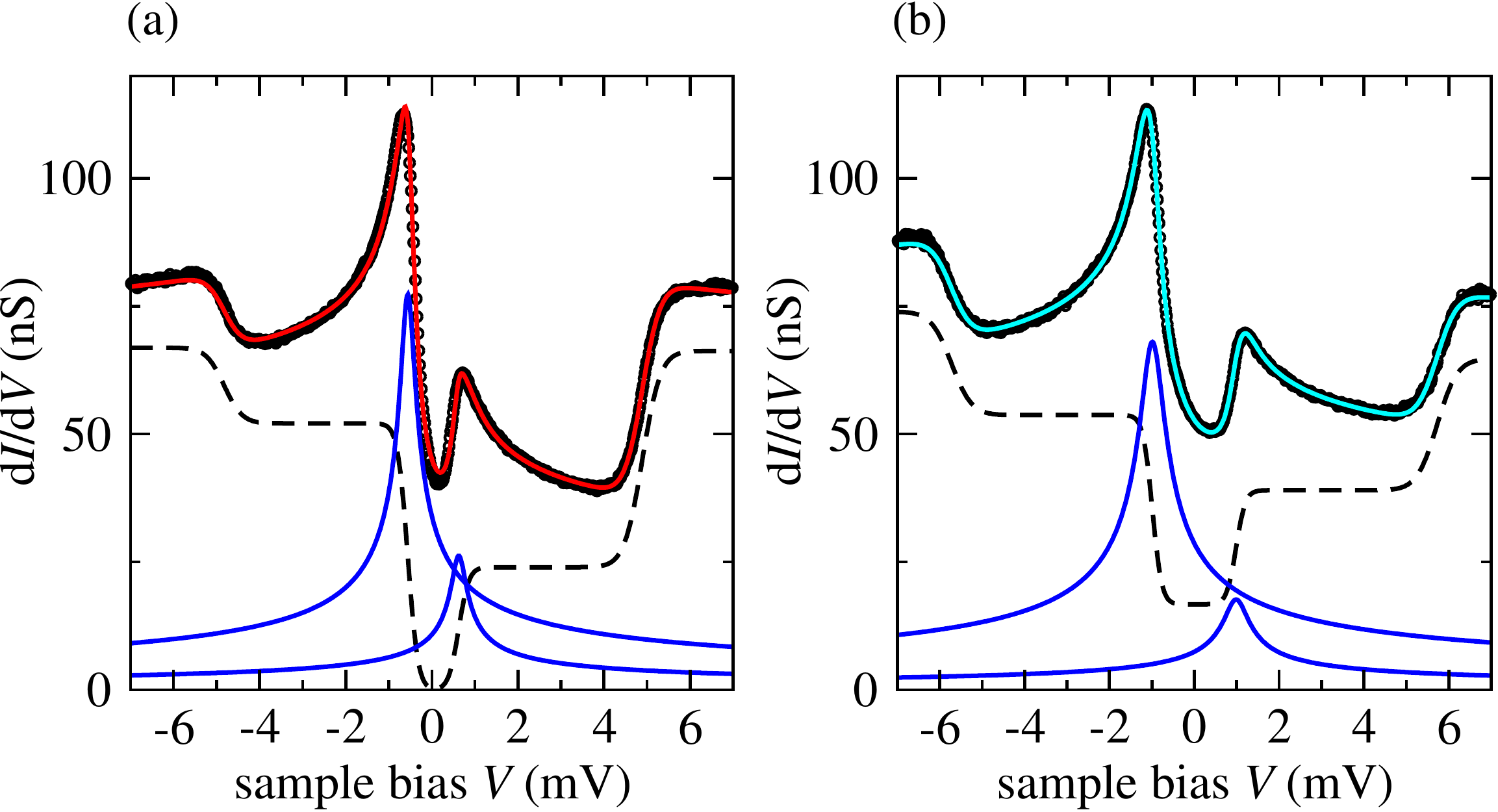}
\caption{\textbf{(a)}, \textbf{(b)}~Experimental data (circles) measured on 
Co atoms on Cu$_2$N at $B=5$~T applied along the vacancy row (a) and along the 
N-row (b). Stabilization setpoint $\sigma = 80$\,nS, $V=+25$\,mV, 
$I=2$\,nA. Upper solid line: best fit using the sum of an asymmetric double 
step function (dashed line) and two Frota functions (lower solid lines). Figure 
adapted from reference \cite{Bergmann15}.}
\label{fig:Co-SP1}
\end{center}
\end{figure}

Utilizing this spin-polarized tip at magnetic fields leads to spectra on 
the Co atoms as displayed for $B=5$~T in figure 
\ref{fig:Co-SP1}. The two spectra differ by the adsorption site of the Co atom 
and therefore correspond to the situation where the magnetic field is 
either applied along (a) or perpendicular (b) to the main anisotropy axis 
(see also figure \ref{fig:Co-CuN-Strong}a). These spectra can be excellently 
fitted by using the sum of two double-step functions with step-energies 
symmetrically around zero-bias and asymmetric step-heights, and two Frota 
peak-functions (equation \ref{equ:Frota}) at the energies of the energetically 
low-lying step, with identical half-width but different intensities. 
Surprisingly, the step-functions, that are due to inelastic spin-flip 
excitations, can be fully described in a second order transport model using the 
standard anisotropy Hamiltonian of equation \ref{eq:Atom-Hamilonian} taking 
into account the spin-polarization governed by equation \ref{equ:brillouin} 
\cite{Bergmann15, Loth10, Loth10b}.

\begin{figure}[tbp]
\begin{center}
\includegraphics[width=0.6\columnwidth]{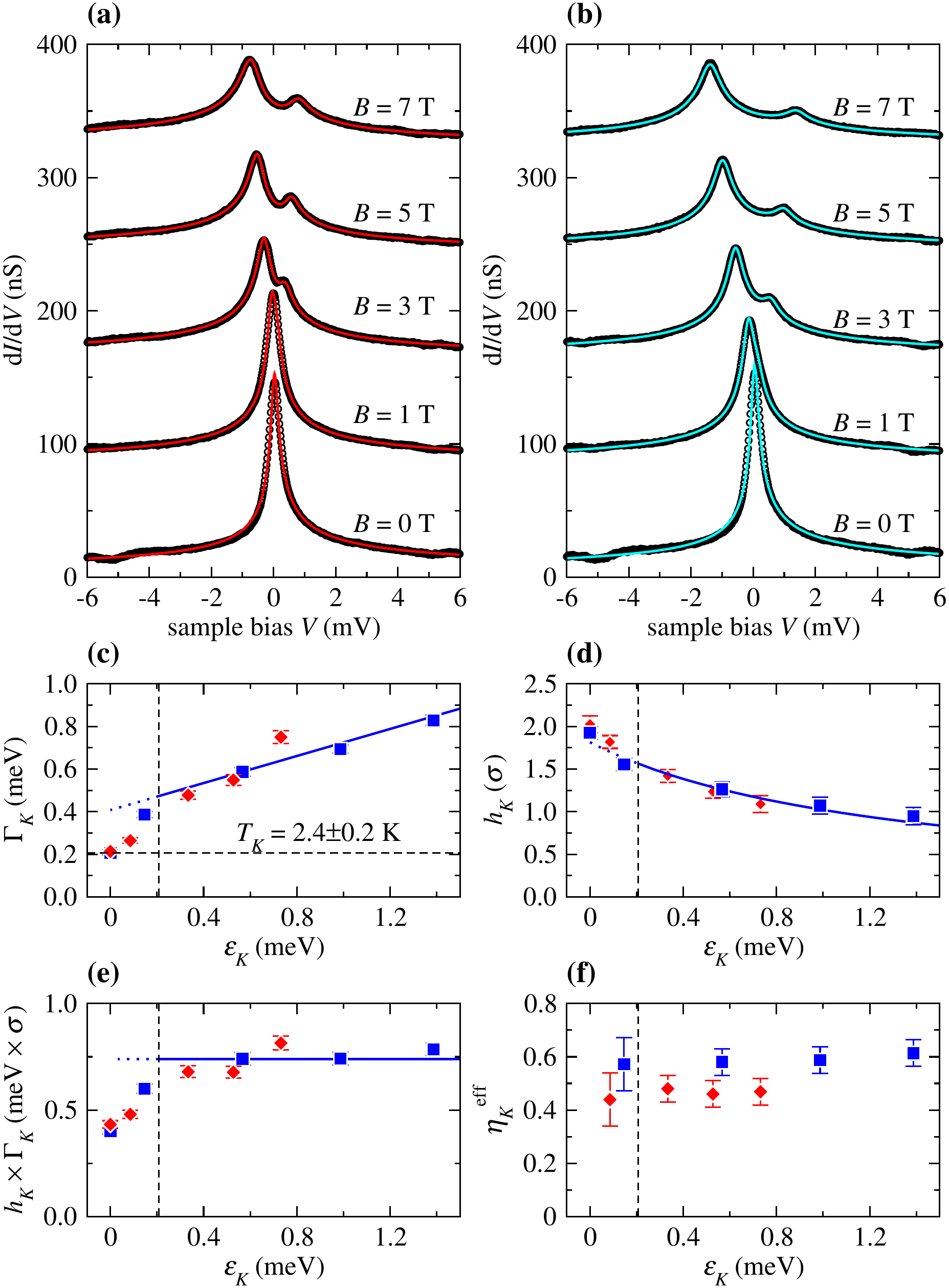}
\caption{The splitting and the polarization of the
Kondo peak. \textbf{(a)}, \textbf{(b)} Kondo-related experimental data 
(circles) and fit with two Frota functions (solid lines) for magnetic fields 
applied along the vacancy row (a) and along the N-row (b). Curves in field 
are shifted vertically for better visualization. \textbf{(c)} Fit results for 
the width $\Gamma_K$, \textbf{(d)} the peak 
height $h_K$, \textbf{(e)} the area $\Gamma_K\times h_K$, and \textbf{(f)} the 
experimental peak asymmetry $\eta^{\rm eff}_K$ plotted against the splitting 
energy $\varepsilon_K$. Error bars indicate the 90\% confidence interval of the 
fit 
and
stem mainly from the interdependence of the parameters (nonzero off-diagonal 
covariance matrix elements). The horizontal and vertical dashed lines in 
(c)--(f) mark the characteristic Kondo energy. Figure 
adapted from reference \cite{Bergmann15}.}
\label{fig:Co-SP2}
\end{center}
\end{figure}

We now continue to evaluate spectra measured at different fields by 
subtracting the inelastic, step-like contributions, and analyzing the Kondo 
peaks using Frota functions as shown in figure \ref{fig:Co-SP2}a and 
\ref{fig:Co-SP2}b.
Interestingly, the behavior of the Kondo peak seems to be related to the 
splitting energy as the data applied along the vacancy row and along the N-row 
fall on top of each other when plotted versus the peak energy $\varepsilon_K$. 
Plotted with this abscissa, the peak height $h_K$ decays with $1/\varepsilon_K$ 
as shown in figure 
\ref{fig:Co-SP2}d.

From the peak width $\Gamma^0_K$ measured at $B=0$ the Kondo temperature is 
extracted to $T_K=2.4\pm0.2$~K, equal to the previously stated 
$T_K=2.3\pm0.3$~K (Figure \ref{fig:Co-CuN-Strong}) \cite{Otte08a}, which 
corresponds to a correlation energy of $k_BT_K=0.21$~meV. When the splitting 
exceeds this energy,
we observe that the spectral weight of the correlated state, i.\,e.\ the area 
$h_k\times\Gamma_K$ underneath the Kondo peak, settles at about twice the zero 
field value and remains afterward constant irrespective of
magnetic field strength up to $7$~T or, equivalently, $\varepsilon_K\approx 
6\times k_BT_K$ 
(Figure~\ref{fig:Co-SP2}e). Presumably, this
stems from the lifting of the spin degeneracy of the Kondo
singlet state at significant field. Note, that the the transport 
measurement, as it is performed here, might introduce additional spectral 
weight created by the hot electrons at finite bias, similar as observed with 
spin-polarized currents on Fe atoms on Cu$_2$N \cite{Loth10b} and 
rationalized with a 3rd order scattering transport model including 
rate-equations \cite{Ternes15}. Here, however, we have not observed any 
strong current dependencies.

In the regime, where $\varepsilon_K > k_BT_K$, our data suggest a linear 
dependence 
of $\Gamma_K$ on $\varepsilon_K$, (Figure \ref{fig:Co-SP2}c) which leads to a 
surprisingly simple equation:
\begin{equation}
\Gamma_K(\varepsilon_K)=(2\pm0.1)\Gamma_K^0+(1\pm0.03)\frac{1}{\pi} 
\varepsilon_K,
\end{equation}
where we relate the linear term to an increased scattering
with bulk electrons which reduces the lifetime \cite{Loth10}. The factor 
2 hints to an equal contribution of correlations induced by the bulk electrons 
and by the biased electrons in the transport experiment when $\varepsilon_K$ 
exceeds the Kondo energy scale.

Finally, figure \ref{fig:Co-SP2}f shows that the effective polarization 
of the Kondo peak $\eta^{\rm eff}_K$ for the two different field directions 
stays approximately constant and is close to $0.5$, with a small systematic 
offset between the two field directions. In order to derive the spin 
polarization of the split Kondo peak $\eta_K$, we need to consider the 
spin-polarization of the tip $\eta_{\rm tip}\approx0.5$ leading to:
\begin{equation}
\eta_K=\frac{\eta^{\rm eff}_K}{\eta_{\rm tip}}\approx1.0. 
\end{equation}

In summary, this experiment shows that the split Kondo state is an 
excellent 
source for spin polarized electrons and might serve as a
magnetic probe in transport measurements, similar to the
fully spin-polarized magnetic field split superconducting
state \cite{Meservey94, Eltschka14}. Experimentally, this could be 
realized, for example, by attaching a magnetic molecule that exhibits a
Kondo resonance to the tip apex. The Kondo resonance
would then act as an energy-dependent spin filter for
quantitative spin-resolved STM measurements.

\section{Coupled spin systems}
\label{sec:Coupled_seystems}

Up to now we have discussed the spectroscopic features due to 
second and third order scattering of electrons on single 
spin systems and the occurrence of correlation effects which lead to the Kondo 
effect. Now we will turn our attention to coupled systems where two or more 
individual spins interact. 
On one hand, this coupling can take place within a single molecule, where 
spin centers are coupled via exchange and superexchange interactions and the
organic ligands control their properties, such as their anisotropies and 
effective spin states \cite{Friedman10}. These molecules form a promising class 
of coupled spin systems called single molecule magnets and have been studied 
quite extensively (for a review see for example \cite{Gatteschi08, Cavallini08, 
Friedman10}). However, up to now few scanning tunneling spin excitation 
spectroscopy measurements have been performed due to the fragility of these 
molecules with complex geometry \cite{Kahle12, Burgess15} which, upon 
adsorption, easily alter their 
magnetic properties \cite{Pineider07, Voss08, Mannini08, Rogez09, Saywell10}. 
In section \ref{sec:Mn12} we will briefly discuss the spectroscopic features 
measured on the prototypical single molecular magnet manganese-12-acetate
\cite{Kahle12}.

On the other hand, the coupling between different spins can be achieved by 
deliberately organizing the individual spins, such as transition metal 
atoms, via lateral or vertical atom manipulation on thin insulating or metal 
substrates. Such experiments revealed, for example, that chains built of 
up to 10 Mn atoms on Cu$_2$N showed a pronounced odd-even behavior that was
clearly visible in the spectroscopic data, which could be rationalized by 
considering the chains as a singular quantum mechanical object in which the Mn 
atoms are strongly coupled via Heisenberg exchange to their next neighbors 
\cite{Hirjibehedin06, Hurley11}, and second-next neighbors \cite{Fernandes09}. 
Furthermore, bistable behavior and spin waves were observed in weakly 
antiferromagnetically coupled Fe chains on Cu$_2$N \cite{Spinelli14, 
Spinelli15a} Here, also ferromagnetically coupled 2D structures containing only 
a dozen Fe atoms demonstrated extremely long living N\'{e}el-states with 
lifetimes reaching hours at low temperatures \cite{Loth12}. Recently, it has 
been shown that the spin state of such structures can be read out by 
their influence onto the spin-dynamics of a near-by spin structure consisting of 
3 Fe atoms \cite{Yan16}. Additionally, first successful attempts have been 
undertaken to use deliberately built spin-structures to explore the 
two-impurity Kondo system \cite{Spinelli15}, the chiral magnetic 
interaction between atomic spin systems \cite{Khajetoorians16}, or the quantum 
criticality of the $xxz$ Heisenberg chain model \cite{Toskovic16}.

In section \ref{sec:S1S12} we will discuss the spectral features 
of a prototypical example: the antiferromagnetic coupling between two spins 
with $S=1/2$ and $S=1$. On such a system which has been recently studied in a 
vertical geometry, i.\,e.\ with the spins attached to tip and sample electrodes 
\cite{Muenks17}, we will show how correlations due to 
higher order scattering influences the spectrum. Then we will turn to the 
experimentally studied Fe--Co (section \ref{sec:Fe-Co}) \cite{Otte09} and 
Co--Co (section \ref{sec:Co-Co}) \cite{Spinelli15} dimers on Cu$_2$N 
whereby the latter is an example for the two-impurity Kondo system 
\cite{Jones87, Jones88, Jones89} which can be deliberately tuned into different 
many-particle correlation phases. Furthermore, we will discuss correlation and 
entanglement in spin chains containing up to 12 spin sites which have been 
experimentally assembled from Fe and Mn atoms on Cu$_2$N \cite{Choi15}.

\subsection{The spectrum of a prototypical molecular magnet}
\label{sec:Mn12}

\begin{figure}[tbp]
\centering
\includegraphics[width=\textwidth]{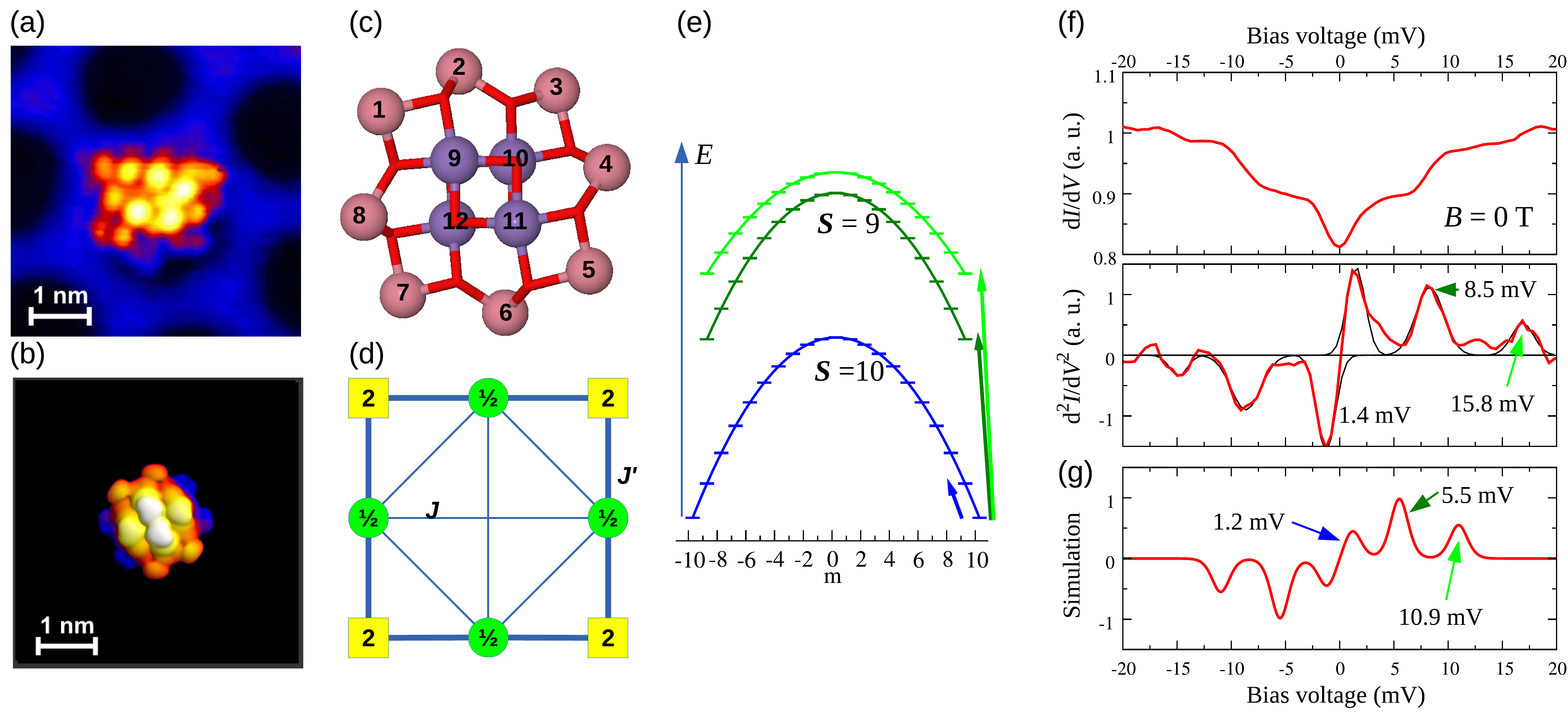}
	\caption{Spin excitations on the prototypical single molecular 
magnet Mn$_{12}$-Acetate$_{16}$. \textbf{(a)} STM constant-current topography 
of a Mn$_{12}$ molecule adsorbed on $h$-BN/Rh(111) measured at $T=1$~K, 
$V=-1$~V, and $I=45$~pA. \textbf{(b)} DFT calculation of the free molecule 
\cite{Kahle12}.  
\textbf{(c)} Magnetic core of the Mn$_{12}$ molecule. The 12 spin-sites are 
indexed. Pink circle correspond to a $S=2$ site, violet circle to a $S=3/2$ 
site. \textbf{(d)} Simplified 8-site model. \textbf{(e)} Schematic energy 
diagram of a Mn$_{12}$ molecule in the giant spin approximation $H = 
DS_z^2$ with $S = 10$ ground state and easy axis
anisotropy $D < 0$ without external magnetic fields applied. Arrows
indicate possible excitations by interaction with the tunneling electron
which obey the spin selection rule $\Delta m = 0, 
\pm1$. \textbf{(f)} Typical $dI/dV$ and $d^2I/dV^2$ spectra ($V = 100$~mV, 
$I = 20$ pA, $T = 1.5$~K) of a Mn$_{12}$ molecule adsorbed on a $h$-BN/Rh(111) 
surface at $B = 0$~T reveal low energy spin-flip excitations which manifest
themself as steps (peaks) in the $dI/dV$ ($d^2I/dV^2$) spectrum. \textbf{(g)} 
Simulated $d^2I/dV^2$ spectra using the 8-spin model. Figure adapted from 
reference \cite{Kahle12}.
}
	\label{fig:Mn12} 
\end{figure}

In single molecular magnets, spin carrying atoms are arranged
within a molecular framework in a way that their magnetic states
can be described as a single giant spin. Manganese-12-acetate (Mn$_{12}$)
is composed of a Mn$_{12}$O$_{12}$ core surrounded by 16 acetate groups and 
represents a prototypical molecular magnet with a total spin $S_T =
10$. Resulting from its relatively large magnetic anisotropy $D$, it has a
magnetization reversal barrier height of $\Delta \varepsilon=DS^2 = 6$~meV in 
bulk,
enough to produce very long spin relaxation times at low temperatures 
\cite{Sessoli93}. Throughout many studies, the immobilization of
Mn$_{12}$ molecules at surfaces has been
found to be difficult, as its fragile structure changes easily upon
deposition, thus altering its magnetic properties \cite{Pineider07, Voss08, 
Mannini08, Rogez09, Saywell10}. 
To circumvent the fragmentation of the molecule during in-vacuo deposition
due to its thermal instability, we use electrospray ion beam deposition 
(ES-IBD) 
as gentle deposition method \cite{Ouyang03, Rauschenbach06, Rauschenbach09, 
Johnson11, Rauschenbach16} to bring Mn$_{12}$ molecules on the well-defined 
ultrathin insulating $h$-BN/Rh(111) surface (Figure \ref{fig:Mn12}a). 

To address the question of whether these molecules still exhibit
their striking magnetic properties, we measured the differential conductance 
$dI/dV$ at low temperature $T = 1.5$~K on top of the
Mn$_{12}$ molecules and observe symmetric features around the
Fermi energy \cite{Kahle12}. The spectra show a step-like structure in
$dI/dV$ which corresponds to peaks in the numerically derivated
$d^2I/dV^2$ as shown in figure \ref{fig:Mn12}f. The innermost step is usually
the most prominent one and can be found at 1--2~meV, while
the outer steps can be observed in a range up to 16~meV. 
To interpret these features we omit for the moment the many
spin nature of the system by using the giant spin approximation
in which $S_T = 10$ is fixed. The magnetic anisotropy is responsible
for the zero-field splitting of the spin eigenstates in the
$z$-projection of the magnetic moment $m_z$ and leads to a degenerate ground 
state for  $\left|S_T,m_z\right\rangle = \left|10,-10\right\rangle$ and 
$\left|10,+10\right\rangle$ (see figure \ref{fig:Mn12}e). The model with fixed 
$S_T$ reduces possible magnetic excitations to changes of $m_z$, explaining the 
inner step of the spectra at $\approx\pm 1.4$~mV as the excitation from 
$\left|S_T,m_z\right\rangle = 
\left|10,\pm10\right\rangle$ to $\left|10,\pm9\right\rangle$. In this 
approximation there are no transitions at higher energy possible which obey the 
conservation of angular momentum, i.\,e.\ which only changes $m_z$ by $\pm1$. 
Note, that this is also true when additional higher order anisotropy terms
are accounted for in the Hamiltonian.

To cover excitations that change $S_T$, we have to go beyond the
giant spin picture. The magnetic core of the Mn$_{12}$
molecule contains  12 Mn atoms which are coupled via superexchange by oxygen 
bridges (Figure \ref{fig:Mn12}c). The eight ferromagetically coupled outer atoms 
are thereby in the Mn$^{3+}$ oxidation state and have an individual spin of $S= 
2$ while the four inner atoms are in the Mn$^{4+}$ state with a spin of $S = 
3/2$ that are also ferromagnetically coupled \cite{Katsnelson99}. Between the 
two ferromagnetically coupled sets of spins a strong antiferromagnetic coupling 
leads to a total spin of $S_T=8\times 2-4\times 3/2=10$.

To calculate the low-energy eigenvalues and state vectors of the coupled spin 
system of the Mn$_{12}$ molecule we use a simplified 8-site Hamiltonian which 
reduces the matrix size with $n^2$ elements from $n=10^8$ to acceptable 
$n=10000$. In this model the 
exchange interaction of the four antiferromagnetically coupled dimers with the 
strongest coupling (1--9, 3--10, 5--11, 7--12 in figure \ref{fig:Mn12}c) are 
approximated by four spins $S = 2-3/2 = 1/2$ (Figure \ref{fig:Mn12}d) 
\cite{Katsnelson99}. This can be done 
because the exchange interaction inside these dimers is much larger than all 
other exchange interactions and larger as the energy range of interest in our 
experiment. Thus, the four $S = 1/2$ ''dimer'' spins interact with each other 
and the remaining four $S = 2$ spins (sites 2, 4, 6, 8 
in figure \ref{fig:Mn12}c).

We regard 3 types of exchange interactions in this system which are determined 
by DFT calculations \cite{Kahle12} : (i) Easy-axis anisotropy on the 
individual spins $H_{\rm ani} = D_z(\hat{S}^i_z)^2$, with $D_z=0.48$~meV as the 
anisotropy 
term which is only relevant at the $S = 2$ sites due to the Kramer's degeneracy 
theorem. (ii) Direct Heisenberg spin-spin interaction $H_{\mathcal{J}}= 
\mathcal{J}_{ij}\left(\hat{\bf S}_i\cdot \hat{\bf S}_j\right)$, which couples 
different 
spin sites isotropically and has in the 8-spin model two distinct strengths: A 
relatively strong ferromagnetic coupling $\mathcal{J}'=-9.3$~meV between the $S 
= 2$ and $S = 1/2$ sites and a weaker coupling $\mathcal{J}=-0.25$~meV between 
the more distanced $S = 2$ sites. (iii) Non-collinear Dzyaloshinsky-Moriya 
interactions \cite{Dzylaloshinskii57, Moriya60} $H_{\rm DM} = 
\mathcal{D}_{ij}\left(\vec{S}_i\times \vec{S}_j\right)$, 
in which the Dzyaloshinsky-Moriya vector parameter 
$\mathcal{D}=(2.1,0,0.1)^T$~meV couples neighboring $S = 2$ and $S = 1/2$ 
sites. 
Note, that similar anisotropy and coupling parameters have been found 
earlier by comparison to electron spin resonance measurements 
\cite{Katsnelson99}. The total Hamiltonian is then diagonalized and the spin 
excitation spectrum is calculated up to second order in the scattering 
elements. 
The resulting spectrum (Figure \ref{fig:Mn12}g) agree well with the model, in 
particular considering that the influence of the substrate was neglected for 
the 
parameters of the Hamiltonian.

\subsection{Coupling between a spin 1 and a spin 1/2}
\label{sec:S1S12}

\begin{figure}[tbp]
\centering
\includegraphics[width=\textwidth]{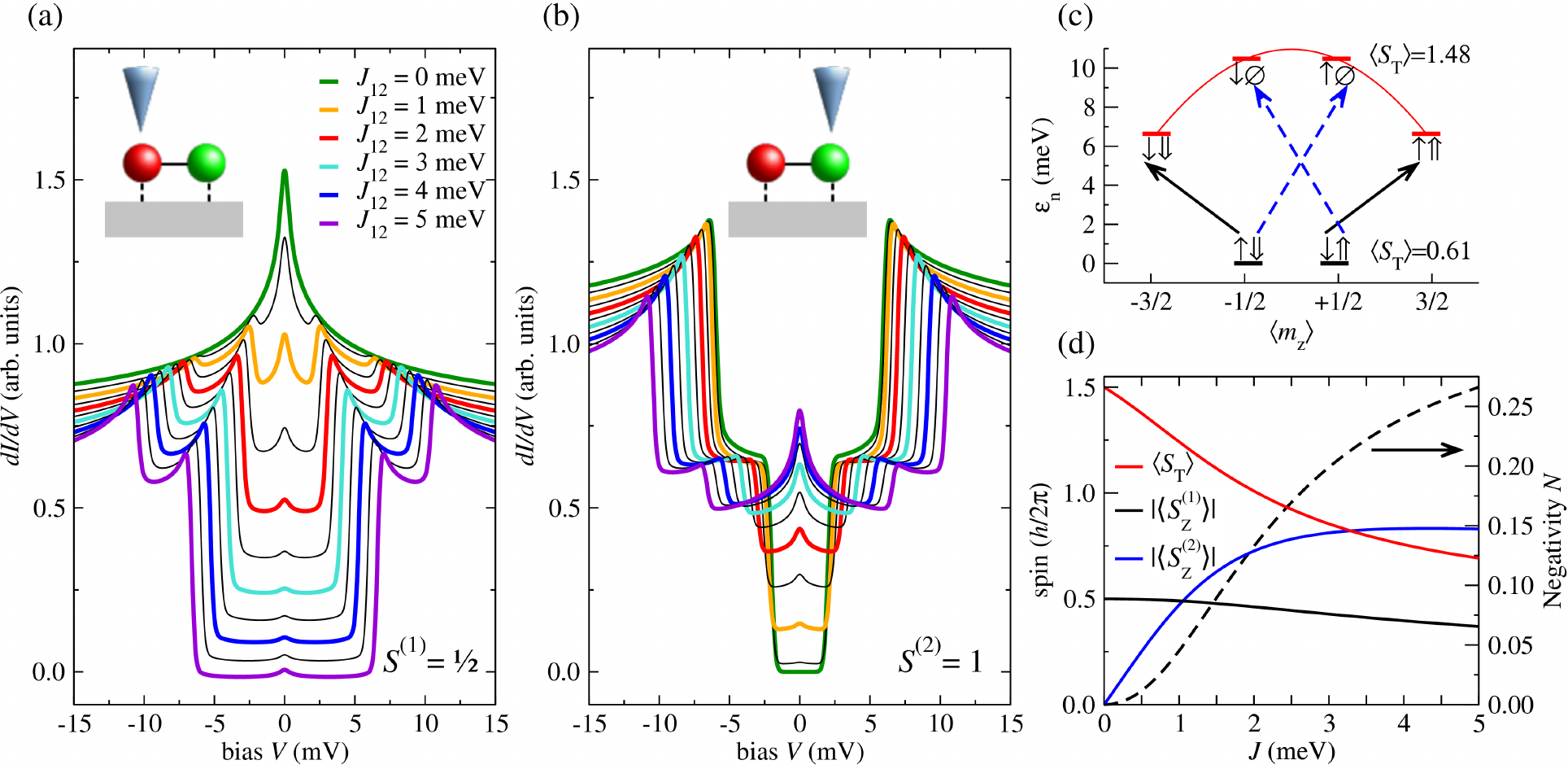}
	\caption{The spectra of an antiferromagnetically  Heisenberg coupled 
dimer consisting of a $S^{(1)}=1/2$ and a $S^{(2)}=1$ system at $T=1$~K and 
$B=0$~T. \textbf{(a)} Simulated spectra on the $S^{(1)}=1/2$ system 
for different coupling strengths. \textbf{(b)} Same as (a) for the $S^{(2)}=1$ 
system with the anisotropy parameters $D=-5$~meV and $E=1$~meV. Both spins are 
coupled to the substrate 
with $J\rho_s=-0.1$. Schematic insets illustrate the probing 
geometry. The spectra are shifted vertically with respect to each other for 
better visibility. \textbf{(c)} Schematic state diagram of the six 
eigenstates at a coupling strength of $\mathcal{J}_{12}=5$~meV. 
The black full arrow indicates the high transition rates on the $S^{(1)}=1/2$ 
spin, the blue dashed arrows on the $S^{(1)}=1$ spin, respectively. 
Arrows indicate the main $m_z$ values of the combined state ($\uparrow=+1/2$,
$\downarrow=-1/2$, $\Uparrow=+1$, $\varnothing=0$, and $\Downarrow=-1$) 
\textbf{(d)} Full lines: Expectation values of the absolute magnetic quantum 
number $\left|\langle S_z^{(i)} \rangle\right|$ on each part of the dimer and 
the total quantum number $\langle S_T \rangle$ of the combined system for one 
of 
the two ground states. Dashed line: Negativity $N$ as a measure of the quantum 
entanglement of the two spin systems.}
	\label{fig:S1-S12} 
\end{figure}

To gain insight into Kondo correlations in coupled structures, we will 
now continue the discussion with a rather small dimer system, which contains 
only two spin centers, one with $S^{(1)}=1/2$ and the other with $S^{(2)}=1$, 
where we want to assume that the degeneracies of the eigenstates are lifted by 
magnetic anisotropy of $D=-5$~meV and $E=1$~meV (see section \ref{sec:CoH}). 
When the isotropic 
antiferromagnetic Heisenberg exchange 
coupling $\mathcal{J}_{12}$ between the two is switched on, we observe new 
step-like increases in the differential conductance that arise from 
spin-flip transitions on the $S^{(1)}=1/2$ species that where absent
before (Figure~\ref{fig:S1-S12}a and b). At small coupling strengths 
$\mathcal{J}_{12}\lesssim1$~meV, the spectrum of the $S^{(2)}=1$ is almost 
unaltered, while at higher coupling the energies and the 
intensities of the transitions change significantly. 
With increased antiferromagnetic coupling the inelastic transitions move to 
higher energy. 
On the spin-$1$ system the 
excitation step at lower energy decreases, while the excitation 
step at higher energy increases in intensity.
In contrast, the step heights on the spin-$1/2$ system 
grow with increased coupling, whereas the energetically lower step is 
always significantly 
stronger then the energetically higher step. Most remarkably, with 
increased coupling the 
intensity of the zero-energy peak on the $S=1/2$ spin diminishes while at the 
same time a zero-energy peak emerges at the $S=1$ site. 
%

To elucidate this behavior, we look at the $(2\times3)$ eigenstates of the 
combined system at a coupling $\mathcal{J}_{12}=|D|=5$~meV 
(Figure~\ref{fig:S1-S12}c). These 
six states can be grouped into a doublet with the total spin expectation value 
of $\langle S_T\rangle\approx1/2$ and $\langle m_z\rangle$ values of 
$\pm1/2$.\footnote{The total spin for a $n$-partite system is calculated as
$\langle S_T\rangle=\sqrt{\langle\Psi| \hat{\bf S}^2_T |\Psi 
\rangle+\frac14}-\frac12$, with $\hat{\bf S}^2_T=\left(\bigotimes_{i=1}^n  
\hat{\bf S}_i\right)^2=
\left(\bigotimes_{i=1}^n\hat{ S}_x^i\right)^2+
\left(\bigotimes_{i=1}^n\hat{ S}_y^i\right)^2+
\left(\bigotimes_{i=1}^n\hat{ S}_z^i\right)^2
$. The total magnetic moment $\langle m_z \rangle$ is just the sum of the 
individual magnetic moments of the constituents.} 
This degenerate ground state doublet is 
separated from the four excited states with $\langle S_T\rangle\approx3/2$ and 
$\langle m_z\rangle$ values of $\pm1/2$ and $\pm3/2$ 
by an energy of about $\frac32\mathcal{J}_{12}$. 
Employing this set of states allows an easy 
rationalization of the observed spectroscopic features: The two symmetric steps 
in the conductance arise due to excitations of the system from the ground 
states 
$|\Psi_{1,2}\rangle=|S_T,m_z\rangle=|1/2,\pm1/2\rangle$ to the excited states 
$|\Psi_{3,4}\rangle=|3/2,\pm1/2\rangle$ and 
$|\Psi_{5,6}\rangle=|3/2,\pm3/2\rangle$, induced by the tunneling electron. 
Note, that in these processes the total spin $S_T$ is not conserved. 
Furthermore, we can attribute the zero-bias peak to scattering processes of 
third and higher orders between the two degenerate ground states 
$|\Psi_{1}\rangle$ and $|\Psi_{2}\rangle$.

The description of the eigenstates with quantum numbers $S_T$ and 
$m_z$ allows an easy understanding of even complex spin systems, but 	
it fails to explain the different spectra of the individual parts of the 
dimer. Here, it is necessary to analyze the contributions to the 
states in the $|m_z^{(1)},m_z^{(2)}\rangle$ basis, for which the main 
contributions are illustrated in figure \ref{fig:S1-S12}c.
The two ground states have the most weight in $|+1/2,-1\rangle$ and 
$|-1/2,+1\rangle$, respectively. Thus, the second order transition matrix 
elements to the low-lying excited states (at an energy of about $6.5$~meV) are 
large, when the tunneling electron interacts with the $S^{(1)}=1/2$ spin. The 
reason is that this transition requires a change of $\Delta m_z=\pm1$ on the 
$S^{(1)}$ spin while the $m_z$ value of the $S^{(2)}$ spin remains unchanged. 
In contrast, these transitions are unfavorable for a tunneling electron 
that interacts with the $S^{(2)}$ spin. However, on this site transitions to 
the high-lying excited states are preferred because they end in states with 
unchanged $m_z^{(1)}$ and a difference between initial and final $m_z^{(2)}$ of 
$\pm1$.   

Interestingly, the coupling has an effect on the spin-1 states similar 
to an applied magnetic field along the $z$-axis. Without coupling, the ground 
state is an antisymmetric superposition of the $m_z=\pm1$ states, as discussed 
in section \ref{sec:CoH}, leading to a total magnetic moment $\langle 
m_z\rangle=0$. The Heisenberg coupling induces a duplication of states that
effectively separates the $m_z$ states and leads for the individual states of 
the $S=1$ subsystem to an effective magnetization (Figure \ref{fig:S1-S12}d). 
At $\mathcal{J}_{12}=5$~meV the absolute magnetization for each ground state 
has reached $\approx0.8$, where the difference to one stems mainly from 
some weight in $m_z=0$. 
Similarly, the 
coupling decreases the average magnetization of the $S=1/2$ subsystem, 
concomitant with the change of the total spin of the bipartite system 
approaching $S_T=1/2$. 

Both effects are a consequence of the emergence of quantum entanglement 
between the two spins which finally allows a description of the eigenstates in 
quantum numbers of the total spin $S_T$ and the total magnetic moment $m_z$. 
There are many different approaches to measure quantum entanglement 
\cite{Plenio07, Horodecki09, Hou11} that are based, for example, on the 
formation or distillation of entanglement \cite{Bennett96}, the entropy 
\cite{Schumacher95, Kim96, Vedral97}, the concurrence \cite{Hill97}, or the 
tangle \cite{Coffman00, Osborne06}. Here, we will restrict ourself to the 
''negativity`` $\mathcal{N}$ \cite{Peres96, Zyczkowski98, Vidal02}, which is 
the sum of the negative eigenvalues $\lambda_j$ of the partially, with respect 
to the subsystem $\Gamma_i$ of the $i$-th spin, transposed density matrix 
$\chi^{\Gamma_i}$:  
\begin{equation}
\mathcal{N}_i(\chi^{\Gamma_i})=\sum_j\frac{|\lambda_j|-\lambda_j}{2},
\label{eq:negativity}
\end{equation} 
with $\chi$ as the total density matrix of the full system. 
The negativity is an excellent measure of the non-separability for spin-$1/2$ 
and spin-$1$ composite quantum systems enabling the quantization of 
non-classical quantum correlations \cite{Horodecki97, Audenaert02}. In the 
dimer system discussed here, the negativity $\mathcal{N}$ increases steadily 
with the coupling $\mathcal{J}_{12}$, reaching $\mathcal{N}\approx0.27$ at  
$\mathcal{J}_{12}=5$~meV, close to the maximal possible entanglement in this 
system of $\mathcal{N}=1/3$ at $\mathcal{J}_{12}\rightarrow\infty$ (Figure 
\ref{fig:S1-S12}d).  

These quantum correlations are the origin for the zero-energy 
peak at the spin-1 system. Na\"ively, scattering between the two groundstates 
should be forbidden because it would require a spin-flip action on both 
subsystems when considering only the ground state components shown in figure 
\ref{fig:S1-S12}c. It is the entanglement that gives weights to other 
components and leads finally to the appearance of logarithmic zero-bias peaks, 
which will diminish at the spin-$1/2$ site and grow at the spin-$1$ site when 
the coupling strengths $J\rho_s^{(1)}$ and $J\rho_s^{(2)}$ of both subsystems 
are equal.

\begin{figure}[tbp]
\centering
\includegraphics[width=\textwidth]{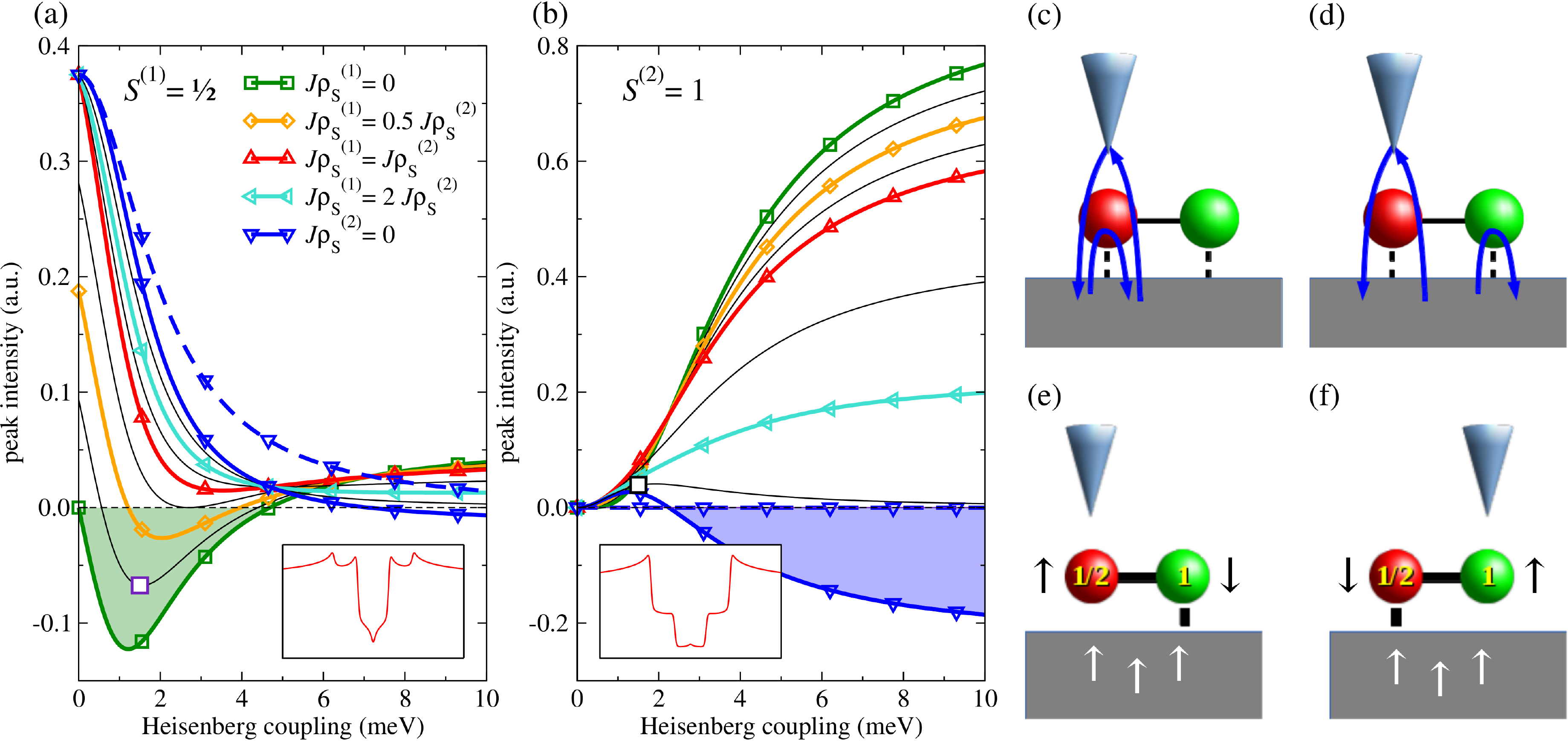}
	\caption{Zero bias peak intensities for a coupled 
dimer consisting of a $S^{(1)}=1/2$ and a $S^{(2)}=1$ spin ($D=-5$~meV, 
$E=1$~meV) with different
Heisenberg interactions $\mathcal{J}_{12}$ and coupling strengths $J\rho_s$ to 
the substrate. 
\textbf{(a)} Peak intensities when the differential conductance is measured on 
the $S^{(1)}=1/2$ spin. \textbf{(b)} Same as (a) when measured on the
$S^{(2)}=1$ spin. Dashed lines: Ising-like 
$\mathcal{J}_{12}=(0;0;J_z)$ coupling at $J\rho_s^{(2)}=0$. 
Insets in (a) and (b) illustrate the spectra at $\mathcal{J}_{12}=1.5$~meV 
(white squares), with $J\rho_s^{(1)}=-0.025$, and $J\rho_s^{(2)}=-0.1$ at 
$T=1$~K. The bias range is from $-15$ to $+15$~meV.
\textbf{(c+d)} In 
third order scattering processes that originate and end in the substrate and 
interact on the probed spin (c) or on any other spin of the coupled 
spin-system (d) have to be accounted for. \textbf{(e)} Schematic of the 
situation illustrated by the green-shaded area in (a): The $S^{(1)}=1/2$ is 
ferromagnetically coupled to the substrate electrons by superexchange via the 
$S^{(2)}=1$ spin, leading to a dip in the spectrum. \textbf{(f)} Same as (e) 
for 
the blue-shaded area in (b).}
	\label{fig:S1S12peak}
\end{figure}

Interestingly, this situation changes when $J\rho_s^{(1)}\neq J\rho_s^{(2)}$. 
Figure \ref{fig:S1S12peak} displays the peak intensities at the sites of the 
two spins for different ratios of coupling strength to the substrate electrons 
and antiferromagnetic Heisenberg interactions.
Surprisingly, when one of the $J\rho_s$ is small compared to the other, dips in 
the spectra can occur at certain Heisenberg coupling strengths (see insets in 
figure \ref{fig:S1S12peak}a and b). Note, that third order contributions can be 
significant even at negligible coupling to the substrate for the corresponding 
spin because of spin-spin interactions of electrons that originate and end in 
the substrate on other spins of the coupled system. 
When the spins are entangled, processes acting on all subsystems have to be 
accounted for and can interfere with each other. Figures \ref{fig:S1S12peak}c 
and \ref{fig:S1S12peak}d schematically illustrate the processes for a dimer: A 
tunneling electron scatters at the spin in proximity of the tip, leaving it in 
an intermediate state. The final state can now be reached either by a 
scattering event on this spin or on the other spin of the coupled system.

A dip-like reduction of the differential conductance at zero bias only occurs 
if one spin is loosely coupled to the substrate and a significant 
exchange interaction is established between the two subsystem. At low enough 
temperature, the states of the strongly coupled spin are antiferromagetically 
correlated with the substrate electrons (see section \ref{sec:Kondo}), while 
the 
weakly coupled spin is ferromagnetically correlated with the substrate bath via 
the Heisenberg interaction (Figure \ref{fig:S1S12peak}e and f). This is 
equivalent to a ferromagnetic Kondo effect, which has been recently proposed to 
emerge for a triple spin system \cite{Baruselli13}. Note, however, that this 
requires a Heisenberg exchange interaction between both spins. If the 
interaction is Ising-like, i.\,e.~couples only one direction of the magnetic 
moments, merely classical correlations occur, which are not sufficient to 
create the entanglement required to observe the Kondo effect for the spin-1 
system (see dashed lines in figure \ref{fig:S1S12peak}a and b). 
However, the results presented here are obtained in the weak coupling limit. 
While it is known that a single spin will become asymptotically free for
$T\rightarrow0$ when ferromagnetically coupled to an electron 
bath \cite{Anderson70, Voit95}, it is not clear what happens in the 
coupled structures discussed here, where one spin is ferromagnetically and the 
other antiferromagetically coupled to the substrate.

Note, that the two spins can be coupled also vertically by having one spin 
center adsorbed to the tip apex and the second one onto the substrate surface. 
In this situation, the antiferromagnetic exchange coupling has been found to be 
proportional to the tunneling coupling which depends exponentially on the 
distance between the two spins on tip and sample \cite{Yan15, Muenks17}. 
Building a similar structure as discussed in this section with a $S=1$ 
spin adsorbed on the sample and a strongly correlated half-integer spin on the 
tip lead to the observation of bias direction dependent step asymmetries at the 
energetically outer steps (blue dashed lines in Fig.~\ref{fig:S1-S12}d). These 
asymmetries enable to directly determine the correlation strength between the 
half-integer spin and the supporting electron bath \cite{Muenks17}.

\subsection{Fe-Co dimers on Cu$_2$N}
\label{sec:Fe-Co}

Dimer spin systems have been studied experimentally for different 
transition metal atoms on metallic and insulating surfaces. For 
Co atoms on Au(111), which show a pronounced Kondo effect with a 
$T_K\approx70$~K, the disappearance of the Kondo effect due to the formation 
of a non-magnetic singlet state was observed only when the two adatoms were 
strongly coupled by placing them on neighboring adsorption sites 
\cite{Chen99}. In a similar experiment using Cu(100), the exchange 
interaction could be varied concomitantly with a change of the spectroscopic 
signature \cite{Wahl07}. More recently, the coupling between two Kondo systems 
was established by attaching one Co atom to the apex of a Au tip and  
having the second Co atom adsorbed on a Au(111) surface \cite{Bork11} or between
metal–molecule complexes in a non-consummate molecular lattice on Au(111) 
\cite{Esat16}. Furthermore, spin-spin interactions have been studied between 
Fe-Fe \cite{Bryant13}, Co-Co \cite{Spinelli15}, and Co-Fe \cite{Otte09} 
adatoms adsorbed on Cu$_2$N. 

\begin{figure}[tbp]
\centering
\includegraphics[width=0.95\textwidth]{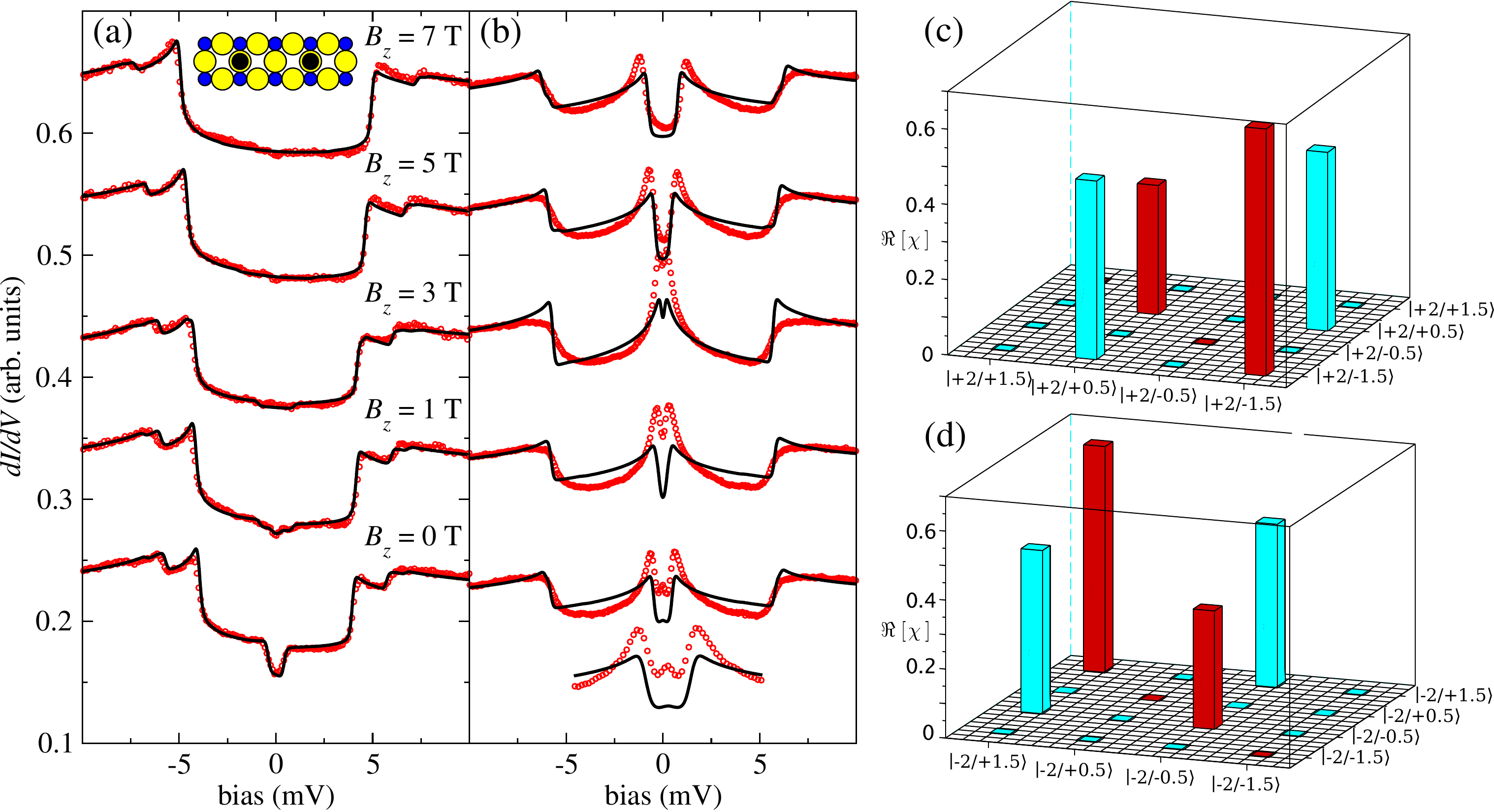}
	\caption{Tunneling spectra measured on a weakly coupled Co-Fe 
dimer on Cu$_2$N. \textbf{(a+b)} Experimental data from reference 
\cite{Otte09} obtained with the tip placed on top of the Fe (a) and Co (b) atom 
of the bipartite system at different fields applied along the main anisotropy 
axis of the Fe measured at $T=550$~mK (red circles). 
The simulations (black lines) are performed using the following parameters 
for the Fe atom: $D=-1.53$~meV, $E=0.31$~meV, $g=2.11$, $J\rho_S=-0.085$, and 
$U=0.35$; for the Co atom: $D=2.7$~meV, $E=0.5$~meV, $g=2.16$, 
$J\rho_S=-0.25$, and $U=0$. Note, that the main anisotropy axis for the Fe is 
along the N-rows, while for the Co it is along the vacancy rows. The isotropic 
Heisenberg exchange coupling between 
both atom was fixed to $\mathcal{J}_{12}=0.16$~meV. Curves are 
vertically offset for clarity. The inset in (a) shows the 
adsorption site of the 3d atoms (black circle) on the Cu$_2$N (Cu 
yellow, N blue circle). The lowest curves in (b) are 
a zoom of the zero-field data measured on the Co atom. 
\textbf{(c + d)} 
Graphical representation of the real part of the reduced density matrices of 
the two lowest, at zero-field degenerated, ground states. The displayed parts 
contain $>95\%$ of the state weights. The labels correspond to the
$| m_{z}^{\rm Fe}, m_{z}^{\rm Co}\rangle$ values.}
	\label{fig:Co-Fe}
\end{figure}

We will revisit the latter Co-Fe dimer system and 
compare it to simulations done in the third order scattering model 
presented in reference \cite{Ternes15}. In the experiment the two spins are 
separated by $0.72$~nm and only weakly coupled via the Cu$_2$N surface (see 
inset of figure \ref{fig:Co-Fe}a), but the measurements of the differential 
conductance on both atoms reveal a change in the spectra compared to the 
individual atoms as discussed in section \ref{sec:Ani-Fe-Mn} and 
\ref{sec:Kondo} 
and shown in figure \ref{fig:Co-Fe}a and b.

Simulating the data by using parameters almost identical to those employed for 
uncoupled single Co and Fe atoms, and an additional isotropic Heisenberg 
interaction of $\mathcal{J}_{12}=0.16$~meV, renders the spectra obtained for 
the 
Fe atom almost perfectly, but only the main features and not all details are 
reproduced for the spectra measured on the Co atom. This stems from the fact, 
that our model only accounts for third order scattering contributions, 
neglecting higher order effects and reaches its limits in systems in the 
strong-coupling Kondo regime (see section \ref{sec:Kondo-limit}). 

At zero field we observe on the Fe atom a reduction of the intensity of the 
low-energy conductance steps compared to the uncoupled atom (Figure 
\ref{fig:spec-Fe-Mn}a). Furthermore, these steps diminish when a field 
along the main anisotropy axis of the Fe atom ($z$-axis) is 
applied, much more quickly than for the uncoupled Fe atom. Nevertheless, the 
overall spectral form is only weakly influenced by the coupling to the Co 
atom. The coupling is also not strong enough to produce a zero-energy feature at
the high spin in this dimer, i.\,e.~the Fe atom, as we would expect from the 
discussion of the 
coupled $S^{(1)}=1/2$ and $S^{(2)}=1$ system (Section \ref{sec:S1S12}). This 
can be 
understood by looking at the density matrices of the two degenerate 
groundstates of the combined system (Figure \ref{fig:Co-Fe}c and d). The 
two groundstates contain only contributions with weights at high $m_z=\pm2$ 
values, inhibiting any scattering between them. Note that a description via
quantum numbers of the total system is not appropriate here due to the small 
coupling strength. Only when the Heisenberg coupling is comparable to the 
anisotropy energies, i.\,e.~at $\mathcal{J}_{12}\gtrsim3$~meV, these quantum 
numbers become a good description of the total system and the emergence of a 
zero-energy peak at the Fe atom is expected, similar to the dimer with 
$S^{(1)}=1/2$ and $S^{(2)}=1$.

Compared to the spectrum measured on the Fe atom, the zero-field spectrum on 
the Co atom is strongly affected by the creation of this bipartite system.
Similar as for the $S^{(1)}=1/2$ and $S^{(2)}=1$ dimer, we observe in the 
experimental data, as well as in the simulation, conductance steps and peaks at 
an energy corresponding to the first excitation energy of the Fe atom. 
Additionally, the remainder of the zero-bias Kondo peak is clearly 
visible, but similar as in the example of the last section, the coupling to 
the Fe atom has already strongly reduced its intensity. At an applied field 
along the easy axes of both atoms, the energetic positions of the two peaks at 
low bias move towards zero and at a critical field of $B_{c}=(g_{\rm 
Co}\mathcal{J}_{12})/(g_0\mu_B)\approx2.6$~T, a 
novel Kondo peak at zero bias emerges. At even higher fields the 
spectra resemble those of an uncoupled Co atom, as shown in Figure 
\ref{fig:Co-CuN}, 
where the presence of the Fe atom has reduced the magnetic field to 
$B_{\rm eff}=B_{\rm ext}-B_c$.

Overall, it is surprising how well the perturbative model can reproduce the 
experimental data for this bipartite systems. Note, that not only does it 
enable us to determine the coupling strength and sign between interacting 
spins, but it also allows us to detect additional non-collinear couplings such 
as the Dzyaloshinskii-Moriya interaction, as it has been recently shown 
\cite{Khajetoorians16}.

\subsection{Co-Co dimers on Cu$_2$N}
\label{sec:Co-Co}

\begin{figure}
\centering
\includegraphics[width=0.75\textwidth]{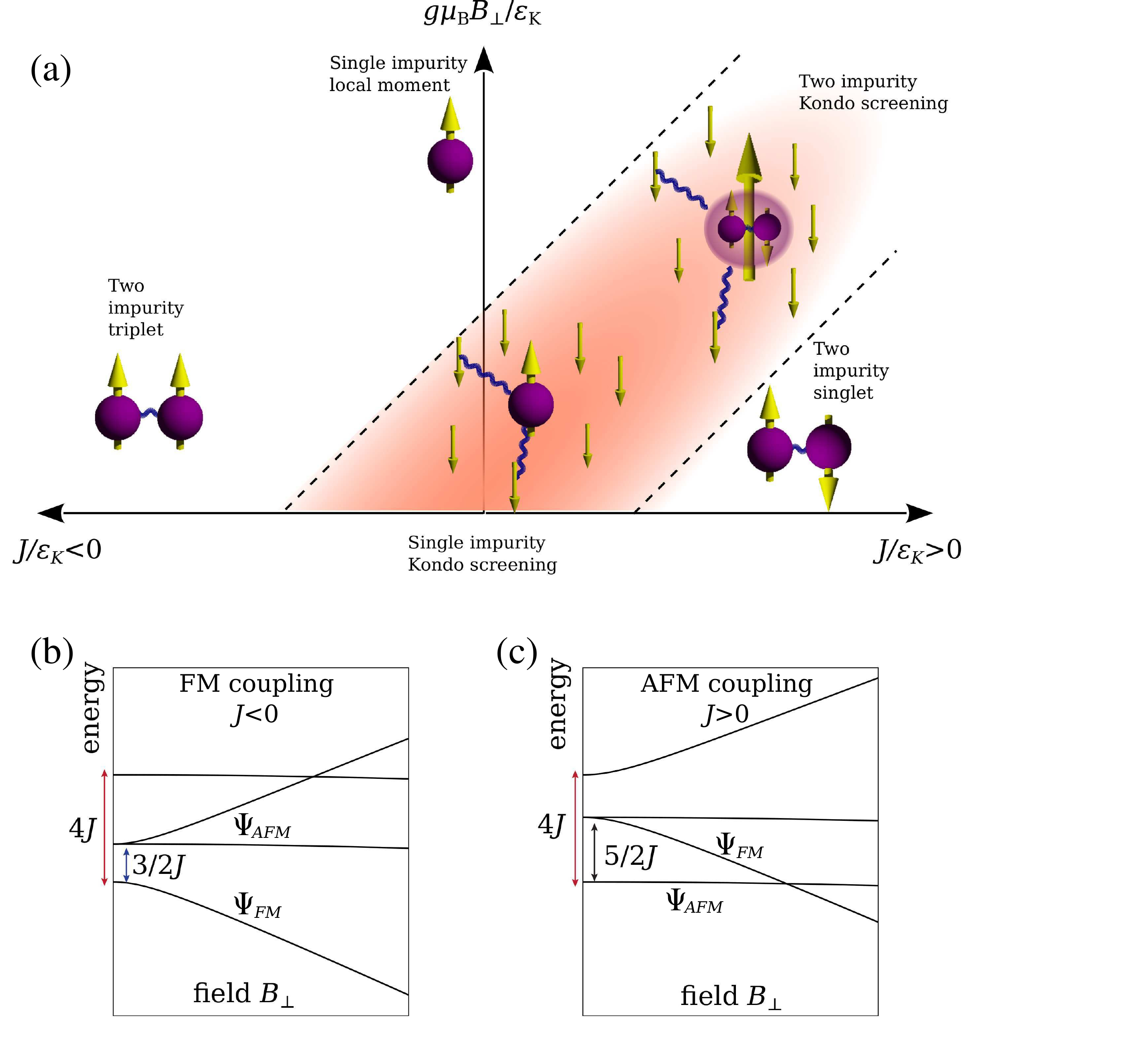}
	\caption{Phase diagram of the two-impurity Kondo problem. 
\textbf{(a)} Schematic phase diagram of two coupled Kondo-screened spins 
with varying interaction strength $\mathcal{J}_{12}$ and external field 
$B_\bot$ transverse to the main anisotropy axis of the individual spins with 
$S=3/2$ and easy plane anisotropy $D>0$. When $|\mathcal{J}_{12}|$ is small 
compared to the characteristic Kondo energy $\Gamma_K$, at $B = 0$ the two 
spins are independently screened by the substrate electrons, while for 
$|\mathcal{J}_{12}|\gg \Gamma_K$ a non-magnetic singlet or a high-spin triplet 
state 
form. For antiferromagnetic coupling, i.\,e.\ $\mathcal{J}_{12}>0$, an applied 
magnetic field can lead to the formation of a new, combined correlated state in 
which both spins are screened. \textbf{(b + c)} Energy versus transverse 
magnetic field $B_{\bot}$ of the four lowest energy states for 
ferromagnetic 
($\mathcal{J}_{12} < 0$) and antiferromagnetic ($\mathcal{J}_{12}>0$) 
coupling, respectively. In the ferromagnetic case (a) the groundstate 
$\psi_{FM}$ does not change with applied field, while in the antiferromagnetic 
case, a state crossing between $\psi_{FM}$ and $\psi_{AFM}$ occurs at a 
critical field $B_c$. Figure adapted from reference \cite{Spinelli15}.}
\label{fig:Co-Co}
\end{figure}

We now turn to the case where two Co atoms with $S=3/2$ are 
coupled via Heisenberg interactions on the Cu$_2$N substrate 
\cite{Spinelli15}. This situation is in particular interesting, because 
at low enough temperature both spin sites form independently a correlated Kondo 
state with the substrate electrons as discussed in section \ref{sec:Kondo}. We 
have seen that the characteristic correlation energy of the Kondo state is 
$\Gamma_K=k_BT_K\approx 0.2$~meV (see section \ref{sec:Kondo-split}). This 
energy scale is of the same order as the Heisenberg exchange coupling strength 
in the previously discussed Fe-Co dimers. 
As illustrated in figure \ref{fig:Co-Co}a, the competition between these two 
effects in combination with an external magnetic field embodies rich physics 
ranging from a correlated singlet or triplet state to complex Kondo states and 
has been of considerable theoretical interest since decades 
\cite{Jayaprakash81, Jones87, Jones88, Jones89, Silva96, Simon05, 
Silva06, Zitko10, Jabben12, Mitchell12}. 
Despite several studies on coupled quantum-dots \cite{Craig04, Baines12} and 
atoms \cite{Bork11, Prueser14, Esat16}, so far few experimental observations 
have been reported of the regime where both interactions have similar strengths 
and are thereby in direct competition with each other.

When the coupling $\mathcal{J}_{12}$ is negligible compared to $\Gamma_K$ and 
the magnetic field is zero, then the two spins are independently Kondo 
screened. 
An applied magnetic field which is strong enough will destroy the correlations 
between the localized spin and the bulk electrons as discussed in detail in 
section \ref{sec:Kondo} leading to a situation where each spin acts like a free 
local magnetic moment.

The situation changes when $\mathcal{J}_{12}$ is at similar order as $\Gamma_K$.
For either sign of $\mathcal{J}_{12}$, the four lowest ground states of the 
combined spins govern now the behavior of the dimer. The zero-field energy 
difference between the lowest and the highest state of the quartet is thereby 
$4\mathcal{J}_{12}$ (see figure \ref{fig:Co-Co}b and \ref{fig:Co-Co}c). In the 
limit of $|\mathcal{J}_{12}|$ being much smaller than the magnetic anisotropy 
$|D|$ of the individual spins, the difference between the ground state and the 
first excited state is either $5/2\mathcal{J}_{12}$ or $3/2\mathcal{J}_{12}$ 
for anti-ferromagnetic (AFM) or ferromagnetic (FM) coupling, respectively. This 
different behavior enables even at zero-field to distinguish clearly from 
differential conductance measurements between AFM and FM coupled dimers 
\cite{Spinelli15}.

In the case of FM coupling, the ground state can be written in the $m_z$ base 
of the two spin sites as 
$\Psi_{\rm FM}=\frac{1}{\sqrt{2}}\left(\left|+\frac12,
-\frac12\right\rangle+\left|-\frac12 ,
+\frac12\right\rangle\right)$, i.\,e.\ as the symmetric high-spin, low-magnetic 
moment triplet state. Any applied external magnetic field will polarize both 
spins moving the system closer to the free local magnetic moment regime as 
can be seen in figure~\ref{fig:Co-Co}b.

More interesting is the case where both spins are AFM coupled. The ground state 
in the $m_z$ base of the two spin sites is the antisymmetric low-spin singlet 
state $\Psi_{\rm AFM}=\frac{1}{\sqrt{2}}\left(\left|+\frac12,
-\frac12\right\rangle-\left|-\frac12 ,
+\frac12\right\rangle\right)$. This singlet state will compete with the many 
electron Kondo singlet between the \emph{individual} spins and the substrate 
electrons. If the coupling $\mathcal{J}_{12}$ is strong enough, the 
dimer-singlet becomes the energetically more favorable ground state and 
correlations with the substrate electrons concomitant with the Kondo peak 
disappear \cite{Spinelli15}. However, as the state diagram in figure 
\ref{fig:Co-Co}c suggests, a magnetic field orthogonal to the main anisotropy 
axis will decrease the energy difference between the ground state singlet and 
the lowest triplet state until the crossing field $B_c = (13/8)\times 
\mathcal{J}_{12}/(g\mu_B B_{\bot})$ is reached, at which the two states become 
degenerate. At this point the two degenerated states $\psi_{\rm AFM}$ and 
$\psi_{\rm FM}$ form the basis for the emergence of a new Kondo state in which 
the substrate electrons are correlated with the \emph{combined} state of the 
dimer \cite{Spinelli15}.

\subsection{Entanglement and the zero-energy peak in spin chains}

Chains of exchange coupled spins have been of interest for studying fundamental 
questions since the early days of quantum mechanics, dating back to the 
exact solution found by H.~Bethe \cite{Bethe31} for the infinite spin-$1/2$ 
chain. Higher dimensional  $\mbox{(anti-)ferromagetically}$ coupled spin 
lattices have led to the development of spin wave models \cite{Anderson52, 
Kubo52}, which are the basis of descriptions of collective bosonic 
excitations, so called magnons, which have been observed in STM measurements, 
for example, on thin Co films on Cu(111) \cite{Balashov06}.

In one-dimensional spin-chains there is a peculiar difference between 
half-integer and integer spins: While infinite half-integer spin chains become 
gap-less \cite{Lieb61}, the energy difference between ground and first excited 
state in integer spins will always be finite, leading to the so called Haldane 
gap \cite{Haldane83, Affleck89}. Interestingly, as long as the magnetic 
anisotropy is small compared to the next-neighbor Heisenberg coupling strength, 
edge-states are expected to appear in finite integer spin-chains of odd lengths 
\cite{Delgado13}. These half-spin degeneracies should reveal themselves by 
symmetric zero-energy Kondo peaks at the end of the chain together with a 
reduced lifetime. Very recently, the lifetime reduction at the ends of an 
antiferromagetically coupled 3-spin Fe chain on Cu$_2$N has been observed 
\cite{Yan15}. Nevertheless, due to the relatively high magnetic anisotropy and 
low next-neighbor coupling strength, the formation of a 
zero-bias peak at the end was impeded. 


Here, we want to explore and model spin-chains starting with the $S=1$ and 
$S=1/2$ dimer of section \ref{sec:Coupled_seystems} where we assume an 
identical coupling $J\rho_0$ of the individual spins with the substrate and, 
for simplicity, a negligible magnetic anisotropy of the $S=1$ spin. An 
antiferromagnetic Heisenberg interaction results then in a zero-bias peak 
mainly at the $S=1$ site similar to the spectra shown in figure 
\ref{fig:S1-S12}. To stay in the interesting $S^T=1/2$ regime,
we expand the dimer with additional $S=1$ sites on the high-spin side, creating 
chains like $(1/2)\leftrightarrow(1)\leftrightarrow(1)\ldots$. Now the 
question arises, where can we find the zero-bias peak for such chains and how 
will it's intensity scale with the different parameters of the chain?

\begin{figure}[tbp]
\centering
\includegraphics[width=0.8\textwidth]{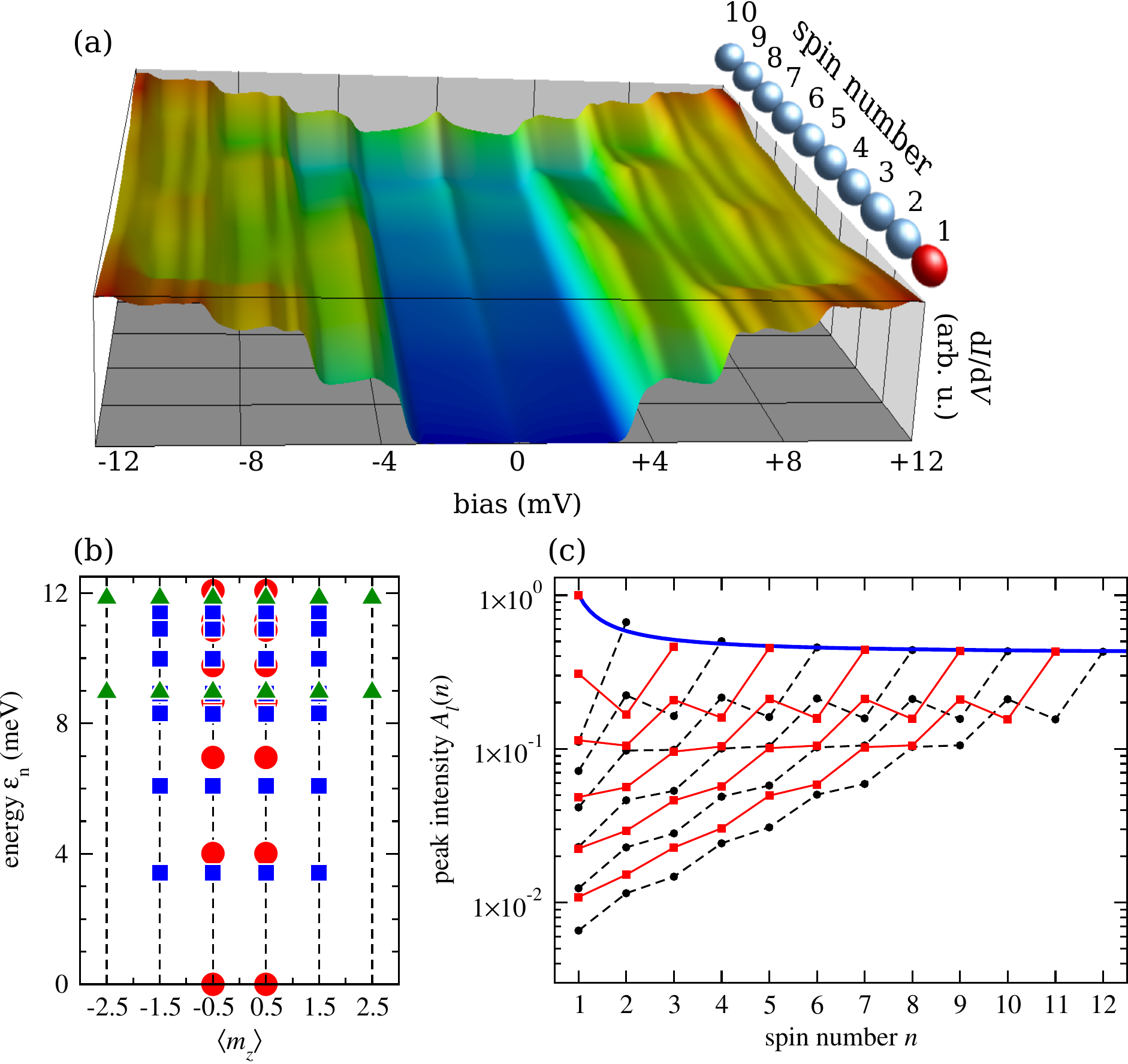}
	\caption{The zero-bias Kondo peak in spin chains without 
magnetic anisotropy. \textbf{(a)} 3D plot of simulated $dI/dV$ spectra along a 
10-spin chain with $S^{(1)}=1/2$ (red sphere), $S^{(2)-(10)}=1$ (gray spheres), 
$J\rho_s=-0.1$ on all spins, and a Heisenberg coupling 
between next neighbors of $\mathcal{J}_{i,i+1}=5$~meV. \textbf{(b)} The energy 
of the lowest 56 eigenstates of the 10-spin chain in the giant spin basis 
$|S_T,m_z\rangle$. Red circles denote states with $S^T=1/2$, blue squares with
$S^T=3/2$, and green triangles with $S^T=5/2$, respectively. \textbf{(c)} 
Zero-bias peak intensity $A_l(n)$ for different chain lengths $l$ with 
$S^{(1)}=1/2$ and all other $S^{(n)}=1$ relative to the peak intensity of a 
single $S=1/2$ spin. Black circles and red squares mark the intensity 
$A_l(n)$ on the $n$-th spin for even and odd chain lengths, respectively. The 
thick blue line is a fit of the peak intensity of the last spin in the chain 
against the chain length $l$.}
\label{fig:chain}
\end{figure}

Figure \ref{fig:chain}a shows the differential conductance 
simulated for the different sites of an 11-spin chain where the first site is
a spin-$1/2$ and all other sites are spin-$1$ without any magnetic anisotropy. 
The coupling between neighboring spins is $\mathcal{J}_{i,i+1}=5$~meV and the 
coupling to the substrate is $J\rho_s=-0.1$ for all spins.
Surprisingly, the zero-bias peak is strongest at the $S=1$ end of 
the chain, extends spatially, and decays towards the $S=1/2$ beginning of the 
chain. 
Furthermore, we observe many step-like inelastic excitations at increasing
energy, which reveal rich features. The probabilities for second and 
third order scattering processes, i.\,e.\ the step heights and the additional 
peak-like structures, oscillate along the chain remarkably differently for 
the various excitation energies. This multitude of possible transitions stems 
from the enormous number of low lying excitations in this many-spin system. 
Figure \ref{fig:chain}b shows the 56 lowest eigenstates with energies 
$\varepsilon_n\leq13$~meV. The groundstate is a doublet with total spin 
$S^T=1/2$ 
and is separated by about $3.4$~meV from the first excited states, a quadruplet 
with $S^T=3/2$. However, there are many eigenstates in the chain that result in 
a total spin of $1/2$ or $3/2$, but with different contributions and weights of 
the individual spins. This is reflected in the oscillatory behavior of the
transition probabilities and stems from the boundaries at the ends of 
the chain and the chain asymmetry, leading to a complex standing wave pattern 
for the coupling to the magnons of the chain \cite{Gauyacq11, Gauyacq11a}.

The amplitude of the zero-energy peak decays 
approximately exponentially along the chains as follows: 
\begin{equation}
A_l(n)\approx A_l(l)\exp\left(\frac{l-n}{\lambda_l}\right).
\label{eq:chain_noD}
\end{equation} 
Here, $A_l(n)$ is the peak amplitude measured at the $n$-th spin of a chain 
with length $l$, and $\lambda_l$ is the decay length, which 
approaches $\lim_{l\rightarrow\infty} \lambda_l = 3$ for long chains. Odd-even 
fluctuations of the peak amplitude are due to some ferromagnetic 
exchange interactions of the spins in the chain with the substrate 
electron bath, similar to what we observed for the dimer (figure 
\ref{fig:S1S12peak}). Interestingly, the maximal zero-energy peak amplitude at 
the last spin of the chains diminishes for longer chains only slowly with 
$A_l(l)\approx0.41 A_1\times l/(l-0.59)$, and thus approaches a final value 
of $\approx 0.41A_1$ for 
long spin chains, with $A_1$ as the peak amplitude of a single $S=1/2$ 
(see figure \ref{fig:chain}c). 

This result is very remarkable. It means that the presence of a spin-$1/2$ at 
the beginning of an ideal, arbitrarily long, $S=1$ chain determines the 
appearance of the zero-bias peak at the opposite end of the chain. Note that 
removing the spin-$1/2$ in chains with \emph{odd} length results in an 
\emph{even} spin-1 chain with a non-magnetic groundstate of $S_T=0$ and a 
finite Haldane gap to the excited states \cite{Haldane83}. Obviously, such a 
spin system cannot have any zero-bias peak. 

Removing the spin-$1/2$ from  chains with \emph{even} length, makes the chain 
an \emph{odd} chain that has a three-fold degenerate groundstate of 
$S^T=1$ and topologically protected edge states which are expected to show a 
weak zero-energy peak at both ends of the chain \cite{Delgado13}. Here too, 
the addition of a single spin-$1/2$ completely changes the properties.

\begin{figure}[tbp]
\centering
\includegraphics[width=\textwidth]{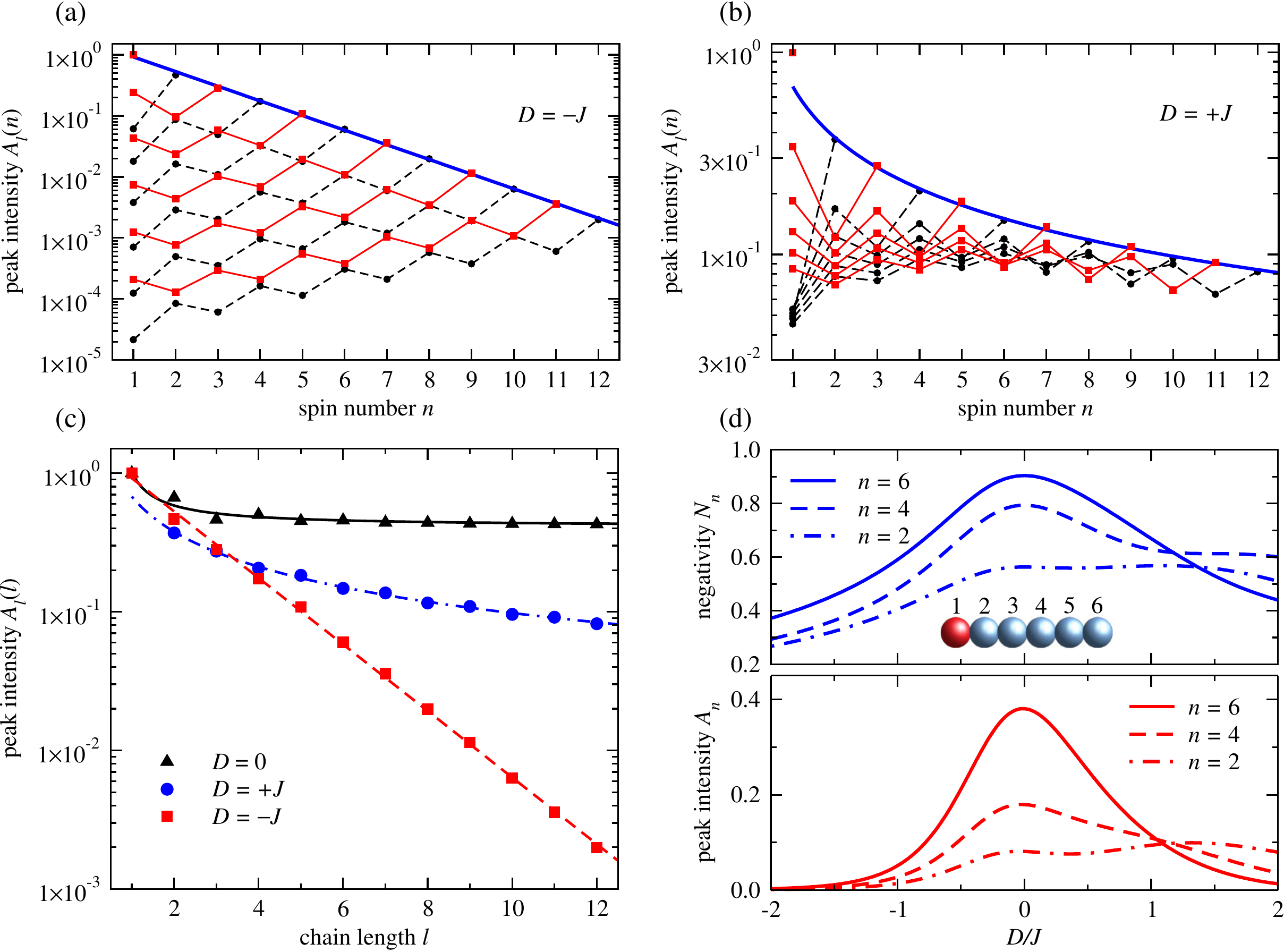}
	\caption{The zero-bias peak and entanglement in spin chains with 
different magnetic anisotropy. \textbf{(a+b)} Zero-bias peak intensity 
$A_n$ for different chain lengths $l$ with $S^{(1)}=1/2$ and all other 
$S^{(n)}=1$ relative to the peak intensity of a single S = 1/2 spin.
In (a) easy axis anisotropy with $D=-\mathcal{J}$, and in (b) easy plane 
anisotropy with $D=+\mathcal{J}$ at the $S=1$ sites is assumed. Black circles 
and red squares mark the intensity $A_l(n)$ on the $n$-th spin for even
and odd chain lengths, respectively. The thick blue lines are fits of the peak 
intensity of the last spin in the chain against the chain length $l$.
\textbf{(c)} Comparison of the decay of the zero-bias peak intensity with the 
chain-length $l$ at the last spin $A_l$ of chains with different anisotropy 
values. The full, dashed-dotted, and dashed line are fits
with the corresponding decay function. \textbf{(d)}~Comparison of the 
negativity $\mathcal{N}_n$ (top panel) and peak intensity $A_n$ (bottom 
panel) of a 6-spin chain at different $D/\mathcal{J}$ ratios.}
	\label{fig:chain_entangle}
\end{figure}

Clearly, the ideal case as discussed above cannot be realized in real 
experiments. The adsorption of the spin-$1$ on a surface will break the 
symmetry inducing some magnetic anisotropy as we have discussed in section 
\ref{sec:CoH}. Assuming an anisotropy of 
the same strength as the Heisenberg interaction between neighboring sites 
leads to much more quickly decaying peak-intensities (figure 
\ref{fig:chain_entangle}a and b).
When the spin-$1$ sites have easy-axis ani\-sotropy ($D<0$) the 
peak-intensities 
drop very quickly and are only about $0.001A_1$ for a $l=12$ chain. 
For the two groundstates the easy-axis anisotropy favors the high $m_z=\pm1$ 
values at each site, and thus is effectively reducing 
any scattering between the groundstates. 
For easy-plane anisotropy ($D>0$), which favors $m_z=0$ values 
at each site, the peak-intensities do not drop as quickly, but the strongest 
peak occurs now no longer at the end of the chain, but is rather smeared out at 
approximately the center of the chain. 

Interestingly, for all three cases discussed, the peak-intensities of the last 
spin of chains with different lengths $l$ follow quite simple algebraic 
relations (see figure \ref{fig:chain_entangle}c):
\begin{eqnarray}
A_l(l)\approx A_1\frac{0.41 l}{l-0.59}, &\quad\mbox{ if }D=0,\nonumber\\
A_l(l)\approx0.68 A_1\times l^{-0.84}, &\quad\mbox{ if 
}D=+\mathcal{J},\nonumber\\
A_l(l)\approx1.60 A_1\times \exp(-0.55 l), &\quad\mbox{ if }D=-\mathcal{J}.
\label{equ:chain_decay}
\end{eqnarray}
Except for $D=0$, all peak intensities decay either with an exponential law or 
an inverse power law. Note, that also for $|D/\mathcal{J}|\neq 1$ decay laws 
exist, which predict that the peak at the end of the 
chain disappears for $l\rightarrow\infty$.

We can understand this behavior if we look at 
the quantum mechanical entanglement inside the chain. For vanishing magnetic 
anisotropy the system is maximally entangled, while any anisotropy reduces the 
chain entanglement. To measure the entanglement inside the chain we calculate 
the negativity $\mathcal{N}_n$  of different sites in the chain with respect to 
the total chain using equation \ref{eq:negativity}. Figure 
\ref{fig:chain_entangle}d shows the results for a small chain of length $l=6$ 
and different relative magnetic anisotropies $D/\mathcal{J}$. For moderately 
small easy-plane anisotropy or for easy-axis anisotropy, the entanglement of 
the total chain is strongest with the last spin, but decays with the 
strength of the magnetic anisotropy. For larger easy-plane anisotropy 
$D/\mathcal{J}\gtrsim1$ the entanglement for spins closer to the beginning of 
the chain grows, and finally becomes larger than the entanglement with the 
chain 
end. Thus, quantum-mechanically, the chain separates in a moderately 
entangled short chain that consists of the spin-$1/2$ and a weakly 
entangled end chain.

Surprisingly, the entanglement measure correlates well with the calculated 
zero-bias peak intensities and positions. Thus, the observation of the Kondo 
peak at such spin chains is a direct measure of the quantum entanglement inside 
these chains. Recently, it was possible to verify this behavior in 
chains constructed atom-by-atom on Cu$_2$N with one Fe atom ($S=2$) at the 
beginning and $2n+1$ Mn atoms ($S=5/2$) \cite{Choi15, Choi16}. The observations 
for these high-spin chains were similar to the calculations presented here. 
Chains up to 
FeMn$_9$ showed a spatially localized Kondo peak at the end of the chain. Due 
to 
the high spin of the chain constituents, odd chains of the form FeMn$_{2n}$ 
showed no zero-bias peak because the groundstate of these chains is 
close to $S^T=2$, the 
spin of the uncoupled Fe atom on Cu$_2$N.


\section{Summary and outlook}
\label{sec:summary}

In this manuscript I have outlined how scanning tunneling spectroscopic 
methods can reveal the rich variety of effects individual and coupled quantum 
spins show when adsorbed on a supporting surface. The local environment 
influences the spins via crystal-field and spin-orbit coupling leading to 
magneto-crystalline anisotropy. Furthermore, we discussed how the exchange 
coupling to substrate electrons can drive the total quantum state into the 
highly correlated Kondo screening phase and how the very same interactions 
lead to renormalization effects influencing the anisotropy and the gyromagnetic 
factor. Not discussed in detail, but these interactions which couple 
the spin to the dissipative bath of the environment, are also of 
crucial importance for the state lifetime, coherence time, and the 
\emph{einselection} of quantum states which generate classical 
behavior \cite{Zurek03, Delgado15}. 

Additionally, we inspected coupled spin systems and their ability to form 
entanglement. I expect that studying these non-classical correlations 
and their imprint in transport measurements, as for example seen in 
Kondo anomalies, will be of great importance for getting a deeper understanding 
of modern complex materials.

We have also seen that applying straight forward perturbation theory to such 
quantum spin systems allows to describe experimentally measured differential 
conductance spectra with very high accuracy. This enables one to
obtain a profound understanding of the physical processes on play and to 
separate single- as well as many-electron effects, where perturbation theory 
and experiment must divert from each other. 

Only during the last few years, both experiment and theory of individual and 
coupled spin systems have made tremendous progress. The versatility of 
low-temperature scanning tunneling measurements led me to believe that we 
should 
expect a multitude of exciting new experiments for the future, which will 
further deepening our fundamental knowledge on quantum systems in general and, 
in particular, quantum magnetism.

\section*{Acknowledgments}

I like to thank the many people which have supported me for this 
manuscript: Kirsten von Bergmann, Oleg Brovko, Deung-Jang  Choi, Katharina 
Franke, Andreas Heinrich, Tobias Herden, Cyrus Hirjibehedin, Peter Jacobson, 
Barbara Jones, Steffen Kahle, Klaus Kern, Alexander Khajetoorians, Gennadi 
Laskin, Sebastian Loth, Chris Lutz, Matthias Muenks, Stephan Rauschenbach, Anna 
Spinelli, Alexander Otte,  Valerie Stepanyuk, Charl\'{e}ne Tonnoir, and Peter 
Wahl.
I am also very grateful for the financial support by the Deutsche 
Forschungsgemeinschaft within the Sonderforschungsbereich at the University of 
Konstanz ”Controlled Nanosystems: Interaction and Interfacing to the Macroscale“ 
(SFB 767).

\bibliographystyle{unsrt}


\end{document}